\documentclass[longauth]{aa}  
\usepackage{array}
\usepackage{graphicx}
\usepackage{txfonts}
\usepackage{natbib,twoopt}
\usepackage{booktabs}
\usepackage{graphicx}
\usepackage{multirow}
\usepackage{subcaption}
\usepackage[colorlinks=true,linkcolor=blue,citecolor=blue,urlcolor=blue]{hyperref}

\usepackage{tikz}
\usetikzlibrary{arrows.meta, positioning, shapes.geometric}
\usepackage{xcolor}
\usepackage{soul}
\usepackage{amsmath}
\usepackage[T1]{fontenc}

\setstcolor{magenta}

\begin{document} 

   \title{The evolutionary history of ultra-compact accreting binaries}

   \subtitle{II. \textit{HST} spectroscopy of two eclipsing AM CVn stars}
   
    \author{
        W.~Yu\inst{1}
        \and A.~F.~Pala\inst{2}
        \and T.~Kupfer\inst{1,3}
        \and B.~T.~G\"{a}nsicke\inst{4}
        \and T.~L.~S.~Wong\inst{5}
        \and D.~Belloni\inst{6}
        \and A.~J.~Brown\inst{1}
        \and D.~Koester\inst{7}
        \and M.~J.~Green\inst{8,9}
        \and J.~van Roestel\inst{10}
        \and M.~R.~Schreiber\inst{11}
        \and W.~A.~Huff\inst{12}
        \and E.~O.~Waagen\inst{13}
        \and J.-L.~Gonz\'{a}lez-Carballo\inst{14}
        \and S.~Bolzoni\inst{15}
        \and C.~Botana-Alb\'{a}\inst{16}
        \and F.~J.~Calvo Fern\'{a}ndez\inst{17}
        \and F.~Campos-Cucarella\inst{18}
        \and R.~Castillo-Garc\'{i}a\inst{19}
        \and D.~Cejudo-Fern\'{a}ndez\inst{20}
        \and D.~De Martino\inst{21}
        \and S.~Dvorak\inst{22}
        \and J.~M.~Fern\'{a}ndez-And\'{u}jar\inst{23}
        \and E.~Fern\'{a}ndez-Ma\~{n}anes\inst{24}
        \and C.~Galdies\inst{25}
        \and F.~Garc\'{i}a-Cuesta\inst{26}
        \and R.~Gonz\'{a}lez Farf\'{a}n\inst{27}
        \and J.-F.~Gout\inst{13}
        \and C.~Knigge\inst{28}
        \and J.-L.~Martin-Velasco\inst{29}
        \and M.~Morales-Aimar\inst{30}
        \and R.~Naves-Nogues\inst{31}
        \and A.~Oksanen\inst{32,13}
        \and R.~Poggiani\inst{33}
        \and G.~Poyner\inst{34}
        \and G.~Ramsay\inst{35}
        \and E.~Reina-Lorenz\inst{36}
        \and P.~Rodr\'{i}guez-Gil\inst{37,38}
        \and F.~D.~Romanov\inst{13}
        \and T.~Rutherford\inst{39,40,41,13}
        \and J.-L.~Salto-Gonz\'{a}lez\inst{42}
        \and J.~Shears\inst{43}
        \and E.~M.~Sion\inst{44}
        \and J.~Solaz\inst{45}
        \and D.~Steeghs\inst{4}
        \and R.~Stephens\inst{13}
        \and R.~Stubbings\inst{46}
        \and P.~Szkody\inst{47}
        \and O.~Toloza\inst{11}
        \and T.~Tordai\inst{48}
        \and G.~Tovmassian\inst{49}
        \and L.~Tremosa-Espasa\inst{50}
        \and B.~Young\inst{51}
     }
    
    \institute{
        Hamburg Observatory, University of Hamburg, Gojenbergsweg 112, 21029 Hamburg, Germany
        \\
        \email{weitian.yu@uni-hamburg.de}
        \and
        European Southern Observatory, Karl-Schwarzschild-Strasse 2, 85748 Garching bei M\"{u}nchen, Germany
        \and
        Department of Physics and Astronomy, Texas Tech University, 2500 Broadway, Lubbock, TX 79409, USA
        \and
        Department of Physics, University of Warwick, Coventry, CV4 7AL, UK
        \and
        The Observatories of the Carnegie Institution for Science, Pasadena, CA 91101, USA
        \and
        International Centre of Supernovae (ICESUN), Yunnan Key Laboratory of Supernova Research, Yunnan Observatories, Chinese Academy of Sciences (CAS), Kunming 650216, China
        \and
        Institut f\"{u}r Theoretische Physik und Astrophysik, University of Kiel, 24098 Kiel, Germany
        \and
        Homer L. Dodge Department of Physics and Astronomy, University of Oklahoma, 440 W. Brooks Street, Norman, OK 73019, USA
        \and
        JILA, University of Colorado and National Institute of Standards and Technology, 440 UCB, Boulder, CO 80309-0440, USA
        \and
        Anton Pannekoek Institute for Astronomy, University of Amsterdam, 1090 GE Amsterdam, The Netherlands
        \and
        Departamento de F\'isica, Universidad T\'ecnica Federico Santa Mar\'ia, Av. Espa\~na 1680, Valpara\'iso, Chile
        \and
        School of Physics, Georgia Institute of Technology, Atlanta, Georgia 30332, USA
        \and
        American Association of Variable Star Observers (AAVSO), 1770 Massachusetts Avenue, Suite 121, Cambridge, MA 02140, USA
        \and
        Observadores de Supernovas (ObSN), Observatorio Cerro del Viento, MPC I84, Pl. Fern\'{a}ndez Pirfano, 3-5A, Badajoz, 06010, Spain
        \and
        AAVSO, c/o Abbey Ridge Observatory, 45 Abbey Rd, Stillwater Lake, NS, B3Z 1R1, Canada
        \and
        ObSN, Observatorio Magalofes, MPC Y85, C/ Feal, 20, Fene, A Coru{\~n}a, 15509, Spain
        \and
        ObSN, Observatorio Irydeo, MPC Z41, C/ Guirnalda, 12, Camarma de Esteruelas, Madrid, 28816, Spain
        \and
        ObSN, Puig d'Agulles Observatory, MPC R74, Passatge Bosc, 1, Vallirana, Barcelona, 08759, Spain
        \and
        ObSN, Chicharronian Tres Cantos Observatory, MPC Z83, Pico de San Pedro, 6, Tres Cantos, Madrid, 28760, Spain
        \and
        ObSN, Observatorio El Gallinero, Camino de las Canteras, 42, El Berrueco, Madrid, 28192, Spain
        \and
        INAF - Osservatorio Astronomico di Capodimonte, Salita Moiariello 16, 80131, Naples, Italy
        \and
        Rolling Hills Observatory, 1643 Nightfall Dr., Clermont, FL 34711, USA
        \and
        ObSN, Observatorio Inmaculada del Molino, C/ Jos\'{e} Laguillo 27, Sevilla, 41003, Spain
        \and
        ObSN, Observatorio Estelia, MPC Y90, C/ Ladines, 12, Ladines, Asturias, 33993, Spain
        \and
        Institute of Earth Systems, University of Malta, Msida, MSD 2080, Malta
        \and
        ObSN, Observatorio LaVara, MPC J38, Barrio La Vara, s/n, Vald\'{e}s, Asturias, 33784, Spain
        \and
        ObSN, Observatorio Uraniborg, MPC Z55, C/ Antequera, 8, \'{E}cija, Sevilla, 41400, Spain
        \and
        School of Physics and Astronomy, University of Southampton, Highfield, Southampton SO17 1BJ, UK
        \and
        ObSN, Observatorio Carpe Noctem, MPC I72, Paseo de la Maliciosa, 11, Collado Mediano, Madrid, 28450, Spain
        \and
        ObSN, Observatorio de Sencelles, MPC K14, Cam\'{i} de Sonfred, 1, Sencelles, Islas Baleares, 07140, Spain
        \and
        ObSN, Observatorio Montcabrer, MPC 213, C/ Jaume Balmes, 24, Cabrils, Barcelona, 08348, Spain
        \and
        Hankasalmi observatory (Jyv\"{a}skyl\"{a}n Sirius ry), Verkkoniementie 30, 40950 Muurame, Finland
        \and
        Dipartimento di Fisica, Universit\`a di Pisa, 56127 Pisa, Italy
        \and
        British Astronomical Association, PO Box 702, Tonbridge TN9 9TX, UK
        \and
        Armagh Observatory \& Planetarium, College Hill, Armagh BT61 9DG, UK
        \and
        ObSN, Observatorio de Masquefa, MPC 232, Av. Can Marcet, 41, Masquefa, Barcelona, 08783, Spain
        \and
        Instituto de Astrof\'{i}sica de Canarias, E-38205 La Laguna, Tenerife, Spain
        \and
        Departamento de Astrof\'{i}sica, Universidad de La Laguna, E-38206 La Laguna, Tenerife, Spain
        \and
        Department of Physics and Astronomy, East Tennessee State University, Johnson City, TN 37614, USA
        \and
        SARA Observatory, Box 70652 ETSU, Johnson City, TN 37614 USA
        \and
        Department of Natural Science and Mathematics, King University, 1350 King College Road, Bristol, TN 37620 USA
        \and
        ObSN, Cal Maciarol m\`{o}dul 8 Observatory, MPC A02, Masia Cal Maciarol, Cam\'{i} de l'Observatori s/n, \`{A}ger, Lleida, 25691, Spain
        \and
        British Astronomical Association Variable Star Section, PO Box 702, Tonbridge, TN9 9TX, UK
        \and
        Department of Astrophysics and Planetary Science, Villanova University, Villanova, PA 19085, USA
        \and
        Escaldarium Observatory, MPC D28, Carrer Cingles de Bert\'{i} 54, 08140, Caldes de Montbui, Spain
        \and
        AAVSO, Tetoora Road Observatory, 2643 Warragul-Korumburra Road, Tetoora Road, 3821, Australia
        \and
        Department of Astronomy, University of Washington, Seattle, WA 98195, USA
        \and
        Polaris Observatory, Hungarian Astronomical Association, Laborc u. 2/c, 1037 Budapest, Hungary
        \and
        Universidad Nacional Aut\'onoma de M\'exico, Instituto de Astronom\'{i}a, A.P. 70-264, 04510, Ciudad de M\'exico, M\'exico
        \and
        ObSN, Observatorio el Sue\~{n}o, MPC C90, C/ Cardenal Vidal i Barraquer n 3, Cambrils, Tarragona, 43850, Spain
        \and
        Perth Observatory, 337 Walnut Rd, Bickley WA 6076, Australia
    }

   \date{Received October 01, 2000; accepted November 01, 2000}
 
\abstract
    {AM\,CVn~stars are ultra-compact binaries in which a white dwarf (WD) accretes helium-rich material from a hydrogen-deficient donor. They are promising mHz gravitational-wave sources and potential progenitors of Type~Ia supernovae. However, their evolutionary pathways remain uncertain. Three formation channels have been proposed: the WD, helium star, and cataclysmic variable channels.}
    {We aim to determine the atmospheric chemical composition of the underlying accreting WDs in two eclipsing AM\,CVn~systems, YZ\,LMi and ZTF\,J0407$-$00, in order to constrain their formation channels.}
    {Using ultraviolet spectra obtained with the \textit{Hubble Space Telescope}, we modelled the atmospheres of the accreting WDs, whose surface compositions are enriched by material transferred from the donor and therefore trace the donor's composition. In particular, the nitrogen-to-carbon abundance ratio provides a diagnostic of the donor's nuclear burning history and hence of the system's formation channel.}
    {We measure the following effective temperatures $T_{\mathrm{eff}} = 24\,400 \pm 1500$\,K and surface gravities $\log g = 8.36 \pm 0.05$ for YZ\,LMi, and $T_{\mathrm{eff}} = 29\,900 \pm 2200$\,K and $\log g = 8.37 \pm 0.05$ for ZTF\,J0407$-$00. The inferred masses of the two WDs are $0.84 \pm 0.10\,\mathrm{M}_\odot$ and $0.86 \pm 0.12\,\mathrm{M}_\odot$, respectively. Carbon is not detected in either system, yielding lower limits on the nitrogen-to-carbon mass ratio of $N/C>117$ and $N/C>185$.}
    {We increase from one to three the number of AM CVn systems whose accreting WDs and chemical abundances have been analysed using detailed UV atmosphere models. The high $N/C$ ratios rule out the helium star formation channel for both systems. The WD channel is favoured by the absence of detectable hydrogen at all observed wavelengths, although the evolved cataclysmic variable channel cannot be conclusively excluded.}


   \maketitle
\section{Introduction}\label{Introduction}    

    AM~Canum\,Venaticorum (AM\,CVn) stars are ultra-compact interacting binaries composed of a white dwarf (WD) accreting matter from a hydrogen-deficient donor. Mass transfer from the Roche-lobe-filling donor forms an accretion disc around the accreting WD. These systems are of fundamental importance for several reasons. First, accreting WDs provide valuable probes for testing models of compact binary evolution \citep{Nelemans_2001, Postnov_2014}. Second, they are potential progenitors of Type~Ia supernovae (SNe~Ia), which play a central role in cosmology. One of several proposed pathways to SNe~Ia is the double-detonation scenario, in which helium accretion onto a WD leads to the detonation of the accreted helium layer, subsequently triggering the detonation of the carbon--oxygen core and a thermonuclear explosion \citep{Fink_2010, Maoz_2014}. Third, they are ideal multi-messenger laboratories. Their short orbital periods (less than 70\,min) make them among the brightest persistent Galactic sources of gravitational waves (GW) in the mHz regime \citep{Nelemans_2004, Kupfer_2024}. For reviews of AM\,CVn~stars, see, e.g. \citet{Nelemans_2005, Solheim_2010}.
    
    The prototype AM\,CVn~star was identified in the 1960s \citep{Smak_1967,Paczynski_1967}. Since the advent of large-scale surveys, the known population has grown to around 150 systems \citep[e.g.][]{Roelofs_2007,Ramsay_2018,Burdge_2020,Pichardo_Marcano_2021,Green_2025,Rodriguez_2025,Kara_2026, Hernandez_Diaz_2026}. This population is expected to increase substantially with ongoing and forthcoming surveys across the electromagnetic spectrum. These include photometric surveys such as \textit{Gaia} \citep{Gaia_2023}, the Gravitational-wave Optical Transient Observer (GOTO; \citealt{Steeghs_2022}), BlackGEM \citep{Groot_2024}, and the Legacy Survey of Space and Time (LSST; \citealt{Ivezic_2019}); spectroscopic surveys such as the Sloan Digital Sky Survey (SDSS; \citealt{Kollmeier_2026}), the Dark Energy Spectroscopic Instrument (DESI; \citealt{Schlafly_2023}), and the 4-metre Multi-Object Spectroscopic Telescope (4MOST; \citealt{de_Jong_2019}); and X-ray surveys such as the \textit{extended ROentgen Survey with an Imaging Telescope Array} (\textit{eROSITA}; \citealt{Predehl_2021}). Moreover, a subset of AM\,CVn~systems is expected to be individually detected by next-generation GW observatories, including \textit{TianQin} \citep{Luo_2016}, \textit{Taiji} \citep{Luo_2021}, the \textit{Laser Interferometer Space Antenna} (\textit{LISA}; \citealt{Amaro-Seoane_2023}), and the \textit{Lunar Gravitational-wave Antenna} (\textit{LGWA}; \citealt{Ajith_2025}). The brightest systems will serve as verification binaries, whose signals are expected to be individually resolved on timescales ranging from $\sim3$~weeks to $\sim48$~months \citep{Roelofs_2006,Kupfer_2015,Chickles_2025,Chickles_2026}. The population as a whole will also contribute to the Galactic GW foreground at mHz frequencies \citep{Nelemans_2004,Kupfer_2018,Kupfer_2024}. Understanding the formation and evolution of AM\,CVn~systems is therefore essential for constraining their population properties and informing population-synthesis models \citep[e.g.][]{Breivik_2020, Korol_2022}, particularly in the emerging era of multi-messenger astronomy.
    
    Nevertheless, the formation channels of AM\,CVn~stars remain poorly constrained. Current theories suggest that these systems originate from close main-sequence binaries that undergo one or more phases of binary interaction, such as stable mass transfer or common-envelope evolution \citep{Nelemans_2001, Podsiadlowski_2003}. Following the first interaction phase, which leads to the formation of a WD, three evolutionary pathways have been proposed. (i) The WD channel, in which a second interaction phase produces a degenerate donor \citep{Paczynski_1967, Webbink_1984, Deloye_2007, Wong_2021, Chen_2022, Munday_2024}; (ii) the helium star (He-star) channel, in which a second interaction phase is again required to produce a semi-degenerate, helium-rich donor \citep{Savonije_1986, Iben_1987, Yungelson_2008, Bauer_2021}; and (iii) the cataclysmic variable (CV) channel, which involves only a single interaction phase. In the latter case, the donor's helium core is exposed as its outer layers are stripped through stable mass transfer \citep{Tutukov_1985, Goliasch_2015, Liu_2021, Belloni_2023, Sarkar_2023a}.
    
    One of the most robust methods for identifying the formation channel of an AM\,CVn~system is to determine the donor's chemical composition \citep{Nelemans_2010}. In particular, the nitrogen-to-carbon abundance ratio ($N/C$) varies significantly with the extent of prior CNO-cycle burning. The accretor's photospheric $N/C$ ratio traces the transferred material, as ongoing accretion replenishes the layer at the WD surface. The expected values differ by more than an order of magnitude for the different formation channels, with $N/C \lesssim 10$ for systems formed through the He-star channel and $N/C \gtrsim 100$ for those formed through either the WD or CV channels. Although previous studies reported an infrared excess that could have potentially been ascribed to the donor \citep{Rivera_Sandoval_2021}, recent \textit{JWST} spectroscopy of AM\,CVns showed that the infrared emission is still significantly affected by the accretion disc \citep{El-Badry_2026}. A direct detection of donors therefore remains missing, limiting our ability to model their spectra and determine their chemical abundances. Instead, the atmosphere of the accreting WDs, enriched by material transferred from the donor, provides an indirect probe of the donor's composition.
    
    Ultraviolet (UV) spectroscopy is essential for studying the atmosphere of the WD in accreting compact binaries. In the optical and infrared wavelength ranges, the accretion disc dominates the system emission. It often outshines the WD, whereas the WD emission peaks in the UV, for the typical effective temperatures of accreting WDs, $T_{\mathrm{eff}} \gtrsim 10\,000$\,K. Moreover, key absorption features of tracer elements such as nitrogen and carbon are primarily found in the UV. Space-based spectroscopy is therefore required to determine the atmospheric parameters accurately, and the \textit{Hubble Space Telescope} (\textit{HST}) is currently the only operational facility capable of delivering such observations.
    
    In this work, we investigate the formation channels of two eclipsing AM\,CVn~systems by constraining their nitrogen-to-carbon abundance ratios ($N/C$). We use ultraviolet spectra obtained with \textit{HST} and the spectral modelling pipeline developed by \citet{Yu_2026} to determine the accreting WDs' effective temperatures ($T_{\mathrm{eff}}$), surface gravities ($\log g$), and chemical abundances. Combined with archival optical photometric studies that provide precise binary parameters, these measurements constrain the donor properties and, consequently, the formation channels of these systems.
    

\section{Targets and observations} \label{Targets_and_observations}

    \subsection{Targets} \label{sub:targets}
        The targets of this study, YZ\,Leonis\,Minoris (also known as SDSS\,J092638.71$+$362402.4; hereafter YZ\,LMi) and ZTF\,J040749.30$-$000716.6 (also known as ASASSN-18dg; hereafter ZTF\,J0407$-$00), are part of a large \textit{HST} Treasury programme (Cycle~29; programme IDs: 16659 and 17401; PI: A.~F.~Pala) targeting 43 compact accreting binaries, including 12 AM\,CVn~systems. An overview of the AM\,CVn~sample is provided by \citet{Yu_2026}, while a comprehensive description of the programme will be given by Pala et al.\ (in prep.). YZ\,LMi and ZTF\,J0407$-$00 are two of the four eclipsing systems in the AM\,CVn~sample. The eclipses enable detailed light-curve modelling, providing constraints on the binary parameters and the WD contribution to the observed flux. Of the remaining two systems, ZTF\,J225237.05$-$051917.4 (hereafter ZTF\,J2252$-$05) was examined as a benchmark to demonstrate the UV spectral modelling pipeline, while Gaia14aae (also known as ASASSN-14cn) is predicted to host a considerably cooler WD ($T_{\mathrm{eff}} \lesssim 15\,000$\,K) and is not included here because modelling its spectrum requires further development of the WD atmosphere models. 
        
        The general astrometric and \textit{Gaia} properties of YZ\,LMi and ZTF\,J0407$-$00 are summarised in Tables~\ref{tab:target_list} and \ref{tab:target_list2}. Both systems have been studied extensively using high-speed optical photometry, yielding precise measurements of their ephemerides, orbital periods ($P_{\mathrm{orb}}$), mass ratios ($q$), inclinations ($i$), and the radii of the accreting WDs ($R_{\mathrm{WD}}$) and donors ($R_{\mathrm{donor}}$). Below, we briefly review the results from previous studies.

            \subsubsection{YZ\,LMi} \label{sub:YZLMi}

                YZ\,LMi is a partially eclipsing AM\,CVn~system with $P_{\mathrm{orb}} = 28.3$\,min \citep{Copperwheat_2011, Green_2026}. It was identified through a visual search of SDSS spectra based on its helium-dominated spectral characteristics \citep{Anderson_2005} and followed-up by \citet{Marsh_2007}. Detailed high-time-resolution photometric observations were later obtained with ULTRACAM on the 4.2-m William Herschel Telescope (WHT), the RISE camera on the 2-m Liverpool Telescope (LT), and the Optical Monitor (OM) onboard \textit{XMM-Newton} \citep{Copperwheat_2011}. These observations yielded precise measurements of $q$, $i$, $R_{\mathrm{WD}}$, and $R_{\mathrm{donor}}$ (Table~\ref{tab:system_parameters}). By fitting the optical-to-UV spectral energy distribution (SED) with synthetic WD atmosphere models, without accounting for interstellar extinction, \citet{Copperwheat_2011} estimated a WD temperature of $T_{\mathrm{eff,WD}} \sim 17\,000\,\mathrm{K}$.

                Subsequent studies have focused on measuring the orbital period derivative, $\dot{P}$, to constrain the evolutionary state of the system. In mass-transferring binaries, $\dot{P}$ reflects the balance between angular-momentum-loss (AML) through GW radiation, which drives orbital contraction, and mass transfer, which expands the orbit. Its sign and magnitude can therefore indicate the direction of orbital evolution and proximity to the period minimum \citep{Munday_2023}. Using eclipse-timing measurements, \citet{Szypryt_2014} and \citet{Schlindwein_2018} derived $\dot{P} = (3.07 \pm 0.56) \times 10^{-13}\,\mathrm{s\,s^{-1}}$ and $(3.2 \pm 0.4) \times 10^{-13}\,\mathrm{s\,s^{-1}}$, respectively. \citet{Green_2026} further improved this measurement using data spanning a 19-year baseline with ULTRACAM on the WHT, HiPERCAM on the 10.4-m Gran Telescopio Canarias (GTC), and ULTRASPEC on the 2.4-m Thai National Telescope (TNT). They derived $\dot{P} \lesssim 10^{-13}\,\mathrm{s\,s^{-1}}$, consistent with zero within $2\sigma$. Throughout this work, we adopt the measurement from \citet{Green_2026}, as it is based on a substantially longer observational baseline and higher-precision data.

            \subsubsection{ZTF\,J0407$-$00} \label{sub:ZTFJ04047}

                ZTF\,J0407$-$00 is a fully eclipsing AM\,CVn~system with $P_{\mathrm{orb}} = 35.4$\,min. It was discovered through a search for deep eclipses in data from the Zwicky Transient Facility (ZTF; \citealt{Bellm_2019}), with subsequent spectroscopic and photometric follow-up observations presented by \citet{van_Roestel_2022}. Spectroscopy obtained with the Low Resolution Imaging Spectrometer (LRIS) on the 10-m Keck~I telescope revealed a helium-dominated spectrum. Subsequent high-speed photometric observations with CHIMERA on the 200-inch (5.1-m) Hale Telescope yielded precise measurements of the binary parameters, similar to those obtained for YZ\,LMi (Table~\ref{tab:system_parameters}). By fitting the SED with a blackbody model, \citet{van_Roestel_2022} estimated a WD effective temperature of $T_{\mathrm{eff,WD}} = 17\,400 \pm 1200\,\mathrm{K}$.

            \begin{table*}
                \caption{Target details and observation logs of the \textit{HST} observation.}
                \label{tab:target_list}
                \centering
                \begin{tabular}{l l l c c c c}
                \toprule
                Target & RA    & DEC   & Ephemeris & Observation date & \textit{HST} orbits & $T_{\mathrm{exp}}$ \\
                       & [deg] & [deg] & [BMJD (TDB)]      & [YYYY-MM-DD]     &            & [s] \\
                \midrule
                YZ\,LMi 
                & 141.6612
                & 36.4007
                & $55469.3164759(6) + 0.019661273247(7)\,E$ 
                & 2021-12-30 
                & 4
                & 9210 \\
                \addlinespace[0.3em]
                \multirow{2}{*}{ZTF\,J0407$-$00} 
                & \multirow{2}{*}{61.9554}
                & \multirow{2}{*}{$-0.1212$}
                & \multirow{2}{*}{$58386.49699(2) + 0.024588123(3)\,E$}
                & 2023-02-10 & $3^{*}$ & 6600 \\
                & & & 
                & 2024-01-11 & 3 & 6358 \\
                \bottomrule
                \end{tabular}
                \tablefoot{
                Ephemerides are adopted from optical high-speed photometry: for YZ\,LMi from \cite{Green_2026} and for ZTF\,J0407$-$00 from \cite{van_Roestel_2022}.
                \tablefoottext{$*$}{The observation of ZTF\,J0407$-$00 was originally planned for five spacecraft orbits in February 2023; however, technical issues resulted in the loss of two orbits. The lost exposures were recovered through subsequent observations obtained in January 2024 (see Sect.~\ref{sub:uv_observations}).}
                }
            \end{table*}

            \begin{table*}
                \centering
                \caption{Additional target information and extinction properties.}
                \label{tab:target_list2}
                \renewcommand{\arraystretch}{1.2}
                \begin{tabular}{llccccc}
                    \toprule
                    Target & \textit{Gaia} DR3 source ID & $(l,b)$ & \textit{Gaia} $G$ & $\varpi$ & $d$  & $A_V$ \\
                           &                    & [deg]   & [mag]    & [mas]    & [pc] & [mag] \\
                    \midrule
                    YZ\,LMi & 798764346831516032 & $(187.508,\,+46.011)$ & $19.269 \pm 0.005$ & $1.23 \pm 0.21$ & $816^{+128}_{-122}$ & $\sim 0.06$ \\
                    \addlinespace[0.3em]
                    ZTF\,J0407$-$00 & 3253845886052817536 & $(191.499,\,-35.713)$ & $19.368 \pm 0.006$ & $1.34 \pm 0.37$ & $809^{+229}_{-191}$ & $\sim 0.48$ \\
                    \bottomrule
                \end{tabular}
                \tablefoot{
                The distances ($d$) are the \textit{Gaia}-based geometric distances reported by \citet{Green_2025}. Extinction values ($A_V$) are adopted from \citet{Edenhofer_2024} at the corresponding distances.
                }
            \end{table*}

    \subsection{Ultraviolet observations} \label{sub:uv_observations}
                 
\begin{figure*}
            \centering
            \includegraphics[width=0.9\textwidth]{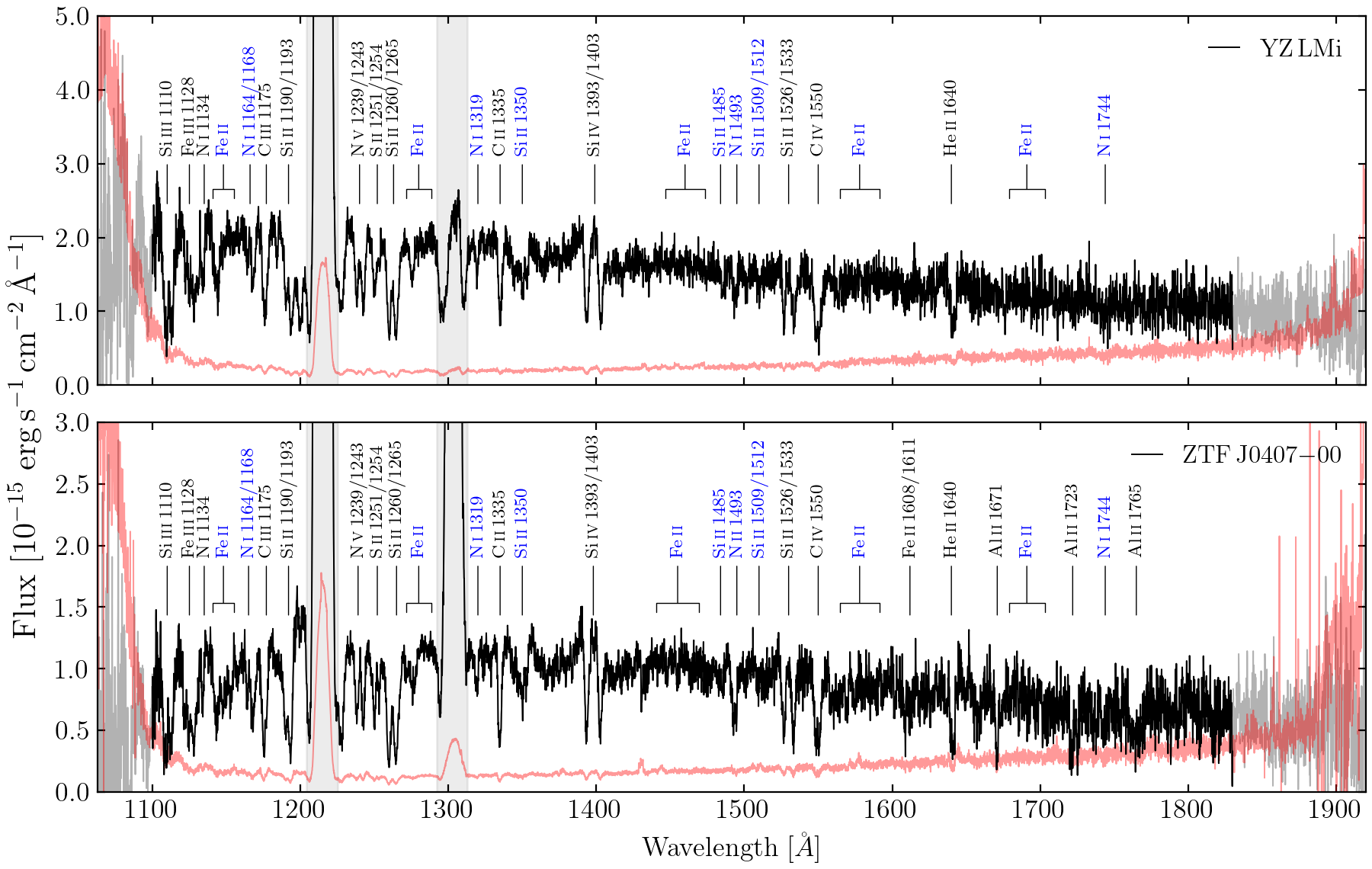}
            \caption{Average \textit{HST}/COS spectra of YZ\,LMi (top) and ZTF\,J0407$-$00 (bottom), with the flux uncertainties shown in red. Regions of low instrumental sensitivity below 1100\,\AA\ and above 1830\,\AA\ are excluded (grey shading). Additional masked regions contain geocoronal emission from Ly$\alpha$ at 1215\,\AA\ and \ion{O}{i} at 1304\,\AA. The identified absorption features are marked and listed in Table~\ref{tab:line_id}. Features originating in the WD atmosphere are labelled in blue, while those associated with the interstellar medium (ISM) or other components are labelled in black. Iron features are indicated in brackets because they generally consist of blends of multiple transitions that cannot be identified individually. The spectra have been smoothed for visual clarity.}
            \label{fig:target_spectra}
        \end{figure*}

\begin{figure*}
            \centering
            \includegraphics[width=0.85\textwidth]{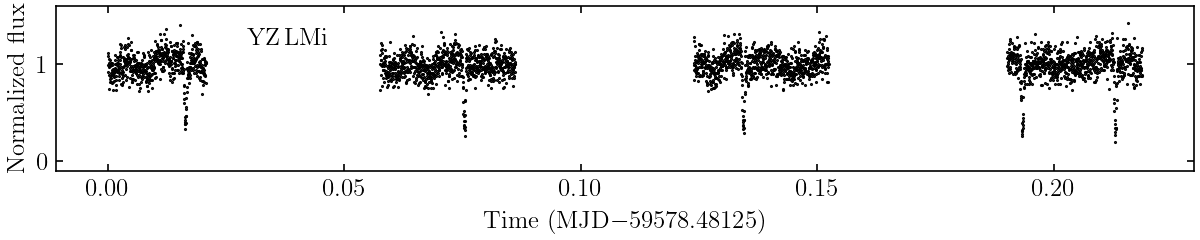}
            \includegraphics[width=0.85\textwidth]{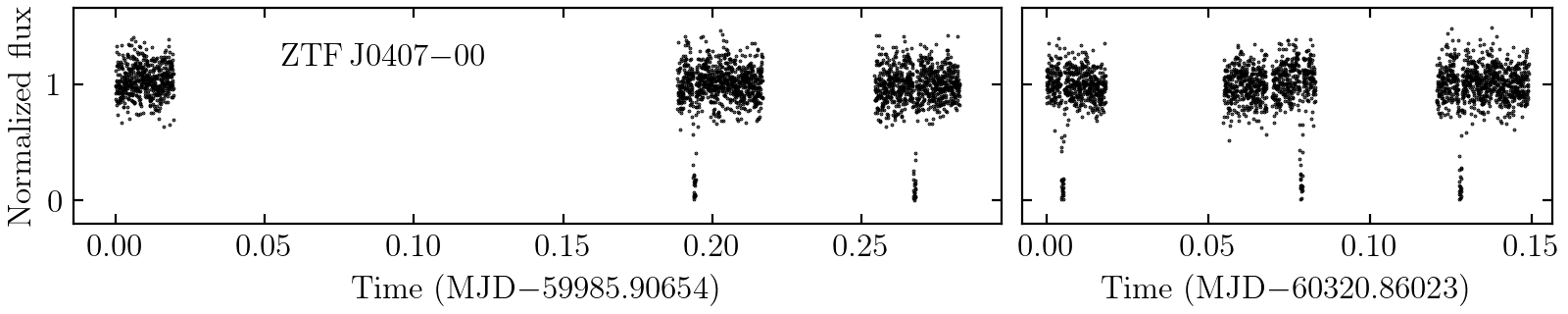}
            \caption{Normalised UV light curves of YZ\,LMi (top) and ZTF\,J0407$-$00 (bottom). The long gap following the first exposure of ZTF\,J0407$-$00 was caused by a technical issue that resulted in the loss of two \textit{HST} spacecraft orbits; see Sect.~\ref{sub:uv_observations} for details.}
            \label{fig:target_lightcurves}
        \end{figure*}

        Both targets were observed with the Cosmic Origins Spectrograph (COS) onboard \textit{HST}, using the G140L grating centred at 800\,\AA. This setup covers the wavelength range 900--2150\,\AA\ with $R \simeq 3000$. The observation log is given in Table~\ref{tab:target_list}. The averaged UV spectra are shown in Fig.~\ref{fig:target_spectra}. The data were reduced using the standard \texttt{CALCOS} pipeline \citep{Green_2012}. All observations were obtained in \texttt{TIME-TAG} mode, enabling the extraction of simultaneous UV light curves (Fig.~\ref{fig:target_lightcurves}). We followed the procedure described by \citet{Pala_2022}, extracting source photons from a predefined detector region and subtracting the background before binning the light curves into 3~s intervals. The photon arrival times were converted to barycentric time using \texttt{JPL Horizons Queries} \citep{Ginsburg_2019}, accounting for the spacecraft's motion.

        For ZTF\,J0407$-$00, the originally scheduled observation of five spacecraft orbits\footnote{\textit{HST} operates in low-Earth orbit with a period of approximately 95\,min, providing about 50\,min of target visibility per orbit.} on February 10, 2023 was interrupted by a technical issue affecting COS, during which the shutter failed to open. This resulted in the loss of two spacecraft orbits and produced the long gap following the first exposure in the bottom panel of Fig.~\ref{fig:target_lightcurves}. To compensate for the lost spacecraft orbits, additional observations were obtained on January 11, 2024. Before combining the two epochs into an average spectrum, we confirmed that their absolute flux levels, continuum slopes, and principal line strengths were statistically consistent within the instrumental calibration uncertainties. We therefore found no evidence of an epoch-dependent effect that would compromise the combined spectrum.

    \subsection{Photometric monitoring} \label{sub:photometric_monitoring}

        Accreting binaries exhibit intrinsic variability, and during disc outbursts their brightness can increase by several magnitudes, potentially posing a risk to the instrument. These outbursts arise from thermal instabilities in the accretion disc that temporarily enhance the mass-transfer rate and luminosity \citep{Pala_2017}. Moreover, during an outburst, the accretion disc dominates the system emission and can obscure the accreting WD, even in the UV. We therefore conducted a photometric monitoring campaign beginning several weeks before each \textit{HST} observation. These supplementary observations were obtained by citizen scientists from the American Association of Variable Star Observers (AAVSO) and the Observadores de Supernovas (ObSN) group, as well as with robotic telescopes operated by Las Cumbres Observatory (LCO; \citealt{Brown_2013}). The monitoring procedure is described by \citet{Pala_2017}, \citet{Tovmassian_2025}, and \citet{Yu_2026}. Both systems were observed by \textit{HST} in quiescence; see \citet{Green_2026} and \citet{van_Roestel_2022} for their recent outburst histories.

\section{Methods} \label{Method}

    \subsection{Light-curve analysis} \label{sub:lc_analysis}

        Modelling the eclipsing UV light curves constrains the contribution of the accreting WD, which subsequently informs the spectral modelling. The light curves were phase-folded using the ephemerides listed in Table~\ref{tab:target_list} and fitted with the built-in \texttt{levmarq} routine in the \texttt{LCURVE} software \citep{Copperwheat_2010}. The code models the individual components of an accreting binary, including the WD, donor, accretion disc, and bright spot, i.e.\ the region where the transferred material impacts the outer edge of the disc.
        
\begin{figure*}
            \centering
            \includegraphics[width=0.48\textwidth]{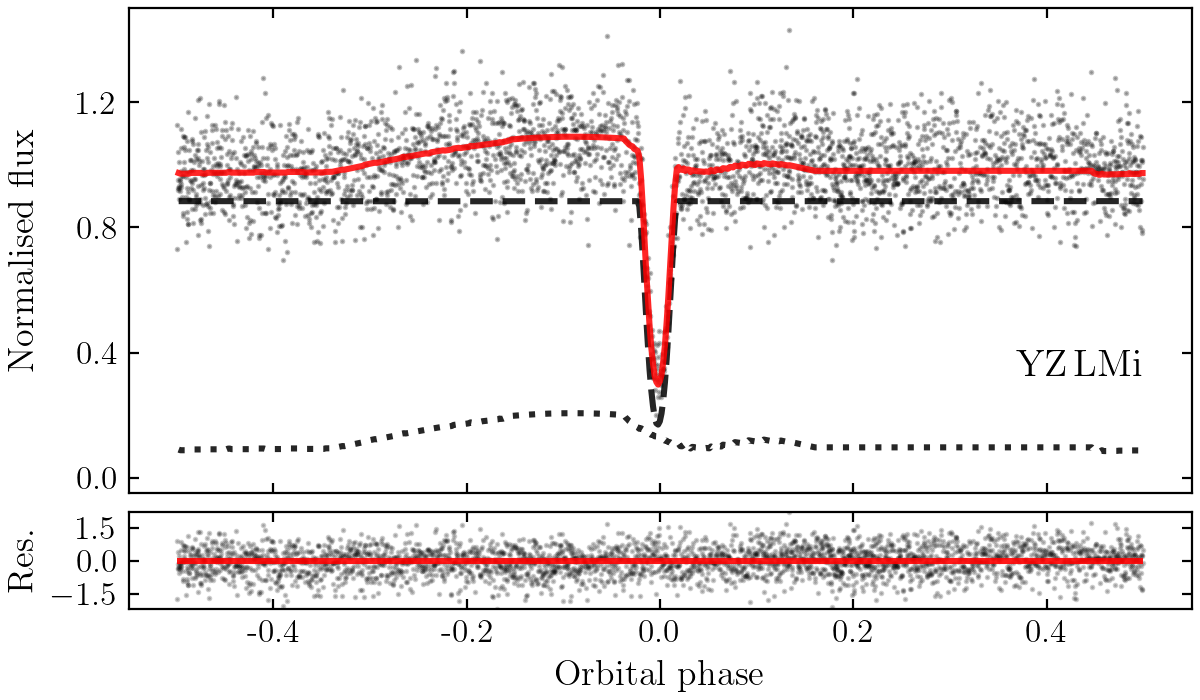}
            \includegraphics[width=0.48\textwidth]{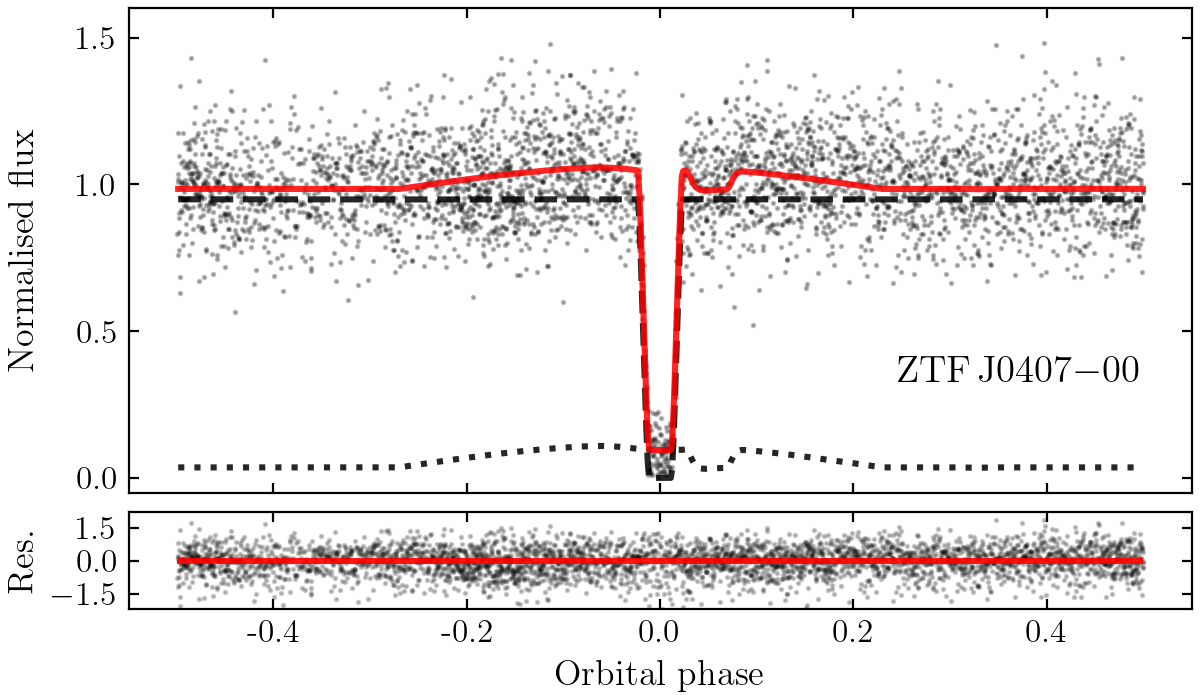}
            \caption{UV light curves of YZ\,LMi (left) and ZTF\,J0407$-$00 (right), extracted from the COS time-tag data, binned into 3\,s intervals, and phase-folded using the orbital periods listed in Table~\ref{tab:target_list}. The observed light curves are shown as black points. The best-fitting model (red) comprises emission from the accreting WD (black dashed line) and the combined contribution from the accretion disc and bright spot (black dotted line). The residuals are shown in the bottom panels.}
            \label{fig:lightcurve_fits}
        \end{figure*}

        We constructed an initial UV light-curve model based on previous high-speed optical photometry, adopting $P_{\mathrm{orb}}$, $i$, $q$, and $R_{\mathrm{WD}}$. For ZTF\,J0407$-$00, these parameters were taken from \citet{van_Roestel_2022}. For YZ\,LMi, we used $P_{\mathrm{orb}}$ from the eclipse-timing analysis by \citet{Green_2026}, who adopted $i$ and $q$ from \citet{Copperwheat_2011}. Since the donor fills its Roche lobe, we assumed $R_{\mathrm{donor}}$ using the Roche-lobe relation from \citet{Eggleton_1983}. The free parameters were \texttt{temp\_disc}, \texttt{temp\_spot}, \texttt{radius\_spot}, \texttt{length\_spot}, \texttt{angle\_spot}, and \texttt{cfrac\_spot}, which describe the disc and bright-spot contributions in the UV and therefore cannot be directly adopted from the optical results (see \citealt{Green_2018} for the definitions of the \texttt{LCURVE} parameters). The best-fitting models and the separate contributions from the WD are shown in Fig.~\ref{fig:lightcurve_fits}. The resulting models indicate that the WD contributes 91\% of the UV flux in YZ\,LMi and 96\% in ZTF\,J0407$-$00. In both systems, the donor is estimated to contribute less than 0.05\% of the total UV flux and is therefore neglected in the spectral modelling.

    \subsection{Spectral analysis} \label{sub:spectral_analysis}
    
        \subsubsection{Ultraviolet spectral-line identification} \label{sub:line_identification}
    
            We restrict the analysis to the wavelength range 1100--1830\,\AA, over which the data have a signal-to-noise ratio ($\mathrm{S/N} \gtrsim 5$) sufficient for our analysis \citep{Gaensicke_2005}. Regions affected by geocoronal Ly$\alpha$~1216\,\AA\ and \ion{O}{i}~1304\,\AA\ emission were masked. The spectra contain features originating from the WD, accretion disc, and interstellar medium (ISM). The identified spectral features are labelled in Fig.~\ref{fig:target_spectra} and listed in Table~\ref{tab:line_id}, together with their possible origins. Their atomic properties, including rest wavelengths and energy levels, were verified using the NIST Atomic Spectra Database (ver.~5.12; \citealt{NIST_2024}).

            We compared the spectra with published ISM line lists and a nearby hydrogen-atmosphere WD with DA spectral type to assess potential interstellar contamination. The comparison with WD\,0920$+$363 is shown in Appendix~\ref{app:spectral_diagnostics} (Fig.~\ref{fig:spectra_compare}); the identified features and their likely origins are listed in Table~\ref{tab:line_id}.

            The metal absorption lines attributed to the WD photosphere are labelled in Fig.~\ref{fig:target_spectra} and include transitions of \ion{N}{i}, \ion{Si}{ii}, and \ion{Fe}{ii}. Most highly ionised lines, including \ion{N}{v}, \ion{Si}{iv}, and \ion{C}{iv} are contaminated by the interstellar medium \citep{Danforth_2016}. We also examined \ion{Si}{iii}~1110\,\AA, \ion{Fe}{iii}~1128\,\AA, and \ion{C}{iii}~1175\,\AA. Their estimated excitation temperatures of $\gtrsim 50\,000$\,K suggest that they arise from a hotter component, such as the boundary layer at the interface between the accretion disc and the WD, rather than from the WD photosphere. They are therefore also excluded from the stellar-atmosphere models.

        \subsubsection{Extinction estimation} \label{sub:extinction}
        
            Extinction must be considered for accurate stellar atmosphere modelling in the UV. The Galactic coordinates and distances of the targets are listed in Table~\ref{tab:target_list2}, with the distances ($d$) adopted from the \textit{Gaia}-based geometric estimates reported by \citet{Green_2025}. The extinction values ($A_V$) of approximately 0.06 and 0.48 for YZ\,LMi and ZTF\,J0407$-$00, respectively, are adopted from the three-dimensional Galactic dust map of \citet{Edenhofer_2024}. We calculated the reddening correction using the extinction law of \citet{Cardelli_1989}, assuming $R_V=3.1$. The targets are faint ($G>19$\,mag), and their parallaxes have relative uncertainties of approximately 15--25\%. The resulting distance uncertainties propagate into the extinction estimates and, consequently, the derived atmospheric parameters. We quantify their impact in Sect.~\ref{sub:uncertainty}.
        
        \subsubsection{Spectral fitting} \label{sub:spectral_fit}
        
            The spectral-fitting procedure largely follows that described in Sect.~3.2 of \citet{Yu_2026} and is summarised below. We fitted the observed UV spectra with synthetic helium-atmosphere WD models \citep{Koester_2010}, allowing the $T_{\mathrm{eff}}$, $\log g$, and individual elemental abundances ($X_i$)\footnote{We denote the abundance of element $i$ as the base-10 logarithm of its particle-number ratio relative to helium. For example, the silicon abundance is given by $X_{\mathrm{Si}} = \log_{10}(N_{\mathrm{Si}}/N_{\mathrm{He}})$, where $N_{\mathrm{Si}}$ and $N_{\mathrm{He}}$ are the particle numbers of silicon and helium, respectively. The chemical profile, denoted by $\{X_i\}$, represents the atmospheric abundances of all elements identified in the spectrum of a given target.} to vary. We adopt a hydrogen abundance of $X_{\mathrm{H}}=-9$, consistent with the absence of detectable hydrogen lines in the optical spectra \citep{Copperwheat_2011, van_Roestel_2022}. \citet{Yu_2026} found that varying $X_{\mathrm{H}}$ from $-9$ to $-6$ has only a minor effect on the atmospheric models. We therefore do not attempt to derive an upper limit on the hydrogen abundance.
        
            For a given $T_{\mathrm{eff}}$ and $R_{\mathrm{WD}}$, we determined $M_{\mathrm{WD}}$ using the mass--radius relation from the Montr\'eal WD cooling models \citep{Bedard_2020}\footnote{\url{https://www.astro.umontreal.ca/~bergeron/CoolingModels/}} and subsequently derived $\log g$. We used $R_{\mathrm{WD}}$ derived from the optical light-curve analyses of \citet{Copperwheat_2011} and \citet{van_Roestel_2022} for YZ\,LMi and ZTF\,J0407$-$00, respectively. The model Eddington flux, $H_\lambda$, comprises contributions from the WD ($H_{\mathrm{WD},\lambda}$) and the non-WD components ($H_{\mathrm{non\mbox{-}WD},\lambda}$). The latter arise primarily from the accretion disc and bright spot, and their combined emission was estimated from the light curve analysis (Sect.~\ref{sub:lc_analysis}) to contribute 9\% of the UV flux in YZ\,LMi and 4\% in ZTF\,J0407$-$00. In the absence of dedicated spectral models for these components, we represent their combined contribution with a constant continuum term in $H_\lambda$. Given their small fractional contribution to the observed flux, the assumed spectral shape has little effect on the results, with alternative prescriptions, such as a blackbody or a power law, yielding comparable results \citep{Pala_2017}. We also included the radial velocity, $v_{\mathrm{rad}}$, as a free parameter to account for the net wavelength shift of the spectrum averaged over the total exposure time due to the binary's systemic velocity. This correction is necessary to align the observed spectral features with the model, as the orbital motion of the WD introduces a Doppler shift during the observations.
        
            The synthetic spectra are given in units of Eddington flux at the stellar surface and can be scaled to the observed flux using $R_{\mathrm{WD}}$ and the distance ($d$). If the distance is known sufficiently well, it can be used to break the degeneracy between $T_{\mathrm{eff}}$ and $\log g$. However, the relatively large uncertainties on the \textit{Gaia} parallaxes of both targets prevent the distances from being reliably obtained by simply inverting the parallaxes and instead require geometric or photogeometric priors \citep{Bailer-Jones_2021,Gentile_Fusillo_2021}. The resulting uncertainties of $>20\%$ in the scaling factor limit its extent to independently constrain $\log g$. We therefore introduced a free scaling factor, $C$, which multiplies the model flux to match the observed flux level. With $R_{\mathrm{WD}}$ fixed from the optical light curve modelling, the WD mass is constrained through the adopted mass--radius relation. More precise distances from the improved astrometry expected with \textit{Gaia} DR4\footnote{\url{https://www.cosmos.esa.int/web/gaia/dr4}} should enable more accurate absolute scaling of the model fluxes, providing a constraint on $M_{\mathrm{WD}}$ independent of optical light curve modelling. The observed spectrum was then modelled as
            \begin{equation}
                \label{eq:flux_components}
                H_{\lambda,\mathrm{obs}} =
                C\,A_\lambda
                \left[
                H_{\mathrm{WD},\lambda}
                (T_{\mathrm{eff}},\log g,v_{\mathrm{rad}},\{X_i\})
                +
                H_{\mathrm{non\mbox{-}WD},\lambda}
                \right].
            \end{equation}
            
            Because generating the synthetic spectra is computationally expensive, we treated the global atmospheric parameters, namely $T_{\mathrm{eff}}$, $v_{\mathrm{rad}}$, and $C$, together with the corresponding $\log g$, separately from the elemental abundances. We first fixed the global parameters and determined $X_i$ for each identified element using isolated spectral regions, thereby minimising the effects of line blending. The resulting abundance profile, $\{X_i\}$, was then held fixed while the global parameters were varied. These steps were repeated iteratively, with the explored parameter ranges progressively narrowed until convergence was reached. In total, we performed seven iterations, comprising four global fits and three elemental-abundance fits. All fits were performed using the Markov chain Monte Carlo (MCMC) sampler implemented in the Python package \texttt{emcee} \citep{Foreman-Macke_2013}. Further details of the MCMC procedure and the posterior distributions from the final global fit are presented in Appendix~\ref{app:mcmc}.

        \subsubsection{Uncertainty estimation} \label{sub:uncertainty}
            Variations in $A_V$ affect the inferred $T_{\mathrm{eff}}$ by altering the slope of the spectral continuum, and consequently influence the derived chemical abundances. This effect is particularly important in the UV, where extinction increases towards shorter wavelengths and reddens the observed spectrum.
            
            The extinction was fixed during the spectral fitting described in Sect.~\ref{sub:spectral_fit}, and its systematic uncertainty was therefore assessed separately. Because the extinction estimates were obtained from distance-dependent three-dimensional dust maps, the relatively large distance uncertainties of both targets translate into substantial uncertainties in $A_V$. We quantified their effect on the atmospheric parameters using a Monte Carlo approach following the same method described in \citet{Pala_2022}. For each target, values of $A_V$ were drawn from a Gaussian distribution whose width was determined from the range of extinction values corresponding to the $1\sigma$ interval of the geometric distance posterior. We compared several available reddening maps \citep[e.g.][]{Green_2019, Lallement_2022, Vergely_2022} and adopted the three-dimensional dust map of \citet{Edenhofer_2024}, which provides improved extinction estimates within 1.25\,kpc. This yields $A_V = 0.06 \pm 0.01$ for YZ\,LMi and $A_V = 0.48 \pm 0.20$ for ZTF\,J0407$-$00. For each sampled value of $A_V$, the spectrum was refitted by $\chi^2$ minimisation using the \texttt{minimize} routine from \texttt{scipy.optimize} \citep{Virtanen_2020}, with at least 2000 Monte Carlo realisations performed for each target.
            
            The uncertainty associated with $R_{\mathrm{WD}}$, adopted from the photometric analyses of \citet{Copperwheat_2011} and \citet{van_Roestel_2022}, was evaluated using the same procedure. The resulting systematic uncertainties were then combined with the statistical uncertainties of the MCMC analysis. The final atmospheric parameters and their uncertainties are reported in Table~\ref{tab:system_parameters}.


\section{Results} \label{Result}

    \subsection{Best-fitting models} \label{sub.best_fitting_models}

\begin{figure*}
        \centering
        \includegraphics[width=0.95\textwidth]{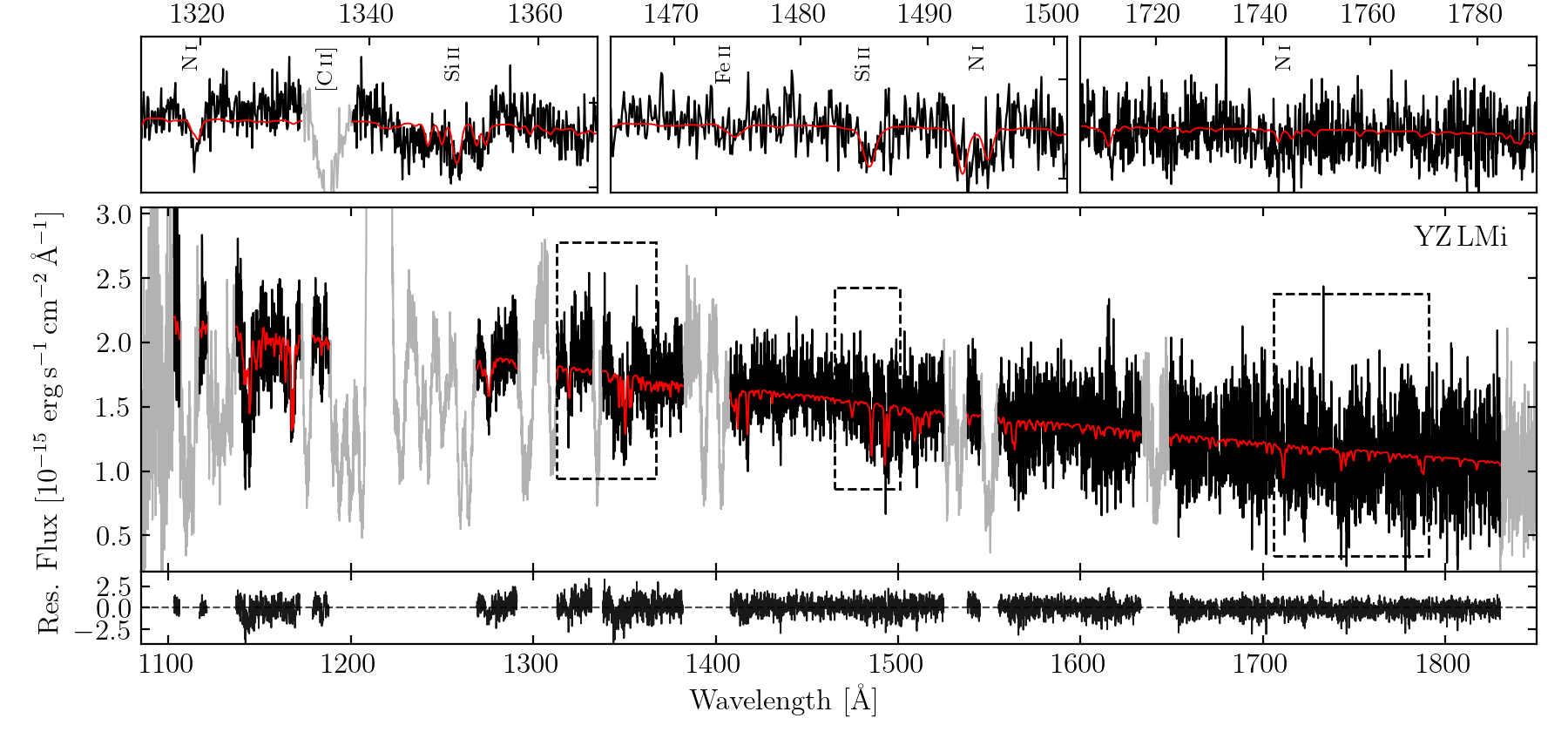}
        \caption{Best-fitting spectrum of YZ\,LMi. Middle panel: Average \textit{HST}/COS UV spectrum (black) and best-fitting WD atmosphere model (red). Spectral regions containing features not attributed to the WD photosphere are shaded in grey and excluded from the fit. Top panels: Enlarged views of the three regions indicated by dashed boxes in the middle panel. Photospheric features of \ion{N}{i}, \ion{Si}{ii}, and \ion{Fe}{ii} are labelled, while identified ISM absorption features are indicated in brackets. The residuals are shown in the bottom panel. The best-fitting model parameters are listed in Table~\ref{tab:system_parameters}.}
        \label{fig:spectral_fit_YZLMi}
    \end{figure*}

\begin{figure*}
        \centering
        \includegraphics[width=0.95\textwidth]{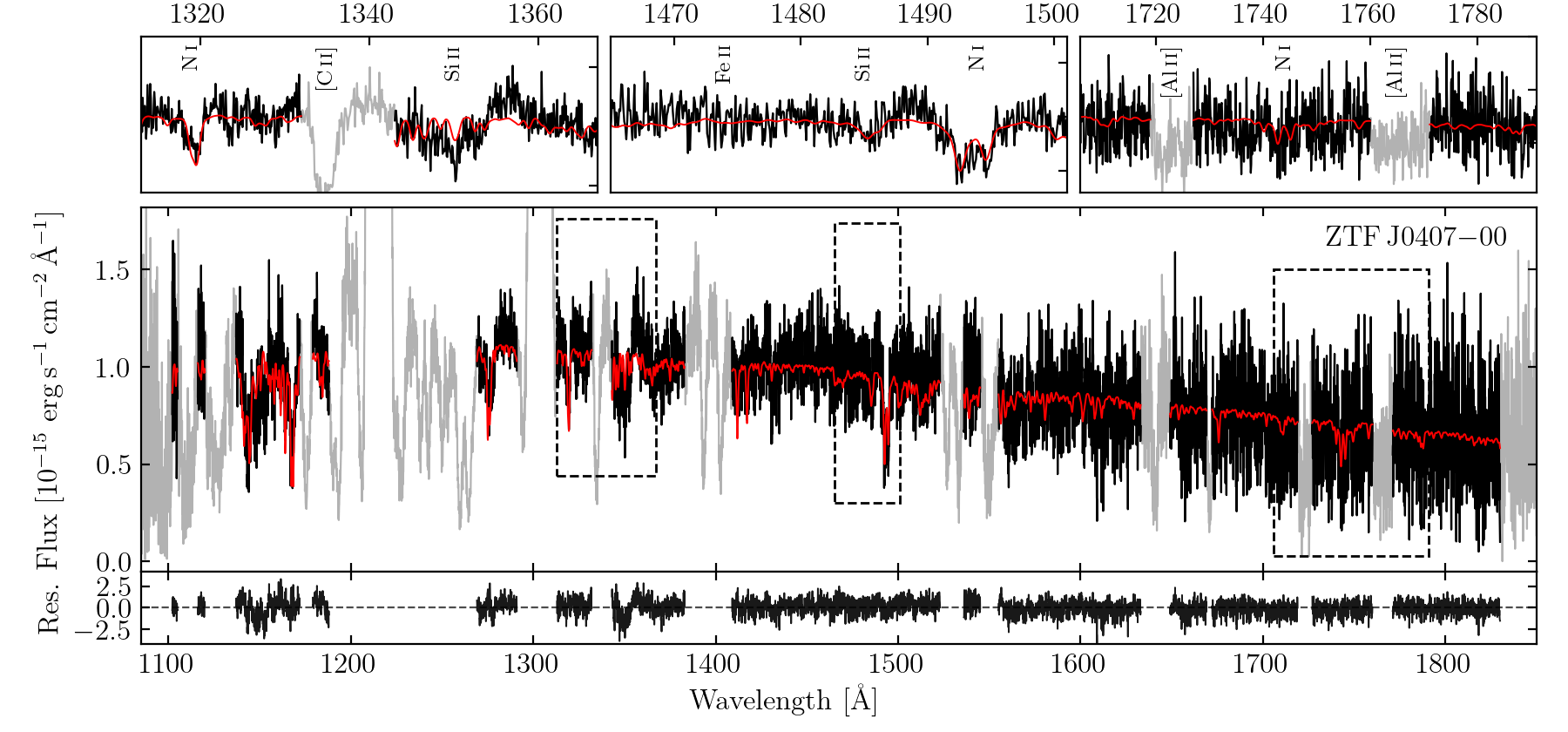}
        \caption{Best-fitting spectrum of ZTF\,J0407$-$00. The panels, colours, labels, and shaded regions follow the same convention as in Fig.~\ref{fig:spectral_fit_YZLMi}. The best-fitting model parameters are listed in Table~\ref{tab:system_parameters}.}
        \label{fig:spectral_fit_ZTFJ0407}
    \end{figure*}

        Figures~\ref{fig:spectral_fit_YZLMi} and \ref{fig:spectral_fit_ZTFJ0407} show the best-fitting models for YZ\,LMi and ZTF\,J0407$-$00, respectively. The corresponding atmospheric and binary parameters are listed in Table~\ref{tab:system_parameters}. 
        
        For YZ\,LMi, the spectral fit yields $T_{\mathrm{eff}} = 24\,400 \pm 1500$\,K and $\log g = 8.36 \pm 0.05$. We adopt $R_{\mathrm{WD}}$ and $q$ from the optical light-curve analysis of \citet{Copperwheat_2011}. Combining $R_{\mathrm{WD}}$ with the mass--radius relation of \citet{Bedard_2020}, we derive a WD mass of $M_{\mathrm{WD}} = 0.84 \pm 0.10\,M_\odot$. Together with $q$, this yields a donor mass of $M_{\mathrm{donor}} = 0.033 \pm 0.008\,M_\odot$. We then calculate an orbital separation of $a = 0.29 \pm 0.02\,R_\odot$ using Kepler's third law. The nitrogen abundance by particle number, $X_{\mathrm{N}} = -3.8 \pm 0.2$, together with the upper limit of $X_{\mathrm{C}} < -6$, implies a nitrogen-to-carbon mass ratio of $N/C > 117$.
        
        For ZTF\,J0407$-$00, the spectral fit yields $T_{\mathrm{eff}} = 29\,900 \pm 2200$\,K and $\log g = 8.37 \pm 0.05$. Similarly, using $R_{\mathrm{WD}}$ and $q$ from the high-speed optical photometry from \citet{van_Roestel_2022}, we derive $M_{\mathrm{WD}} = 0.86 \pm 0.12\,M_\odot$, $M_{\mathrm{donor}} = 0.021 \pm 0.006\,M_\odot$, and $a = 0.34 \pm 0.03\,R_\odot$. Following the same procedure as for YZ\,LMi, the nitrogen abundance of $X_{\mathrm{N}} = -2.5 \pm 0.3$, together with the upper limit of $X_{\mathrm{C}} < -5$, implies $N/C > 185$.
 
    \subsection{Constraint on the carbon abundance} \label{constrain_carbon_abundance}
\begin{figure*}
        \centering
        \includegraphics[width=0.49\textwidth]{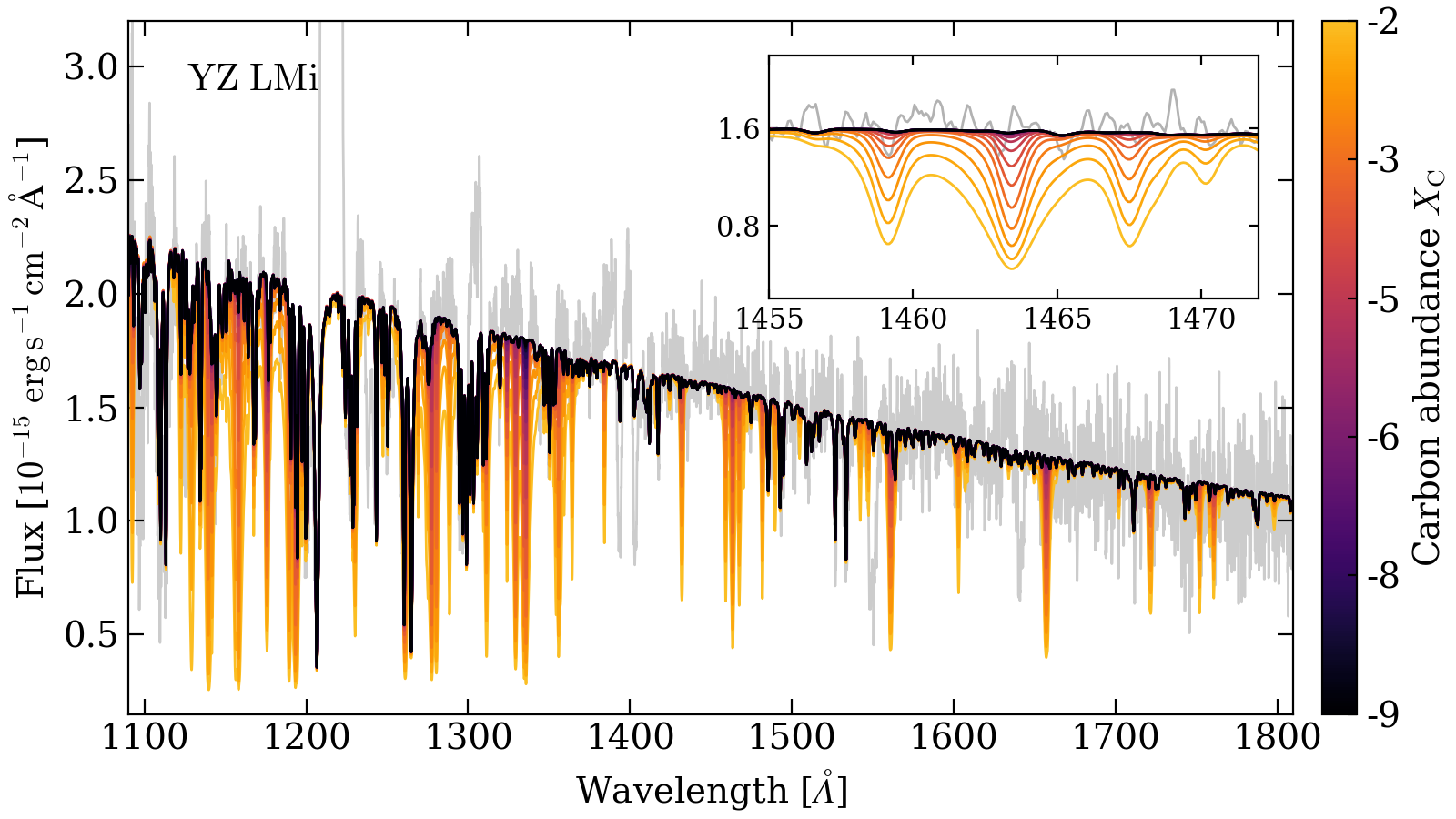}
        \includegraphics[width=0.49\textwidth]{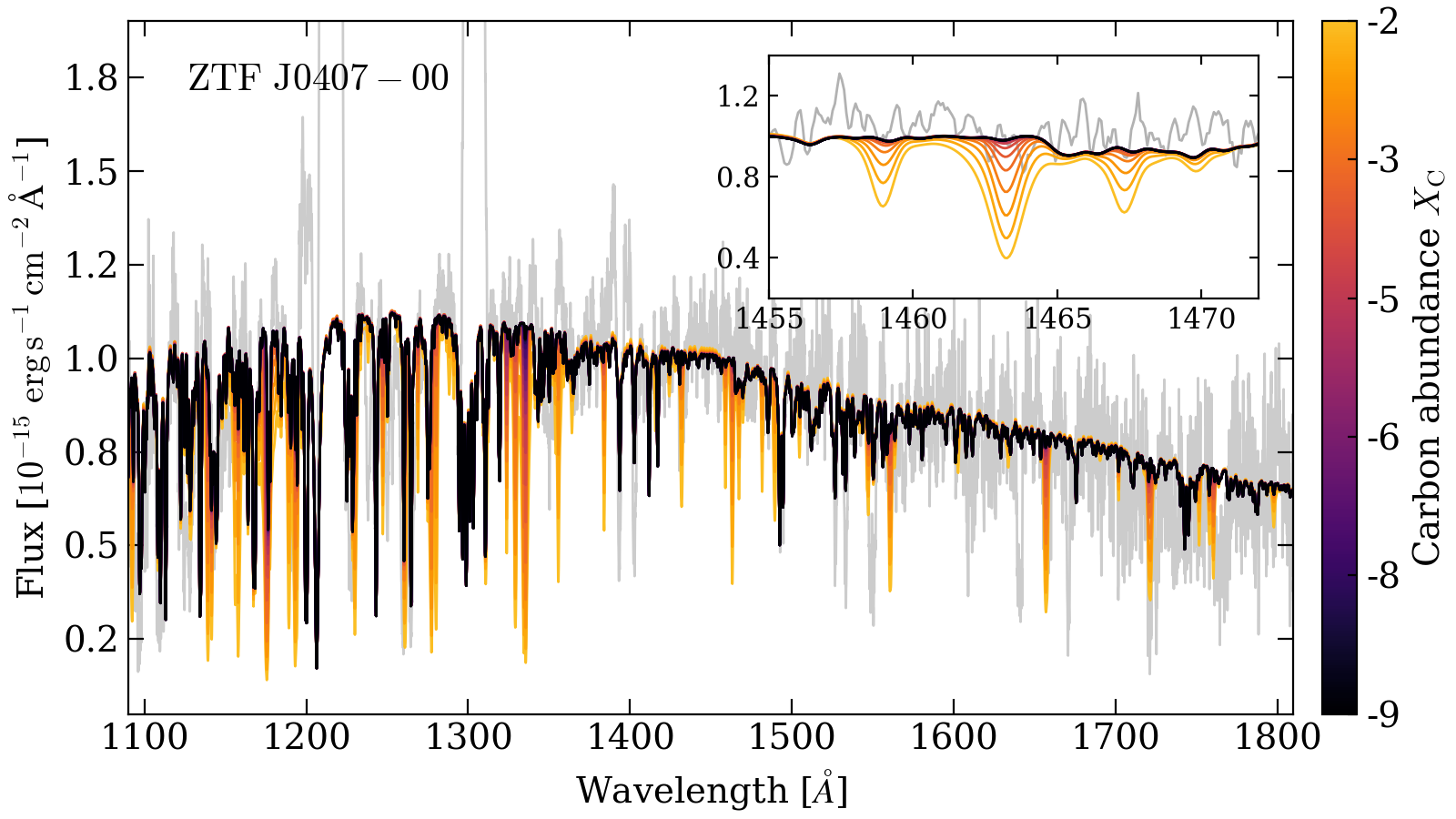}
        \caption{Constraints on the carbon abundances of YZ\,LMi (left) and ZTF\,J0407$-$00 (right). The observed spectra are shown in grey, while the coloured curves show synthetic spectra calculated from the best-fitting atmospheric models with carbon abundances ranging from $X_{\mathrm{C}}=-8.0$ to $-2.0$ in steps of 0.5. The inset panels show the diagnostic region around the predicted \ion{C}{i}~1459/1463/1468\,\AA\ triplet feature. Its absence implies $X_{\mathrm{C}}<-6.0$ for YZ\,LMi and $X_{\mathrm{C}}<-5.0$ for ZTF\,J0407$-$00.}
        \label{fig:Carbon_estimation}
    \end{figure*}

        As discussed in Sect.~\ref{sub:line_identification}, the \ion{C}{ii}~1335\,\AA\ feature is affected by interstellar absorption. No other carbon feature in the observed spectra is sufficiently strong to provide a reliable abundance measurement, indicating that both atmospheres are carbon poor. Given the importance of carbon as a product of nuclear burning and as a tracer of AM\,CVn evolutionary channels \citep{Nelemans_2010}, we nevertheless sought to constrain its abundance. To this end, we inspected the spectral region containing the theoretically predicted \ion{C}{i}~1459/1463/1468\,\AA\ triplet. With $E_i \approx 1.3$\,eV, corresponding to a characteristic excitation temperature of $\sim2\times10^4$\,K, these transitions are expected to arise preferentially in the WD photosphere. We generated a grid of synthetic spectra based on the best-fitting atmospheric models described in Sect.~\ref{sub:spectral_fit}, varying only the carbon abundance, and compared these with the observations to determine an upper limit for each target. The carbon-abundance constraints are illustrated in Fig.~\ref{fig:Carbon_estimation}, where synthetic spectra with carbon abundances ranging from $X_{\mathrm{C}}=-8.0$ to $-2.0$ in steps of 0.5 are compared with the observations. The absence of the theoretical \ion{C}{i}~1459/1463/1468\,\AA\ triplet implies upper limits of $X_{\mathrm{C}}<-6.0$ for YZ\,LMi and $X_{\mathrm{C}}<-5.0$ for ZTF\,J0407$-$00.

    \begin{table}
        \caption{System parameters of YZ\,LMi and ZTF\,J0407$-$00.}
        \label{tab:system_parameters}
        \centering
        \normalsize
        \begin{tabular}{lll}
            \toprule
            Parameter                          & YZ\,LMi                  & ZTF\,J0407$-$00        \\
            \midrule
            $P_{\mathrm{orb}}$ [min]           & $28.312233476(10)^{a}$   & $35.4068971(43)^{b}$     \\
            $\dot{P}$ [$\mathrm{s\,s^{-1}}$]   & $\lesssim 10^{-13\,a}$   & --                     \\
            $q$                                & $0.040 \pm 0.002^{c}$    & $0.024 \pm 0.004^{b}$  \\
            $i$ [deg]                          & $82.7 \pm 0.2^{c}$       & $87.0 \pm 1.0^{b}$     \\
            $a$ [$R_\odot$]                    & $0.29 \pm 0.02$          & $0.34 \pm 0.03$        \\
            $v_{\mathrm{rad}}$ [km\,s$^{-1}$]  & $-19 \pm 5$              & $20 \pm 8$             \\
            \midrule
            $T_{\mathrm{eff}}$ [K]             & $24\,400 \pm 1500$       & $29\,900 \pm 2200$     \\
            $\log g$ [cgs]                     & $8.36 \pm 0.05$          & $8.37 \pm 0.05$        \\
            $M_{\mathrm{WD}}$ [$M_\odot$]      & $0.84 \pm 0.10$          & $0.86 \pm 0.12$        \\
            $R_{\mathrm{WD}}$ [$R_\odot$]      & $0.010 \pm 0.001^{c}$    & $0.010 \pm 0.001^{b}$  \\
            $X_{\mathrm{N}}$                   & $-3.8 \pm 0.2$           & $-2.5 \pm 0.3$         \\
            $X_{\mathrm{Si}}$                  & $-4.6 \pm 0.2$           & $-4.4 \pm 0.3$         \\
            $X_{\mathrm{Fe}}$                  & $-4.8 \pm 0.2$           & $-3.7 \pm 0.3$         \\
            $X_{\mathrm{C}}$                   & $<-6$                    & $<-5$                  \\
            $N/C$                              & $>117$                   & $>185$                 \\
            \midrule
            $M_{\mathrm{donor}}$ [$M_\odot$]   & $0.033 \pm 0.008$        & $0.021 \pm 0.006$      \\
            $R_{\mathrm{donor}}$ [$R_\odot$]   & $0.047 \pm 0.001^{c}$    & $0.044 \pm 0.002^{b}$  \\
            \bottomrule
        \end{tabular}
        \tablefoot{
        $^{a}$Values adopted from the eclipse-timing analysis of \citet{Green_2026}.
        $^{b}$Values adopted from \citet{van_Roestel_2022}.
        $^{c}$Values adopted from \citet{Copperwheat_2011}.
        }
    \end{table}


\section{Discussion} \label{Discussion}

    \subsection{Comparison with optical results}
        \label{sub.compare_with_optical}
    
            We compare the WD masses and effective temperatures derived from our ultraviolet spectral fitting with those obtained from previous optical analyses. For YZ\,LMi, our derived WD mass of $M_{\mathrm{WD}} = 0.84 \pm 0.10\,M_{\odot}$ is consistent with the value of $M_{\mathrm{WD}} = 0.85 \pm 0.04\,M_{\odot}$ obtained by \citet{Copperwheat_2011} from $g$-band light-curve fitting. In contrast, our measured effective temperature, $T_{\mathrm{eff}} = 24\,400 \pm 1500$\,K, is significantly higher than their estimate of $T_{\mathrm{eff}} \simeq 17\,000$\,K based on the optical and ultraviolet SED. This discrepancy most likely arises from differences in the spectral modelling. Our analysis employs WD atmosphere models with accurately measured component contributions and  an appropriate chemical composition, whereas \citet{Copperwheat_2011} used the simplified synthetic spectra of \citet{Gaensicke_1995}. Furthermore, we adopt a non-zero extinction of $A_V = 0.06 \pm 0.01$ from the updated dust map of \citet{Edenhofer_2024}, whereas extinction was not included in the analysis by \citet{Copperwheat_2011}.
    
            Similarly, for ZTF\,J0407$-$00, compared to the result of \citet{van_Roestel_2022}, our derived mass of $M_{\mathrm{WD}} = 0.86 \pm 0.12\,M_{\odot}$ is consistent within the uncertainties with the value of $M_{\mathrm{WD}} = 0.79 \pm 0.06\,M_{\odot}$. The effective temperature inferred from our ultraviolet spectral fit, $T_{\mathrm{eff}} = 29\,900 \pm 2200$\,K, is again substantially higher than the optical estimate of $T_{\mathrm{eff}} = 17\,400 \pm 1200$\,K. As for YZ\,LMi, this discrepancy most likely arises from the differences in the spectral models and adopted extinction. In particular, \citet{van_Roestel_2022} modelled the WD as a blackbody, which provides only a simplified description of its SED and does not account for the detailed wavelength-dependent emission of a WD atmosphere. They also adopted $E(B-V) \simeq 0.14$ from \citet{Green_2019}, whereas we adopt $A_V = 0.48 \pm 0.20$ from \citet{Edenhofer_2024}, corresponding to $E(B-V) \simeq 0.16 \pm 0.07$ for $R_V = 3.1$. We emphasise, however, that the large uncertainties in the parallax and distance of ZTF\,J0407$-$00 propagate into the extinction estimate and, consequently, contribute to the uncertainty in the inferred $T_{\mathrm{eff}}$. Improved astrometry from the forthcoming \textit{Gaia} DR4 \citep{Gaia_2023} should provide tighter constraints on these parameters.

    \subsection{Comparison with stellar evolutionary models} \label{sub.compare_with_model}

        As discussed in Sect.~\ref{Introduction}, three main formation channels have been proposed for AM\,CVn~systems: the WD, He-star, and CV channels. Here, we compare the measured properties of YZ\,LMi and ZTF\,J0407$-$00 with theoretical evolutionary models to assess their possible formation histories. We also include ZTF\,J2252$-$05, the only other AM\,CVn~system for which a detailed UV spectral analysis of the accreting WD has been performed \citep{Yu_2026}.

        \subsubsection{Donor mass and radius} \label{sub:compare_donor_M_R}

\begin{figure}
                \centering
                \includegraphics[width=\columnwidth]{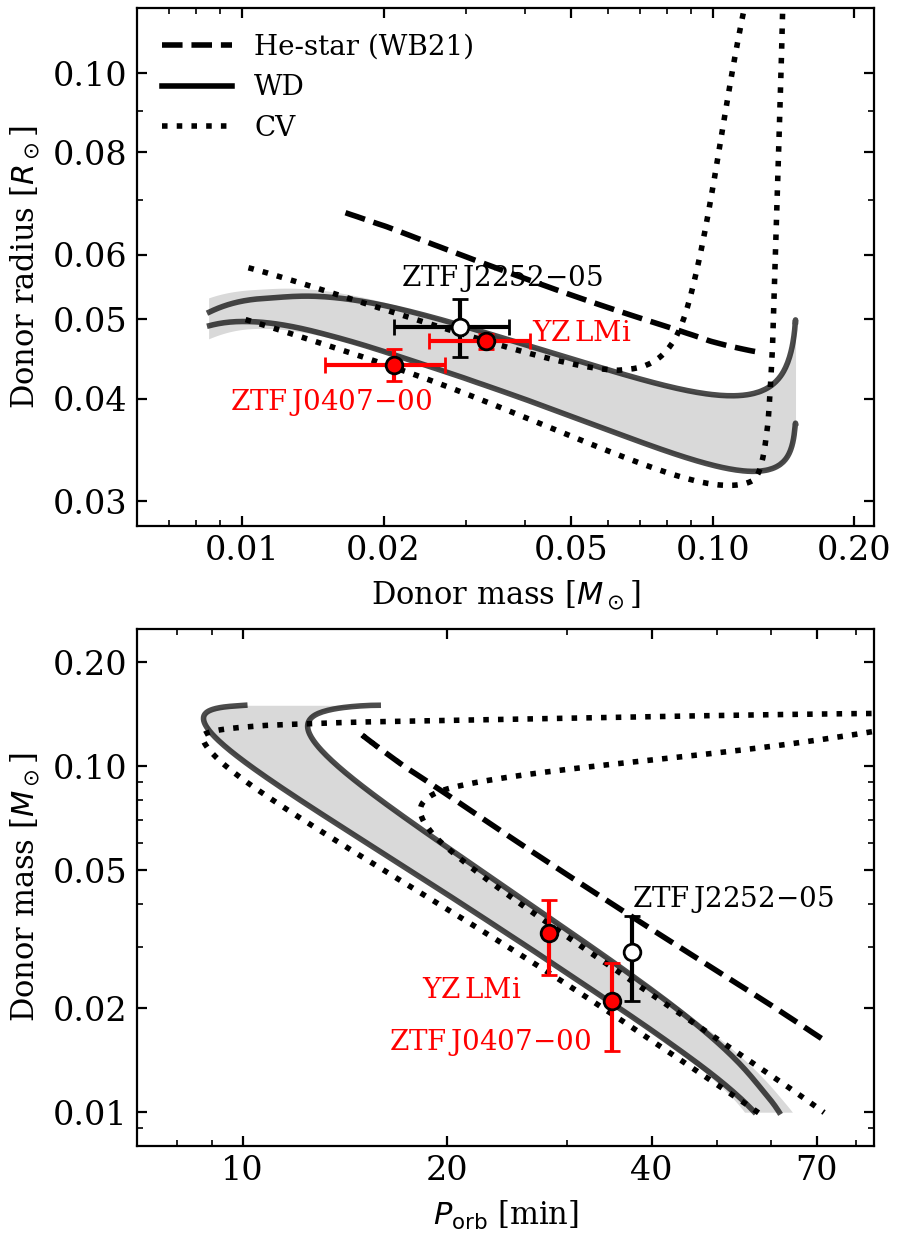}
                \caption{Comparison of the inferred donor properties of ZTF\,J2252$-$05, YZ\,LMi, and ZTF\,J0407$-$00 with \texttt{MESA} evolutionary tracks. The upper panel shows $R_{\mathrm{donor}}$ versus $M_{\mathrm{donor}}$; the lower panel shows $M_{\mathrm{donor}}$ versus $P_{\mathrm{orb}}$. The dashed curve shows the He-star channel track of \citet{Wong_2021} (labelled ``WB21''). The grey shaded region spans the WD channel models, with two representative tracks shown as solid lines. The dotted curves show CV channel tracks selected to match the systems' inferred donor properties.}
                \label{fig.MESA_R_M}
            \end{figure}

            We compare the measured donor masses and radii with AM\,CVn evolutionary models computed using Modules for Experiments in Stellar Astrophysics (\texttt{MESA}; \citealt{Paxton_2011, Paxton_2013, Paxton_2015, Paxton_2018, Paxton_2019}), in Fig.~\ref{fig.MESA_R_M}. The He-star track is adopted from \citet{Wong_2021} and assumes adiabatic evolution once the donor mass falls below $0.2\,M_\odot$. We further describe the WD and CV channel models below.

            The WD channel models are generated following the methods of \citet{Wong_2023}. The donor initially loses mass approximately adiabatically, at nearly constant central specific entropy, during the short-period phase of high mass transfer. The subsequent evolution follows the magnetic-braking (MB) prescription of \citet[][hereafter RVJ]{Rappaport_1983}. We explored initial WD masses of $M_{\mathrm{WD},0}=0.7$--$0.8\,M_\odot$ and RVJ boosting factors of 3--10. This prescription was found in good agreement with the observed WD luminosity by Wong et al. (in prep.). The grey shaded region in Fig.~\ref{fig.MESA_R_M} shows the resulting parameter space, while the two solid lines illustrate models with a boosting factor of six and $M_{\mathrm{WD},0}=0.7\,M_\odot$ (lower) and $0.8\,M_\odot$ (upper).

            The CV channel models follow \citet{Belloni_2023}, with the subsequent evolution depending primarily on the helium-core mass at the onset of mass transfer. These models adopt the MB prescription of \citet{Van_2019}. We selected the WD and CV channel tracks based on the derived donor properties of the targets. Both sets of models are subject to uncertainties in the adopted MB prescriptions; a detailed comparison of these prescriptions is beyond the scope of this work, and we refer the reader to Wong et al. (in prep.) for further discussion.

            The WD and CV channel tracks are consistent with the measured donor masses and radii of both systems within the current uncertainties. The measurements appear to disfavour the He-star track, although uncertainties in the model preclude a firm conclusion from the mass--radius comparison alone. The current donor mass--radius constraints therefore do not robustly distinguish among the formation channels. More precise measurements of the donor properties could help potentially discriminate between the evolved CV, WD, and He-star channels.

        \subsubsection{Accreting WD effective temperature} \label{sub:compare_WD_teff} 

\begin{figure}
                \centering
                \includegraphics[width=\columnwidth]{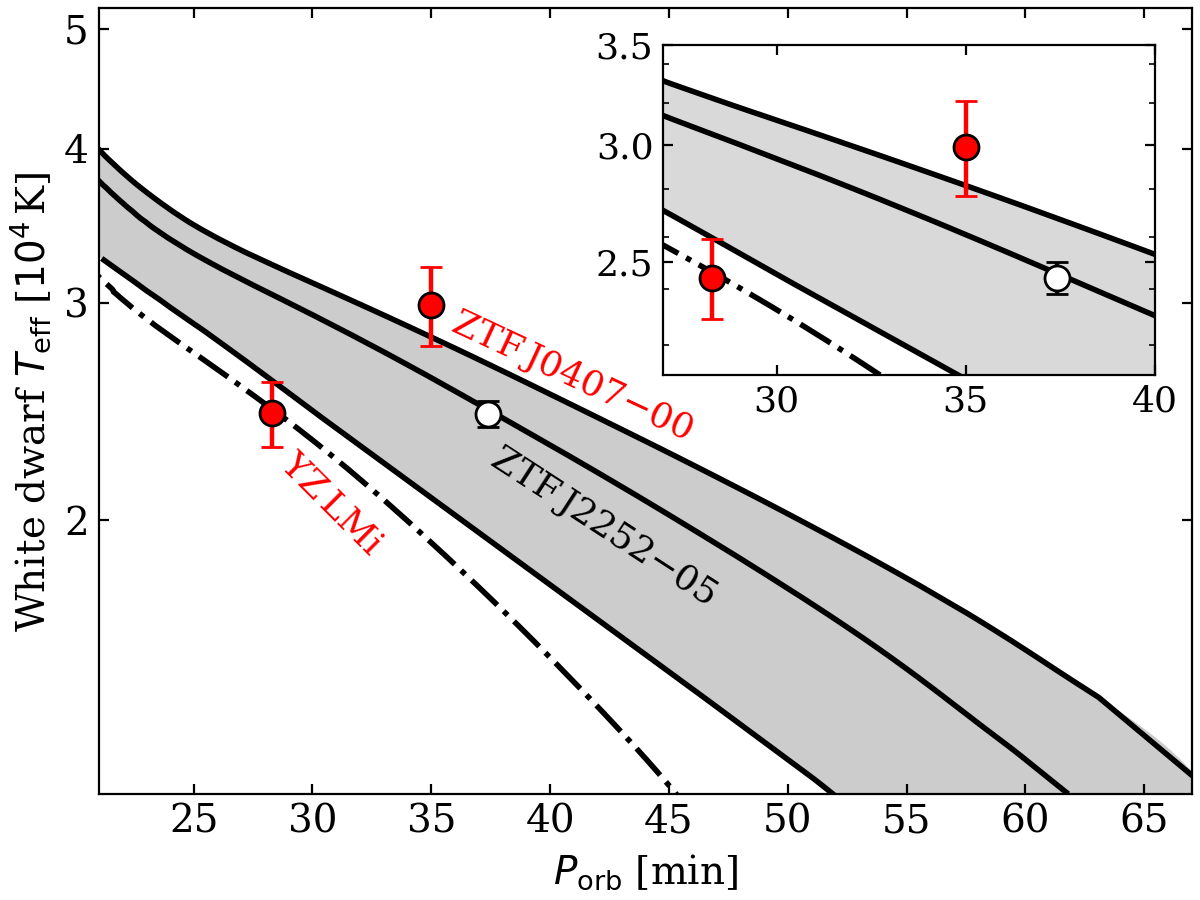}
                    \caption{Measured WD $T_{\mathrm{eff}}$ values compared with \texttt{MESA} evolutionary tracks for the WD channel. The dash-dotted curve shows evolution driven solely by GW radiation. The grey shaded region spans models with $M_{\mathrm{WD},0}=0.7$--$0.8\,M_\odot$ and RVJ MB boosting factors of 3--10. The three solid black curves are illustrative tracks consistent with the observations of ZTF\,J0407$-$00, ZTF\,J2252$-$05, and YZ\,LMi, from top to bottom, respectively.}
                \label{fig.MESA_teff}
            \end{figure}

            The thermal state of the accreting WD provides an additional probe of the secular evolution of AM\,CVn~systems. At short orbital periods, rapid accretion heats the WD through compression, linking its luminosity to the mass-transfer rate \citep{Bildsten_2006}. As the binary evolves towards longer periods and the mass-transfer rate decreases, cooling eventually dominates. The effective temperature at a given orbital period therefore depends on both the accretion history and the rate of orbital evolution \citep{Wong_2021}. In Fig.~\ref{fig.MESA_teff}, we compare our measured $T_{\mathrm{eff}}$ values with evolutionary tracks computed using \texttt{MESA} for the WD channel.

            The reference track assumes AML driven solely by GW radiation. We also show the WD channel model grid described in Sect.~\ref{sub:compare_donor_M_R}, spanning $M_{\mathrm{WD},0}=0.7$--$0.8\,M_\odot$ and RVJ boosting factors of 3--10 (grey shaded region in Fig.~\ref{fig.MESA_teff}). Three illustrative tracks are shown as solid black lines. From top to bottom, they have $M_{\mathrm{WD},0}=0.8\,M_\odot$ with a boosting factor of 10, a boosting factor of 6, and $M_{\mathrm{WD},0}=0.7\,M_\odot$ with a boosting factor of 3. These tracks are consistent with the measured temperatures of ZTF\,J0407$-$00, ZTF\,J2252$-$05, and YZ\,LMi, respectively.
            
            For YZ\,LMi, the measured $T_{\mathrm{eff}}=24\,400\pm1500$\,K is consistent with the GW-driven track and tracks that include relatively weak MB. An independent constraint on the current AML comes from the long-term eclipse-timing analysis of \citet{Green_2026}, who measured $\dot{P}\lesssim10^{-13}\,\mathrm{s\,s^{-1}}$ and excluded substantial additional AML beyond GW radiation. The measured $\dot{P}$ is somewhat smaller than the secular value predicted by purely GW-driven evolution, possibly reflecting short-term variations in the eclipse timings. Together, $T_{\mathrm{eff}}$ and $\dot{P}$ favour present-day evolution dominated by GW radiation, consistent with the WD channel, and disfavour the strong MB invoked in CV channel models \citep{Belloni_2023}.
            
            For ZTF\,J0407$-$00, the higher temperature of $T_{\mathrm{eff}}=29\,900\pm2200$\,K suggests a different thermal or accretion history. In the models shown in Fig.~\ref{fig.MESA_teff}, additional AML accelerates the orbital evolution and leaves less time for the WD to cool, producing a higher $T_{\mathrm{eff}}$ at a given orbital period. The measured temperature may therefore indicate enhanced AML, either recently or at the present epoch, relative to the GW-only case. Temperature alone cannot determine the AML history, however, because it also depends on the WD's earlier accretion and thermal evolution. Without an independent measurement of $\dot{P}$ for ZTF\,J0407$-$00, we cannot robustly distinguish among the AML prescriptions.
            
        \subsubsection{Formation-channel identification} \label{sub:channel_identification}
        
            The chemical composition of the WD atmosphere provides a strong diagnostic of the AM\,CVn formation channel because it is enriched by material transferred from the donor \citep{Nelemans_2010}. In particular, $N/C$ differs by more than an order of magnitude between the WD and He-star models considered here. A variant of the He-star channel proposed by \citet{Sarkar_2023b} assumes the donor is stripped before the onset of helium burning. However, the formation of the required post-common-envelope configurations remains uncertain. Their common-envelope treatment assumes a fixed envelope-binding parameter and requires either additional energy sources or an alternative common-envelope prescription, represented by effective common-envelope efficiencies substantially exceeding unity. As the viability and occurrence rate of these progenitors have not yet been established with self-consistent binary-evolution calculations, we do not consider this scenario as a separate fourth formation channel here. 
            
\begin{figure}
                \centering
                \includegraphics[width=\columnwidth]{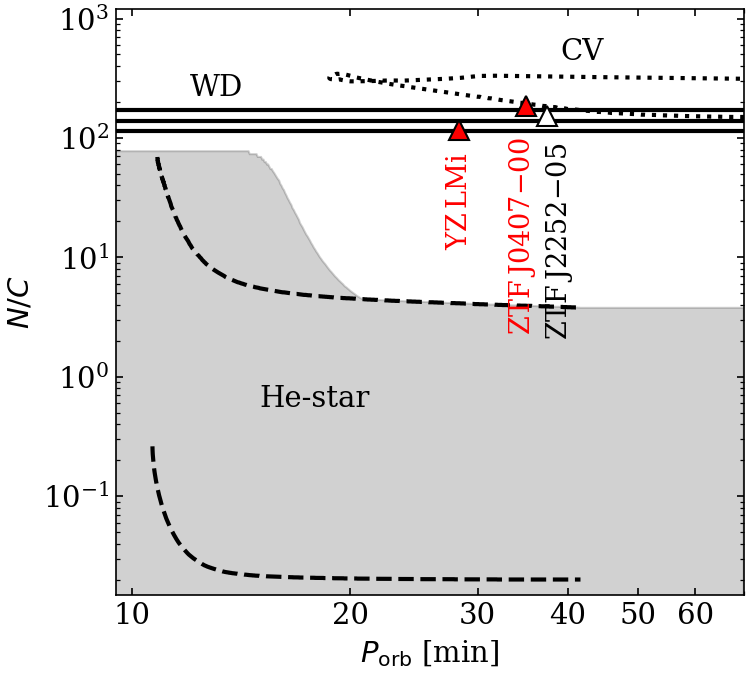}
                \caption{Measured nitrogen-to-carbon abundance ratios of YZ\,LMi ($N/C > 117$, red), ZTF\,J0407$-$00 ($N/C > 185$, red), and ZTF\,J2252$-$05 ($N/C > 153$, black; adopted from \citealt{Yu_2026}) compared with theoretical predictions for the three AM\,CVn formation channels. The WD- and He-star-channel models are taken from \citet{Nelemans_2010}. WD donor tracks with progenitor masses of 1.0, 1.5, and 2.0\,$M_\odot$ are shown as solid lines from top to bottom. The shaded region marks the range expected for He-star donors, bounded by two example tracks with initial orbital periods of 20 and 60\,min (upper and lower dashed lines, respectively). The CV channel model is adopted from \citet{Yu_2026}.} 
                \label{fig.abundance_ratios}
            \end{figure}

            In Fig.~\ref{fig.abundance_ratios}, we compare the measured $N/C$ of YZ\,LMi and ZTF\,J0407$-$00 with theoretical predictions. The WD and He-star channel models are taken from \citet{Nelemans_2010}, based on stellar evolutionary calculations and MB prescriptions from \citet{Rappaport_1983}, \citet{Eggleton_2002}, and \citet{Podsiadlowski_2002}. The CV channel model is adopted from \citet{Yu_2026}, following a computation similar to that of \citet{Belloni_2023}. By comparing the measured constraints for YZ\,LMi ($P_{\mathrm{orb}} = 28.3$\,min, $N/C > 117$) and ZTF\,J0407$-$00 ($P_{\mathrm{orb}} = 35.4$\,min, $N/C > 185$) with these models, we can rule out the He-star channel for both systems. 
            
            Because the derived $X_{\mathrm{C}}$ values are upper limits, the inferred $N/C$ ratios are lower limits. Lower values of $X_{\mathrm{C}}$ would place both systems at higher $N/C$ in Fig.~\ref{fig.abundance_ratios}, further disfavouring the canonical He-star formation channel. At the orbital period of YZ\,LMi, this channel predicts $N/C \lesssim 4.1$, requiring $X_{\mathrm{C}} \gtrsim -4.3$ for $X_{\mathrm{N}}=-3.8$. For ZTF\,J0407$-$00, the predicted $N/C \lesssim 3.9$ requires $X_{\mathrm{C}} \gtrsim -3.0$, given $X_{\mathrm{N}} = -2.5$. In both systems, the required carbon abundances exceed our upper limits by more than an order of magnitude and would produce clearly detectable carbon absorption lines in the observed spectra, as illustrated in Fig.~\ref{fig:Carbon_estimation}. The absence of these features therefore makes the canonical He-star channel difficult to reconcile with the observations.
                    
            Among the three proposed AM\,CVn formation channels, our abundance constraints strongly disfavour the He-star channel because of the high inferred $N/C$ ratios. Distinguishing between the WD and CV channels is more difficult. Traditionally, the CV channel has been expected to leave a detectable amount of hydrogen in the system, whereas no hydrogen has been identified in the optical spectra of either target \citep{Podsiadlowski_2003, Solheim_2010}. However, more recent evolutionary calculations that include enhanced MB can produce AM\,CVn~systems with hydrogen abundances as low as $X_{\mathrm{H}} \lesssim -6$, potentially below the observational detection limit \citep{Belloni_2023}.
            
            For ZTF\,J0407$-$00, its relatively high $T_{\mathrm{eff}}$ may indicate a high long-term mass-transfer rate and, consequently, an angular-momentum-loss rate exceeding that expected from GW radiation alone. Without a measurement of $\dot{P}$, however, this interpretation remains uncertain.
            
            For YZ\,LMi, the long-term orbital evolution provides an additional constraint. The eclipse-timing study of \citet{Green_2026} found no significant period variation, with $\dot{P} \lesssim 10^{-13}\,\mathrm{s\,s^{-1}}$. This is consistent with an evolutionary track dominated by GW-driven AML and disfavour CV channel models that require substantially enhanced MB. Taken together, the chemical abundances and orbital evolution therefore favour the WD channel for YZ\,LMi.
        
\section{Summary and conclusions} \label{Summary_conclusions}

    We present the analysis of \textit{HST}/COS spectroscopic data of the eclipsing AM\,CVn~systems YZ\,LMi and ZTF\,J0407$-$00. Their UV spectra are fitted with synthetic helium-atmosphere models to determine the WD effective temperatures, surface gravities, and chemical abundances.
    
    For YZ\,LMi, we measure $T_{\mathrm{eff}} = 24\,400 \pm 1500$\,K and $\log g = 8.36 \pm 0.05$, while for ZTF\,J0407$-$00 we obtain $T_{\mathrm{eff}} = 29\,900 \pm 2200$\,K and $\log g = 8.37 \pm 0.05$. Combined with the WD radii and mass ratios inferred from optical eclipse modelling, these measurements imply WD masses of $0.84 \pm 0.10\,M_\odot$ and $0.86 \pm 0.12\,M_\odot$, and donor masses of $0.033 \pm 0.008\,M_\odot$ and $0.021 \pm 0.006\,M_\odot$, respectively. The absence of detectable photospheric carbon features yields $N/C$ mass-ratio limits of $>117$ for YZ\,LMi and $>185$ for ZTF\,J0407$-$00. The differences between these UV-based temperatures and previous optical or broadband estimates further emphasise the importance of UV spectroscopy for accurately characterising AM\,CVn accreting WDs.
    
    The measured donor masses and radii are consistent with evolutionary tracks from all three proposed formation channels and therefore do not provide a strong test to distinguish among different formation channels. In contrast, the high $N/C$ ratios strongly disfavour the He-star channel for both systems. For YZ\,LMi, the chemical abundances and the long-term constraint on $\dot{P}$ from \citet{Green_2026} favour the WD channel and an evolution dominated by GW-driven angular-momentum loss. The WD channel is also favoured for ZTF\,J0407$-$00, although its comparatively high WD temperature may indicate an enhanced mass-transfer rate, and the CV channel cannot yet be excluded.
    
    Together with ZTF\,J2252$-$05 \citep{Yu_2026}, these results increase the number of AM\,CVn~systems with detailed UV atmosphere modelling from one to three. Expanding this sample will be essential for establishing the diversity of accreting WD compositions and determining the relative contributions of the different formation channels. Such empirical constraints are particularly important because AM\,CVn~systems are expected to contribute both as individually resolvable verification sources and a Galactic foreground for \textit{LISA}. Electromagnetic observations can therefore provide valuable guidance for population-synthesis studies \citep[e.g.][]{Breivik_2020, Korol_2022} and improve predictions for the Galactic AM\,CVn~population.


\begin{acknowledgements}
We thank Stuart Littlefair and Elena Mason for valuable insights and discussions. 

This research is based on observations made with the NASA/ESA \textit{Hubble Space Telescope} obtained from the Space Telescope Science Institute, which is operated by the Association of Universities for Research in Astronomy, Inc., under NASA contract NAS5–26555. These observations are associated with program 16659. 

We acknowledge with thanks the variable star observations from the AAVSO International Database contributed by observers worldwide and used in this research. We thank the members of the Spanish Observers of Supernovae (ObSN) group for their valuable photometric contributions. 

This research was supported by Deutsche Forschungsgemeinschaft  (DFG, German Research Foundation) under Germany’s Excellence Strategy - EXC 2121 "Quantum Universe" – 390833306. Co-funded by the European Union (ERC, CompactBINARIES, 101078773). Views and opinions expressed are however those of the author(s) only and do not necessarily reflect those of the European Union or the European Research Council. Neither the European Union nor the granting authority can be held responsible for them. 

This work is supported by the National Natural Science Foundation of China (NSFC, Nos. W2632018, 12288102, 12125303, 12525304), the Strategic Priority Research Program of the Chinese Academy of Sciences (grant Nos. XDB1160303, XDB1160300, XDB1160000, XDB1160200, XDB1160201), the CAS Project for Young Scientists in Basic Research (YSBR-148), the Yunnan Revitalization Talent Support Program ``YunLing Scholar'' project, International Centre of Supernovae (ICESUN), Yunnan Key Laboratory of Supernova Research (No. 202505AV340004),  the New Cornerstone Science Foundation through the XPLORER PRIZE, and the Yunnan Revitalization Talent Support Program-Science \& Technology Champion Project (No. 202305AB350003).

MRS thanks for support from the research unit FOR2990 (eRO-STEP: Stellar endpoints with eROSITA) funded by the Deutsche Forschungsgemeinschaft and the internal research project PI\_LIR\_26\_07 at USM.

DdM acknowledges support from INAF Astrofisica Fondamentale 2025.

OT acknowledges Proyectos Internos USM 2025 - PI\_LII\_2025\_03 and FONDECYT iniciación 11260733.

GT was supported by the grant  IN108426 from the Programa de Apoyo a Proyectos de Investigaci\'on e Innovaci\'on Tecnol\'ogica (DGAPA-PAPIIT).

\end{acknowledgements}

\bibliographystyle{aa}
\bibliography{references}

@ARTICLE{Amaro-Seoane_2023,
       author = {{Amaro-Seoane}, Pau and {Andrews}, Jeff and {Arca Sedda}, Manuel and {Askar}, Abbas and {Baghi}, Quentin and {Balasov}, Razvan and {Bartos}, Imre and {Bavera}, Simone S. and {Bellovary}, Jillian and {Berry}, Christopher P.~L. and {Berti}, Emanuele and {Bianchi}, Stefano and {Blecha}, Laura and {Blondin}, St{\'e}phane and {Bogdanovi{\'c}}, Tamara and {Boissier}, Samuel and {Bonetti}, Matteo and {Bonoli}, Silvia and {Bortolas}, Elisa and {Breivik}, Katelyn and {Capelo}, Pedro R. and {Caramete}, Laurentiu and {Cattorini}, Federico and {Charisi}, Maria and {Chaty}, Sylvain and {Chen}, Xian and {Chru{\'s}li{\'n}ska}, Martyna and {Chua}, Alvin J.~K. and {Church}, Ross and {Colpi}, Monica and {D'Orazio}, Daniel and {Danielski}, Camilla and {Davies}, Melvyn B. and {Dayal}, Pratika and {De Rosa}, Alessandra and {Derdzinski}, Andrea and {Destounis}, Kyriakos and {Dotti}, Massimo and {Du{\c{t}}an}, Ioana and {Dvorkin}, Irina and {Fabj}, Gaia and {Foglizzo}, Thierry and {Ford}, Saavik and {Fouvry}, Jean-Baptiste and {Franchini}, Alessia and {Fragos}, Tassos and {Fryer}, Chris and {Gaspari}, Massimo and {Gerosa}, Davide and {Graziani}, Luca and {Groot}, Paul and {Habouzit}, Melanie and {Haggard}, Daryl and {Haiman}, Zoltan and {Han}, Wen-Biao and {Istrate}, Alina and {Johansson}, Peter H. and {Khan}, Fazeel Mahmood and {Kimpson}, Tomas and {Kokkotas}, Kostas and {Kong}, Albert and {Korol}, Valeriya and {Kremer}, Kyle and {Kupfer}, Thomas and {Lamberts}, Astrid and {Larson}, Shane and {Lau}, Mike and {Liu}, Dongliang and {Lloyd-Ronning}, Nicole and {Lodato}, Giuseppe and {Lupi}, Alessandro and {Ma}, Chung-Pei and {Maccarone}, Tomas and {Mandel}, Ilya and {Mangiagli}, Alberto and {Mapelli}, Michela and {Mathis}, St{\'e}phane and {Mayer}, Lucio and {McGee}, Sean and {McKernan}, Berry and {Miller}, M. Coleman and {Mota}, David F. and {Mumpower}, Matthew and {Nasim}, Syeda S. and {Nelemans}, Gijs and {Noble}, Scott and {Pacucci}, Fabio and {Panessa}, Francesca and {Paschalidis}, Vasileios and {Pfister}, Hugo and {Porquet}, Delphine and {Quenby}, John and {Ricarte}, Angelo and {R{\"o}pke}, Friedrich K. and {Regan}, John and {Rosswog}, Stephan and {Ruiter}, Ashley and {Ruiz}, Milton and {Runnoe}, Jessie and {Schneider}, Raffaella and {Schnittman}, Jeremy and {Secunda}, Amy and {Sesana}, Alberto and {Seto}, Naoki and {Shao}, Lijing and {Shapiro}, Stuart and {Sopuerta}, Carlos and {Stone}, Nicholas C. and {Suvorov}, Arthur and {Tamanini}, Nicola and {Tamfal}, Tomas and {Tauris}, Thomas and {Temmink}, Karel and {Tomsick}, John and {Toonen}, Silvia and {Torres-Orjuela}, Alejandro and {Toscani}, Martina and {Tsokaros}, Antonios and {Unal}, Caner and {V{\'a}zquez-Aceves}, Ver{\'o}nica and {Valiante}, Rosa and {van Putten}, Maurice and {van Roestel}, Jan and {Vignali}, Christian and {Volonteri}, Marta and {Wu}, Kinwah and {Younsi}, Ziri and {Yu}, Shenghua and {Zane}, Silvia and {Zwick}, Lorenz and {Antonini}, Fabio and {Baibhav}, Vishal and {Barausse}, Enrico and {Bonilla Rivera}, Alexander and {Branchesi}, Marica and {Branduardi-Raymont}, Graziella and {Burdge}, Kevin and {Chakraborty}, Srija and {Cuadra}, Jorge and {Dage}, Kristen and {Davis}, Benjamin and {de Mink}, Selma E. and {Decarli}, Roberto and {Doneva}, Daniela and {Escoffier}, Stephanie and {Gandhi}, Poshak and {Haardt}, Francesco and {Lousto}, Carlos O. and {Nissanke}, Samaya and {Nordhaus}, Jason and {O'Shaughnessy}, Richard and {Portegies Zwart}, Simon and {Pound}, Adam and {Schussler}, Fabian and {Sergijenko}, Olga and {Spallicci}, Alessandro and {Vernieri}, Daniele and {Vigna-G{\'o}mez}, Alejandro},
        title = "{Astrophysics with the Laser Interferometer Space Antenna}",
      journal = {Living Reviews in Relativity},
         year = 2023,
        month = dec,
       volume = {26},
       number = {1},
          eid = {2},
        pages = {2},
          doi = {10.1007/s41114-022-00041-y},
    archivePrefix = {arXiv},
       eprint = {2203.06016},
    primaryClass = {gr-qc},
       adsurl = {https://ui.adsabs.harvard.edu/abs/2023LRR....26....2A}
    }

@ARTICLE{Ajith_2025,
       author = {{Ajith}, Parameswaran and {Seoane}, Pau Amaro and {Arca Sedda}, Manuel and {Arcodia}, Riccardo and {Badaracco}, Francesca and {Banerjee}, Biswajit and {Belgacem}, Enis and {Benetti}, Giovanni and {Benetti}, Stefano and {Bobrick}, Alexey and {Bonforte}, Alessandro and {Bortolas}, Elisa and {Braito}, Valentina and {Branchesi}, Marica and {Burrows}, Adam and {Cappellaro}, Enrico and {Della Ceca}, Roberto and {Chakraborty}, Chandrachur and {Subrahmanya}, Shreevathsa Chalathadka and {Coughlin}, Michael W. and {Covino}, Stefano and {Derdzinski}, Andrea and {Doshi}, Aayushi and {Falanga}, Maurizio and {Foffa}, Stefano and {Franchini}, Alessia and {Frigeri}, Alessandro and {Futaana}, Yoshifumi and {Gerberding}, Oliver and {Gill}, Kiranjyot and {Di Giovanni}, Matteo and {Giudice}, Ines Francesca and {Giustini}, Margherita and {Gl{\"a}ser}, Philipp and {Harms}, Jan and {van Heijningen}, Joris and {Iacovelli}, Francesco and {Kavanagh}, Bradley J. and {Kawamura}, Taichi and {Kenath}, Arun and {Keppler}, Elisabeth-Adelheid and {Kobayashi}, Chiaki and {Komatsu}, Goro and {Korol}, Valeriya and {Krishnendu}, N.~V. and {Kumar}, Prayush and {Longo}, Francesco and {Maggiore}, Michele and {Mancarella}, Michele and {Maselli}, Andrea and {Mastrobuono-Battisti}, Alessandra and {Mazzarini}, Francesco and {Melandri}, Andrea and {Melini}, Daniele and {Menina}, Sabrina and {Miniutti}, Giovanni and {Mitra}, Deeshani and {Mor{\'a}n-Fraile}, Javier and {Mukherjee}, Suvodip and {Muttoni}, Niccol{\`o} and {Olivieri}, Marco and {Onori}, Francesca and {Papa}, Maria Alessandra and {Patat}, Ferdinando and {Perali}, Andrea and {Piran}, Tsvi and {Piranomonte}, Silvia and {Pol}, Alberto Roper and {Pookkillath}, Masroor C. and {Prasad}, R. and {Prasad}, Vaishak and {De Rosa}, Alessandra and {Chowdhury}, Sourav Roy and {Serafinelli}, Roberto and {Sesana}, Alberto and {Severgnini}, Paola and {Stallone}, Angela and {Tissino}, Jacopo and {Tkal{\v{c}}i{\'c}}, Hrvoje and {Tomasella}, Lina and {Toscani}, Martina and {Vartanyan}, David and {Vignali}, Cristian and {Zaccarelli}, Lucia and {Zeoli}, Morgane and {Zuccarello}, Luciano},
        title = "{The Lunar Gravitational-wave Antenna: mission studies and science case}",
      journal = {\jcap},
         year = 2025,
        month = jan,
       volume = {2025},
       number = {1},
          eid = {108},
        pages = {108},
          doi = {10.1088/1475-7516/2025/01/108},
archivePrefix = {arXiv},
       eprint = {2404.09181},
 primaryClass = {gr-qc},
       adsurl = {https://ui.adsabs.harvard.edu/abs/2025JCAP...01..108A}
}

@ARTICLE{Anderson_2005,
       author = {{Anderson}, Scott F. and {Haggard}, Daryl and {Homer}, Lee and {Joshi}, Nikhil R. and {Margon}, Bruce and {Silvestri}, Nicole M. and {Szkody}, Paula and {Wolfe}, Michael A. and {Agol}, Eric and {Becker}, Andrew C. and {Henden}, Arne and {Hall}, Patrick B. and {Knapp}, Gillian R. and {Richmond}, Michael W. and {Schneider}, Donald P. and {Stinson}, Gregory and {Barentine}, J.~C. and {Brewington}, Howard J. and {Brinkmann}, J. and {Harvanek}, Michael and {Kleinman}, S.~J. and {Krzesinski}, Jurek and {Long}, Dan and {Neilsen}, Jr., Eric H. and {Nitta}, Atsuko and {Snedden}, Stephanie A.},
        title = "{Ultracompact AM Canum Venaticorum Binaries from the Sloan Digital Sky Survey: Three Candidates Plus the First Confirmed Eclipsing System}",
      journal = {\aj},
         year = 2005,
        month = nov,
       volume = {130},
       number = {5},
        pages = {2230-2236},
          doi = {10.1086/491587},
archivePrefix = {arXiv},
       eprint = {astro-ph/0506730},
 primaryClass = {astro-ph},
       adsurl = {https://ui.adsabs.harvard.edu/abs/2005AJ....130.2230A}
}

@ARTICLE{Bailer-Jones_2021,
       author = {{Bailer-Jones}, C.~A.~L. and {Rybizki}, J. and {Fouesneau}, M. and {Demleitner}, M. and {Andrae}, R.},
        title = "{Estimating Distances from Parallaxes. V. Geometric and Photogeometric Distances to 1.47 Billion Stars in Gaia Early Data Release 3}",
      journal = {\aj},
         year = 2021,
        month = mar,
       volume = {161},
       number = {3},
          eid = {147},
        pages = {147},
          doi = {10.3847/1538-3881/abd806},
archivePrefix = {arXiv},
       eprint = {2012.05220},
 primaryClass = {astro-ph.SR},
       adsurl = {https://ui.adsabs.harvard.edu/abs/2021AJ....161..147B}
}

@ARTICLE{Bauer_2021,
       author = {{Bauer}, Evan B. and {Kupfer}, Thomas},
        title = "{Phases of Mass Transfer from Hot Subdwarfs to White Dwarf Companions and Their Photometric Properties}",
      journal = {\apj},
         year = 2021,
        month = dec,
       volume = {922},
       number = {2},
          eid = {245},
        pages = {245},
          doi = {10.3847/1538-4357/ac25f0},
archivePrefix = {arXiv},
       eprint = {2106.13297},
 primaryClass = {astro-ph.SR},
       adsurl = {https://ui.adsabs.harvard.edu/abs/2021ApJ...922..245B}
}

@ARTICLE{Bedard_2020,
       author = {{B{\'e}dard}, A. and {Bergeron}, P. and {Brassard}, P. and {Fontaine}, G.},
        title = "{On the Spectral Evolution of Hot White Dwarf Stars. I. A Detailed Model Atmosphere Analysis of Hot White Dwarfs from SDSS DR12}",
      journal = {\apj},
         year = 2020,
        month = oct,
       volume = {901},
       number = {2},
          eid = {93},
        pages = {93},
          doi = {10.3847/1538-4357/abafbe},
archivePrefix = {arXiv},
       eprint = {2008.07469},
 primaryClass = {astro-ph.SR},
       adsurl = {https://ui.adsabs.harvard.edu/abs/2020ApJ...901...93B}
}

@ARTICLE{Bellm_2019,
       author = {{Bellm}, Eric C. and {Kulkarni}, Shrinivas R. and {Graham}, Matthew J. and {Dekany}, Richard and {Smith}, Roger M. and {Riddle}, Reed and {Masci}, Frank J. and {Helou}, George and {Prince}, Thomas A. and {Adams}, Scott M. and {Barbarino}, C. and {Barlow}, Tom and {Bauer}, James and {Beck}, Ron and {Belicki}, Justin and {Biswas}, Rahul and {Blagorodnova}, Nadejda and {Bodewits}, Dennis and {Bolin}, Bryce and {Brinnel}, Valery and {Brooke}, Tim and {Bue}, Brian and {Bulla}, Mattia and {Burruss}, Rick and {Cenko}, S. Bradley and {Chang}, Chan-Kao and {Connolly}, Andrew and {Coughlin}, Michael and {Cromer}, John and {Cunningham}, Virginia and {De}, Kishalay and {Delacroix}, Alex and {Desai}, Vandana and {Duev}, Dmitry A. and {Eadie}, Gwendolyn and {Farnham}, Tony L. and {Feeney}, Michael and {Feindt}, Ulrich and {Flynn}, David and {Franckowiak}, Anna and {Frederick}, S. and {Fremling}, C. and {Gal-Yam}, Avishay and {Gezari}, Suvi and {Giomi}, Matteo and {Goldstein}, Daniel A. and {Golkhou}, V. Zach and {Goobar}, Ariel and {Groom}, Steven and {Hacopians}, Eugean and {Hale}, David and {Henning}, John and {Ho}, Anna Y.~Q. and {Hover}, David and {Howell}, Justin and {Hung}, Tiara and {Huppenkothen}, Daniela and {Imel}, David and {Ip}, Wing-Huen and {Ivezi{\'c}}, {\v{Z}}eljko and {Jackson}, Edward and {Jones}, Lynne and {Juric}, Mario and {Kasliwal}, Mansi M. and {Kaspi}, S. and {Kaye}, Stephen and {Kelley}, Michael S.~P. and {Kowalski}, Marek and {Kramer}, Emily and {Kupfer}, Thomas and {Landry}, Walter and {Laher}, Russ R. and {Lee}, Chien-De and {Lin}, Hsing Wen and {Lin}, Zhong-Yi and {Lunnan}, Ragnhild and {Giomi}, Matteo and {Mahabal}, Ashish and {Mao}, Peter and {Miller}, Adam A. and {Monkewitz}, Serge and {Murphy}, Patrick and {Ngeow}, Chow-Choong and {Nordin}, Jakob and {Nugent}, Peter and {Ofek}, Eran and {Patterson}, Maria T. and {Penprase}, Bryan and {Porter}, Michael and {Rauch}, Ludwig and {Rebbapragada}, Umaa and {Reiley}, Dan and {Rigault}, Mickael and {Rodriguez}, Hector and {van Roestel}, Jan and {Rusholme}, Ben and {van Santen}, Jakob and {Schulze}, S. and {Shupe}, David L. and {Singer}, Leo P. and {Soumagnac}, Maayane T. and {Stein}, Robert and {Surace}, Jason and {Sollerman}, Jesper and {Szkody}, Paula and {Taddia}, F. and {Terek}, Scott and {Van Sistine}, Angela and {van Velzen}, Sjoert and {Vestrand}, W. Thomas and {Walters}, Richard and {Ward}, Charlotte and {Ye}, Quan-Zhi and {Yu}, Po-Chieh and {Yan}, Lin and {Zolkower}, Jeffry},
        title = "{The Zwicky Transient Facility: System Overview, Performance, and First Results}",
      journal = {\pasp},
         year = 2019,
        month = jan,
       volume = {131},
       number = {995},
        pages = {018002},
          doi = {10.1088/1538-3873/aaecbe},
archivePrefix = {arXiv},
       eprint = {1902.01932},
 primaryClass = {astro-ph.IM},
       adsurl = {https://ui.adsabs.harvard.edu/abs/2019PASP..131a8002B}
}

@ARTICLE{Belloni_2023,
       author = {{Belloni}, Diogo and {Schreiber}, Matthias R.},
        title = "{Reversing the verdict: Cataclysmic variables could be the dominant progenitors of AM CVn binaries after all}",
      journal = {\aap},
         year = 2023,
        month = oct,
       volume = {678},
          eid = {A34},
        pages = {A34},
          doi = {10.1051/0004-6361/202347047},
archivePrefix = {arXiv},
       eprint = {2305.19312},
 primaryClass = {astro-ph.SR},
       adsurl = {https://ui.adsabs.harvard.edu/abs/2023A&A...678A..34B}
}

@ARTICLE{Bildsten_2006,
       author = {{Bildsten}, Lars and {Townsley}, Dean M. and {Deloye}, Christopher J. and {Nelemans}, Gijs},
        title = "{The Thermal State of the Accreting White Dwarf in AM Canum Venaticorum Binaries}",
      journal = {\apj},
         year = 2006,
        month = mar,
       volume = {640},
       number = {1},
        pages = {466-473},
          doi = {10.1086/500080},
archivePrefix = {arXiv},
       eprint = {astro-ph/0510652},
 primaryClass = {astro-ph},
       adsurl = {https://ui.adsabs.harvard.edu/abs/2006ApJ...640..466B}
}

@ARTICLE{Brown_2013,
       author = {{Brown}, T.~M. and {Baliber}, N. and {Bianco}, F.~B. and {Bowman}, M. and {Burleson}, B. and {Conway}, P. and {Crellin}, M. and {Depagne}, {\'E}. and {De Vera}, J. and {Dilday}, B. and {Dragomir}, D. and {Dubberley}, M. and {Eastman}, J.~D. and {Elphick}, M. and {Falarski}, M. and {Foale}, S. and {Ford}, M. and {Fulton}, B.~J. and {Garza}, J. and {Gomez}, E.~L. and {Graham}, M. and {Greene}, R. and {Haldeman}, B. and {Hawkins}, E. and {Haworth}, B. and {Haynes}, R. and {Hidas}, M. and {Hjelstrom}, A.~E. and {Howell}, D.~A. and {Hygelund}, J. and {Lister}, T.~A. and {Lobdill}, R. and {Martinez}, J. and {Mullins}, D.~S. and {Norbury}, M. and {Parrent}, J. and {Paulson}, R. and {Petry}, D.~L. and {Pickles}, A. and {Posner}, V. and {Rosing}, W.~E. and {Ross}, R. and {Sand}, D.~J. and {Saunders}, E.~S. and {Shobbrook}, J. and {Shporer}, A. and {Street}, R.~A. and {Thomas}, D. and {Tsapras}, Y. and {Tufts}, J.~R. and {Valenti}, S. and {Vander Horst}, K. and {Walker}, Z. and {White}, G. and {Willis}, M.},
        title = "{Las Cumbres Observatory Global Telescope Network}",
      journal = {\pasp},
         year = 2013,
        month = sep,
       volume = {125},
       number = {931},
        pages = {1031},
          doi = {10.1086/673168},
archivePrefix = {arXiv},
       eprint = {1305.2437},
 primaryClass = {astro-ph.IM},
       adsurl = {https://ui.adsabs.harvard.edu/abs/2013PASP..125.1031B}
}

@ARTICLE{Breivik_2020,
       author = {{Breivik}, Katelyn and {Coughlin}, Scott and {Zevin}, Michael and {Rodriguez}, Carl L. and {Kremer}, Kyle and {Ye}, Claire S. and {Andrews}, Jeff J. and {Kurkowski}, Michael and {Digman}, Matthew C. and {Larson}, Shane L. and {Rasio}, Frederic A.},
        title = "{COSMIC Variance in Binary Population Synthesis}",
      journal = {\apj},
         year = 2020,
        month = jul,
       volume = {898},
       number = {1},
          eid = {71},
        pages = {71},
          doi = {10.3847/1538-4357/ab9d85},
archivePrefix = {arXiv},
       eprint = {1911.00903},
 primaryClass = {astro-ph.HE},
       adsurl = {https://ui.adsabs.harvard.edu/abs/2020ApJ...898...71B}
}

@ARTICLE{Burdge_2020,
       author = {{Burdge}, Kevin B. and {Prince}, Thomas A. and {Fuller}, Jim and {Kaplan}, David L. and {Marsh}, Thomas R. and {Tremblay}, Pier-Emmanuel and {Zhuang}, Zhuyun and {Bellm}, Eric C. and {Caiazzo}, Ilaria and {Coughlin}, Michael W. and {Dhillon}, Vik S. and {Gaensicke}, Boris and {Rodr{\'\i}guez-Gil}, Pablo and {Graham}, Matthew J. and {Hermes}, JJ and {Kupfer}, Thomas and {Littlefair}, S.~P. and {Mr{\'o}z}, Przemek and {Phinney}, E.~S. and {van Roestel}, Jan and {Yao}, Yuhan and {Dekany}, Richard G. and {Drake}, Andrew J. and {Duev}, Dmitry A. and {Hale}, David and {Feeney}, Michael and {Helou}, George and {Kaye}, Stephen and {Mahabal}, Ashish. A. and {Masci}, Frank J. and {Riddle}, Reed and {Smith}, Roger and {Soumagnac}, Maayane T. and {Kulkarni}, S.~R.},
        title = "{A Systematic Search of Zwicky Transient Facility Data for Ultracompact Binary LISA-detectable Gravitational-wave Sources}",
      journal = {\apj},
         year = 2020,
        month = dec,
       volume = {905},
       number = {1},
          eid = {32},
        pages = {32},
          doi = {10.3847/1538-4357/abc261},
archivePrefix = {arXiv},
       eprint = {2009.02567},
 primaryClass = {astro-ph.SR},
       adsurl = {https://ui.adsabs.harvard.edu/abs/2020ApJ...905...32B}
}

@ARTICLE{Cardelli_1989,
       author = {{Cardelli}, Jason A. and {Clayton}, Geoffrey C. and {Mathis}, John S.},
        title = "{The Relationship between Infrared, Optical, and Ultraviolet Extinction}",
      journal = {\apj},
         year = 1989,
        month = oct,
       volume = {345},
        pages = {245},
          doi = {10.1086/167900},
       adsurl = {https://ui.adsabs.harvard.edu/abs/1989ApJ...345..245C}
}

@ARTICLE{Chen_2022,
       author = {{Chen}, Hai-Liang and {Chen}, Xuefei and {Han}, Zhanwen},
        title = "{Evolution of AM CVn Binaries with White Dwarf Donors}",
      journal = {\apj},
         year = 2022,
        month = aug,
       volume = {935},
       number = {1},
          eid = {9},
        pages = {9},
          doi = {10.3847/1538-4357/ac7fec},
archivePrefix = {arXiv},
       eprint = {2207.04592},
 primaryClass = {astro-ph.SR},
       adsurl = {https://ui.adsabs.harvard.edu/abs/2022ApJ...935....9C}
}

@ARTICLE{Chickles_2025,
       author = {{Chickles}, Emma T. and {Burdge}, Kevin B. and {Chakraborty}, Joheen and {Dhillon}, Vik S. and {Draghis}, Paul and {Munday}, James and {Rappaport}, Saul A. and {Tonry}, John and {Bauer}, Evan B. and {Brown}, Alex J. and {Castro}, Noel and {Chakrabarty}, Deepto and {Dyer}, Martin and {El-Badry}, Kareem and {Frebel}, Anna and {Furesz}, Gabor and {Garbutt}, James and {Green}, Matthew J. and {Householder}, Aaron and {Hughes}, Scott A. and {Jarvis}, Daniel and {Kara}, Erin and {Kennedy}, Mark R. and {Kerry}, Paul and {Littlefair}, Stuart P. and {McCormac}, James and {Mo}, Geoffrey and {Ng}, Mason and {Parsons}, Steven and {Pelisoli}, Ingrid and {Pike}, Eleanor and {Prince}, Thomas A. and {Ricker}, George R. and {van Roestel}, Jan and {Sahman}, David and {Shen}, Ken J. and {Simcoe}, Robert A. and {Tremblay}, Pier-Emmanuel and {Vanderburg}, Andrew and {Wong}, Tin Long Sunny},
        title = "{A Gravitational-wave-detectable Candidate Type Ia Supernova Progenitor}",
      journal = {\apj},
         year = 2025,
        month = jul,
       volume = {987},
       number = {2},
          eid = {206},
        pages = {206},
          doi = {10.3847/1538-4357/add34c},
archivePrefix = {arXiv},
       eprint = {2411.19916},
 primaryClass = {astro-ph.SR},
       adsurl = {https://ui.adsabs.harvard.edu/abs/2025ApJ...987..206C}
}

@ARTICLE{Chickles_2026,
       author = {{Chickles}, Emma T. and {Chakraborty}, Joheen and {Burdge}, Kevin B. and {Dhillon}, Vik S. and {Draghis}, Paul and {El-Badry}, Kareem and {Green}, Matthew J. and {Householder}, Aaron and {Hughes}, Sarah and {Layden}, Christopher and {Littlefair}, Stuart P. and {Munday}, James and {Pelisoli}, Ingrid and {Redden}, Maya S. and {Tonry}, John and {van Roestel}, Jan and {Angile}, Francesco Elio and {Brown}, Alex J. and {Castro Segura}, Noel and {Dinsmore}, Jack and {Dyer}, Martin and {Furesz}, Gabor and {Gabutti}, Michelle and {Garbutt}, James and {Garc{\'\i}a-Mej{\'\i}a}, Juliana and {Jarvis}, Daniel and {Kennedy}, Mark R. and {Kerry}, Paul and {McCormac}, James and {Mo}, Geoffrey and {Osip}, Dave and {Parsons}, Steven and {Pike}, Eleanor and {Piotrowski}, John J. and {Romani}, Roger W. and {Sahman}, David and {Simcoe}, Rob},
        title = "{An Eclipsing 8.56 Minutes Orbital Period Mass-transferring Binary}",
      journal = {\apj},
         year = 2026,
        month = apr,
       volume = {1000},
       number = {2},
          eid = {237},
        pages = {237},
          doi = {10.3847/1538-4357/ae4871},
archivePrefix = {arXiv},
       eprint = {2601.07925},
 primaryClass = {astro-ph.SR},
       adsurl = {https://ui.adsabs.harvard.edu/abs/2026ApJ..1000..237C}
}

@ARTICLE{Copperwheat_2010,
       author = {{Copperwheat}, C.~M. and {Marsh}, T.~R. and {Dhillon}, V.~S. and {Littlefair}, S.~P. and {Hickman}, R. and {G{\"a}nsicke}, B.~T. and {Southworth}, J.},
        title = "{Physical properties of IP Pegasi: an eclipsing dwarf nova with an unusually cool white dwarf}",
      journal = {\mnras},
         year = 2010,
        month = mar,
       volume = {402},
       number = {3},
        pages = {1824-1840},
          doi = {10.1111/j.1365-2966.2009.16010.x},
archivePrefix = {arXiv},
       eprint = {0911.1637},
 primaryClass = {astro-ph.SR},
       adsurl = {https://ui.adsabs.harvard.edu/abs/2010MNRAS.402.1824C}
}

@ARTICLE{Copperwheat_2011,
       author = {{Copperwheat}, C.~M. and {Marsh}, T.~R. and {Littlefair}, S.~P. and {Dhillon}, V.~S. and {Ramsay}, G. and {Drake}, A.~J. and {G{\"a}nsicke}, B.~T. and {Groot}, P.~J. and {Hakala}, P. and {Koester}, D. and {Nelemans}, G. and {Roelofs}, G. and {Southworth}, J. and {Steeghs}, D. and {Tulloch}, S.},
        title = "{SDSS J0926+3624: the shortest period eclipsing binary star}",
      journal = {\mnras},
         year = 2011,
        month = jan,
       volume = {410},
       number = {2},
        pages = {1113-1129},
          doi = {10.1111/j.1365-2966.2010.17508.x},
archivePrefix = {arXiv},
       eprint = {1008.1907},
 primaryClass = {astro-ph.SR},
       adsurl = {https://ui.adsabs.harvard.edu/abs/2011MNRAS.410.1113C}
}

@ARTICLE{Danforth_2016,
       author = {{Danforth}, Charles W. and {Keeney}, Brian A. and {Tilton}, Evan M. and {Shull}, J. Michael and {Stocke}, John T. and {Stevans}, Matthew and {Pieri}, Matthew M. and {Savage}, Blair D. and {France}, Kevin and {Syphers}, David and {Smith}, Britton D. and {Green}, James C. and {Froning}, Cynthia and {Penton}, Steven V. and {Osterman}, Steven N.},
        title = "{An HST/COS Survey of the Low-redshift Intergalactic Medium. I. Survey, Methodology, and Overall Results}",
      journal = {\apj},
         year = 2016,
        month = feb,
       volume = {817},
       number = {2},
          eid = {111},
        pages = {111},
          doi = {10.3847/0004-637X/817/2/111},
archivePrefix = {arXiv},
       eprint = {1402.2655},
 primaryClass = {astro-ph.CO},
       adsurl = {https://ui.adsabs.harvard.edu/abs/2016ApJ...817..111D}
}

@ARTICLE{Deloye_2007,
       author = {{Deloye}, Christopher J. and {Taam}, Ronald E. and {Winisdoerffer}, Christophe and {Chabrier}, Gilles},
        title = "{The thermal evolution of the donors in AM Canum Venaticorum binaries}",
      journal = {\mnras},
         year = 2007,
        month = oct,
       volume = {381},
       number = {2},
        pages = {525-542},
          doi = {10.1111/j.1365-2966.2007.12262.x},
archivePrefix = {arXiv},
       eprint = {0708.0220},
 primaryClass = {astro-ph},
       adsurl = {https://ui.adsabs.harvard.edu/abs/2007MNRAS.381..525D}
}

@ARTICLE{de_Jong_2019,
       author = {{de Jong}, R.~S. and {Agertz}, O. and {Berbel}, A.~A. and {Aird}, J. and {Alexander}, D.~A. and {Amarsi}, A. and {Anders}, F. and {Andrae}, R. and {Ansarinejad}, B. and {Ansorge}, W. and {Antilogus}, P. and {Anwand-Heerwart}, H. and {Arentsen}, A. and {Arnadottir}, A. and {Asplund}, M. and {Auger}, M. and {Azais}, N. and {Baade}, D. and {Baker}, G. and {Baker}, S. and {Balbinot}, E. and {Baldry}, I.~K. and {Banerji}, M. and {Barden}, S. and {Barklem}, P. and {Barth{\'e}l{\'e}my-Mazot}, E. and {Battistini}, C. and {Bauer}, S. and {Bell}, C.~P.~M. and {Bellido-Tirado}, O. and {Bellstedt}, S. and {Belokurov}, V. and {Bensby}, T. and {Bergemann}, M. and {Bestenlehner}, J.~M. and {Bielby}, R. and {Bilicki}, M. and {Blake}, C. and {Bland-Hawthorn}, J. and {Boeche}, C. and {Boland}, W. and {Boller}, T. and {Bongard}, S. and {Bongiorno}, A. and {Bonifacio}, P. and {Boudon}, D. and {Brooks}, D. and {Brown}, M.~J.~I. and {Brown}, R. and {Br{\"u}ggen}, M. and {Brynnel}, J. and {Brzeski}, J. and {Buchert}, T. and {Buschkamp}, P. and {Caffau}, E. and {Caillier}, P. and {Carrick}, J. and {Casagrande}, L. and {Case}, S. and {Casey}, A. and {Cesarini}, I. and {Cescutti}, G. and {Chapuis}, D. and {Chiappini}, C. and {Childress}, M. and {Christlieb}, N. and {Church}, R. and {Cioni}, M.-R.~L. and {Cluver}, M. and {Colless}, M. and {Collett}, T. and {Comparat}, J. and {Cooper}, A. and {Couch}, W. and {Courbin}, F. and {Croom}, S. and {Croton}, D. and {Daguis{\'e}}, E. and {Dalton}, G. and {Davies}, L.~J.~M. and {Davis}, T. and {de Laverny}, P. and {Deason}, A. and {Dionies}, F. and {Disseau}, K. and {Doel}, P. and {D{\"o}scher}, D. and {Driver}, S.~P. and {Dwelly}, T. and {Eckert}, D. and {Edge}, A. and {Edvardsson}, B. and {Youssoufi}, D.~E. and {Elhaddad}, A. and {Enke}, H. and {Erfanianfar}, G. and {Farrell}, T. and {Fechner}, T. and {Feiz}, C. and {Feltzing}, S. and {Ferreras}, I. and {Feuerstein}, D. and {Feuillet}, D. and {Finoguenov}, A. and {Ford}, D. and {Fotopoulou}, S. and {Fouesneau}, M. and {Frenk}, C. and {Frey}, S. and {Gaessler}, W. and {Geier}, S. and {Gentile Fusillo}, N. and {Gerhard}, O. and {Giannantonio}, T. and {Giannone}, D. and {Gibson}, B. and {Gillingham}, P. and {Gonz{\'a}lez-Fern{\'a}ndez}, C. and {Gonzalez-Solares}, E. and {Gottloeber}, S. and {Gould}, A. and {Grebel}, E.~K. and {Gueguen}, A. and {Guiglion}, G. and {Haehnelt}, M. and {Hahn}, T. and {Hansen}, C.~J. and {Hartman}, H. and {Hauptner}, K. and {Hawkins}, K. and {Haynes}, D. and {Haynes}, R. and {Heiter}, U. and {Helmi}, A. and {Aguayo}, C.~H. and {Hewett}, P. and {Hinton}, S. and {Hobbs}, D. and {Hoenig}, S. and {Hofman}, D. and {Hook}, I. and {Hopgood}, J. and {Hopkins}, A. and {Hourihane}, A. and {Howes}, L. and {Howlett}, C. and {Huet}, T. and {Irwin}, M. and {Iwert}, O. and {Jablonka}, P. and {Jahn}, T. and {Jahnke}, K. and {Jarno}, A. and {Jin}, S. and {Jofre}, P. and {Johl}, D. and {Jones}, D. and {J{\"o}nsson}, H. and {Jordan}, C. and {Karovicova}, I. and {Khalatyan}, A. and {Kelz}, A. and {Kennicutt}, R. and {King}, D. and {Kitaura}, F. and {Klar}, J. and {Klauser}, U. and {Kneib}, J.-P. and {Koch}, A. and {Koposov}, S. and {Kordopatis}, G. and {Korn}, A. and {Kosmalski}, J. and {Kotak}, R. and {Kovalev}, M. and {Kreckel}, K. and {Kripak}, Y. and {Krumpe}, M. and {Kuijken}, K. and {Kunder}, A. and {Kushniruk}, I. and {Lam}, M.~I. and {Lamer}, G. and {Laurent}, F. and {Lawrence}, J. and {Lehmitz}, M. and {Lemasle}, B. and {Lewis}, J. and {Li}, B. and {Lidman}, C. and {Lind}, K. and {Liske}, J. and {Lizon}, J.-L. and {Loveday}, J. and {Ludwig}, H.-G. and {McDermid}, R.~M. and {Maguire}, K. and {Mainieri}, V. and {Mali}, S. and {Mandel}, H.},
        title = "{4MOST: Project overview and information for the First Call for Proposals}",
      journal = {The Messenger},
         year = 2019,
        month = mar,
       volume = {175},
        pages = {3-11},
          doi = {10.18727/0722-6691/5117},
archivePrefix = {arXiv},
       eprint = {1903.02464},
 primaryClass = {astro-ph.IM},
       adsurl = {https://ui.adsabs.harvard.edu/abs/2019Msngr.175....3D}
}

@ARTICLE{Edenhofer_2024,
       author = {{Edenhofer}, Gordian and {Zucker}, Catherine and {Frank}, Philipp and {Saydjari}, Andrew K. and {Speagle}, Joshua S. and {Finkbeiner}, Douglas and {En{\ss}lin}, Torsten A.},
        title = "{A parsec-scale Galactic 3D dust map out to 1.25 kpc from the Sun}",
      journal = {\aap},
         year = 2024,
        month = may,
       volume = {685},
          eid = {A82},
        pages = {A82},
          doi = {10.1051/0004-6361/202347628},
archivePrefix = {arXiv},
       eprint = {2308.01295},
 primaryClass = {astro-ph.GA},
       adsurl = {https://ui.adsabs.harvard.edu/abs/2024A&A...685A..82E}
}

@ARTICLE{Eggleton_1983,
       author = {{Eggleton}, P.~P.},
        title = "{Aproximations to the radii of Roche lobes.}",
      journal = {\apj},
         year = 1983,
        month = may,
       volume = {268},
        pages = {368-369},
          doi = {10.1086/160960},
       adsurl = {https://ui.adsabs.harvard.edu/abs/1983ApJ...268..368E}
}

@ARTICLE{Eggleton_2002,
       author = {{Eggleton}, Peter P. and {Kiseleva-Eggleton}, Ludmila},
        title = "{The Evolution of Cool Algols}",
      journal = {\apj},
         year = 2002,
        month = aug,
       volume = {575},
       number = {1},
        pages = {461-473},
          doi = {10.1086/341215},
archivePrefix = {arXiv},
       eprint = {astro-ph/0209201},
 primaryClass = {astro-ph},
       adsurl = {https://ui.adsabs.harvard.edu/abs/2002ApJ...575..461E}
}

@ARTICLE{El-Badry_2026,
       author = {{El-Badry}, Kareem and {Rodriguez}, Antonio C. and {Green}, Matthew J. and {Burdge}, Kevin B.},
        title = "{JWST observations of three long-period AM CVn binaries: detection of the donors and hints of magnetically truncated disks}",
      journal = {The Open Journal of Astrophysics},
         year = 2026,
        month = feb,
       volume = {9},
        pages = {57856},
          doi = {10.33232/001c.157856},
archivePrefix = {arXiv},
       eprint = {2601.00945},
 primaryClass = {astro-ph.SR},
       adsurl = {https://ui.adsabs.harvard.edu/abs/2026OJAp....957856E}
}

@ARTICLE{Fink_2010,
       author = {{Fink}, M. and {R{\"o}pke}, F.~K. and {Hillebrandt}, W. and {Seitenzahl}, I.~R. and {Sim}, S.~A. and {Kromer}, M.},
        title = "{Double-detonation sub-Chandrasekhar supernovae: can minimum helium shell masses detonate the core?}",
      journal = {\aap},
         year = 2010,
        month = may,
       volume = {514},
          eid = {A53},
        pages = {A53},
          doi = {10.1051/0004-6361/200913892},
archivePrefix = {arXiv},
       eprint = {1002.2173},
 primaryClass = {astro-ph.SR},
       adsurl = {https://ui.adsabs.harvard.edu/abs/2010A&A...514A..53F}
}

@ARTICLE{Foreman-Macke_2013,
       author = {{Foreman-Mackey}, Daniel and {Hogg}, David W. and {Lang}, Dustin and {Goodman}, Jonathan},
        title = "{emcee: The MCMC Hammer}",
      journal = {\pasp},
         year = 2013,
        month = mar,
       volume = {125},
       number = {925},
        pages = {306},
          doi = {10.1086/670067},
archivePrefix = {arXiv},
       eprint = {1202.3665},
 primaryClass = {astro-ph.IM},
       adsurl = {https://ui.adsabs.harvard.edu/abs/2013PASP..125..306F}
}

@ARTICLE{Gaensicke_1995,
       author = {{G{\"a}nsicke}, B.~T. and {Beuermann}, K. and {de Martino}, D.},
        title = "{The white dwarf in AM Herculis.}",
      journal = {\aap},
         year = 1995,
        month = nov,
       volume = {303},
        pages = {127},
       adsurl = {https://ui.adsabs.harvard.edu/abs/1995A&A...303..127G}
}

@ARTICLE{Gaensicke_2005,
       author = {{G{\"a}nsicke}, Boris T. and {Szkody}, Paula and {Howell}, Steve B. and {Sion}, Edward M.},
        title = "{Hubble Space Telescope STIS Observations of the Accreting White Dwarfs in BW Sculptoris, BC Ursae Majoris, and SW Ursae Majoris}",
      journal = {\apj},
         year = 2005,
        month = aug,
       volume = {629},
       number = {1},
        pages = {451-460},
          doi = {10.1086/431271},
archivePrefix = {arXiv},
       eprint = {astro-ph/0504382},
 primaryClass = {astro-ph},
       adsurl = {https://ui.adsabs.harvard.edu/abs/2005ApJ...629..451G}
}

@ARTICLE{Gaia_2023,
       author = {{Gaia Collaboration} and {Vallenari}, A. and {Brown}, A.~G.~A. and {Prusti}, T. and {de Bruijne}, J.~H.~J. and {Arenou}, F. and {Babusiaux}, C. and {Biermann}, M. and {Creevey}, O.~L. and {Ducourant}, C. and {Evans}, D.~W. and {Eyer}, L. and {Guerra}, R. and {Hutton}, A. and {Jordi}, C. and {Klioner}, S.~A. and {Lammers}, U.~L. and {Lindegren}, L. and {Luri}, X. and {Mignard}, F. and {Panem}, C. and {Pourbaix}, D. and {Randich}, S. and {Sartoretti}, P. and {Soubiran}, C. and {Tanga}, P. and {Walton}, N.~A. and {Bailer-Jones}, C.~A.~L. and {Bastian}, U. and {Drimmel}, R. and {Jansen}, F. and {Katz}, D. and {Lattanzi}, M.~G. and {van Leeuwen}, F. and {Bakker}, J. and {Cacciari}, C. and {Casta{\~n}eda}, J. and {De Angeli}, F. and {Fabricius}, C. and {Fouesneau}, M. and {Fr{\'e}mat}, Y. and {Galluccio}, L. and {Guerrier}, A. and {Heiter}, U. and {Masana}, E. and {Messineo}, R. and {Mowlavi}, N. and {Nicolas}, C. and {Nienartowicz}, K. and {Pailler}, F. and {Panuzzo}, P. and {Riclet}, F. and {Roux}, W. and {Seabroke}, G.~M. and {Sordo}, R. and {Th{\'e}venin}, F. and {Gracia-Abril}, G. and {Portell}, J. and {Teyssier}, D. and {Altmann}, M. and {Andrae}, R. and {Audard}, M. and {Bellas-Velidis}, I. and {Benson}, K. and {Berthier}, J. and {Blomme}, R. and {Burgess}, P.~W. and {Busonero}, D. and {Busso}, G. and {C{\'a}novas}, H. and {Carry}, B. and {Cellino}, A. and {Cheek}, N. and {Clementini}, G. and {Damerdji}, Y. and {Davidson}, M. and {de Teodoro}, P. and {Nu{\~n}ez Campos}, M. and {Delchambre}, L. and {Dell'Oro}, A. and {Esquej}, P. and {Fern{\'a}ndez-Hern{\'a}ndez}, J. and {Fraile}, E. and {Garabato}, D. and {Garc{\'\i}a-Lario}, P. and {Gosset}, E. and {Haigron}, R. and {Halbwachs}, J.-L. and {Hambly}, N.~C. and {Harrison}, D.~L. and {Hern{\'a}ndez}, J. and {Hestroffer}, D. and {Hodgkin}, S.~T. and {Holl}, B. and {Jan{\ss}en}, K. and {Jevardat de Fombelle}, G. and {Jordan}, S. and {Krone-Martins}, A. and {Lanzafame}, A.~C. and {L{\"o}ffler}, W. and {Marchal}, O. and {Marrese}, P.~M. and {Moitinho}, A. and {Muinonen}, K. and {Osborne}, P. and {Pancino}, E. and {Pauwels}, T. and {Recio-Blanco}, A. and {Reyl{\'e}}, C. and {Riello}, M. and {Rimoldini}, L. and {Roegiers}, T. and {Rybizki}, J. and {Sarro}, L.~M. and {Siopis}, C. and {Smith}, M. and {Sozzetti}, A. and {Utrilla}, E. and {van Leeuwen}, M. and {Abbas}, U. and {{\'A}brah{\'a}m}, P. and {Abreu Aramburu}, A. and {Aerts}, C. and {Aguado}, J.~J. and {Ajaj}, M. and {Aldea-Montero}, F. and {Altavilla}, G. and {{\'A}lvarez}, M.~A. and {Alves}, J. and {Anders}, F. and {Anderson}, R.~I. and {Anglada Varela}, E. and {Antoja}, T. and {Baines}, D. and {Baker}, S.~G. and {Balaguer-N{\'u}{\~n}ez}, L. and {Balbinot}, E. and {Balog}, Z. and {Barache}, C. and {Barbato}, D. and {Barros}, M. and {Barstow}, M.~A. and {Bartolom{\'e}}, S. and {Bassilana}, J.-L. and {Bauchet}, N. and {Becciani}, U. and {Bellazzini}, M. and {Berihuete}, A. and {Bernet}, M. and {Bertone}, S. and {Bianchi}, L. and {Binnenfeld}, A. and {Blanco-Cuaresma}, S. and {Blazere}, A. and {Boch}, T. and {Bombrun}, A. and {Bossini}, D. and {Bouquillon}, S. and {Bragaglia}, A. and {Bramante}, L. and {Breedt}, E. and {Bressan}, A. and {Brouillet}, N. and {Brugaletta}, E. and {Bucciarelli}, B. and {Burlacu}, A. and {Butkevich}, A.~G. and {Buzzi}, R. and {Caffau}, E. and {Cancelliere}, R. and {Cantat-Gaudin}, T. and {Carballo}, R. and {Carlucci}, T. and {Carnerero}, M.~I. and {Carrasco}, J.~M. and {Casamiquela}, L. and {Castellani}, M. and {Castro-Ginard}, A. and {Chaoul}, L. and {Charlot}, P. and {Chemin}, L. and {Chiaramida}, V. and {Chiavassa}, A. and {Chornay}, N. and {Comoretto}, G. and {Contursi}, G. and {Cooper}, W.~J. and {Cornez}, T. and {Cowell}, S. and {Crifo}, F. and {Cropper}, M. and {Crosta}, M. and {Crowley}, C. and {Dafonte}, C. and {Dapergolas}, A. and {David}, M. and {David}, P. and {de Laverny}, P. and {De Luise}, F. and {De March}, R.},
        title = "{Gaia Data Release 3. Summary of the content and survey properties}",
      journal = {\aap},
         year = 2023,
        month = jun,
       volume = {674},
          eid = {A1},
        pages = {A1},
          doi = {10.1051/0004-6361/202243940},
archivePrefix = {arXiv},
       eprint = {2208.00211},
 primaryClass = {astro-ph.GA},
       adsurl = {https://ui.adsabs.harvard.edu/abs/2023A&A...674A...1G}
}

@ARTICLE{Gentile_Fusillo_2021,
       author = {{Gentile Fusillo}, N.~P. and {Tremblay}, P.-E. and {Cukanovaite}, E. and {Vorontseva}, A. and {Lallement}, R. and {Hollands}, M. and {G{\"a}nsicke}, B.~T. and {Burdge}, K.~B. and {McCleery}, J. and {Jordan}, S.},
        title = "{A catalogue of white dwarfs in Gaia EDR3}",
      journal = {\mnras},
         year = 2021,
        month = dec,
       volume = {508},
       number = {3},
        pages = {3877-3896},
          doi = {10.1093/mnras/stab2672},
archivePrefix = {arXiv},
       eprint = {2106.07669},
 primaryClass = {astro-ph.SR},
       adsurl = {https://ui.adsabs.harvard.edu/abs/2021MNRAS.508.3877G}
}

@ARTICLE{Ginsburg_2019,
       author = {{Ginsburg}, Adam and {Sip{\H{o}}cz}, Brigitta M. and {Brasseur}, C.~E. and {Cowperthwaite}, Philip S. and {Craig}, Matthew W. and {Deil}, Christoph and {Guillochon}, James and {Guzman}, Giannina and {Liedtke}, Simon and {Lian Lim}, Pey and {Lockhart}, Kelly E. and {Mommert}, Michael and {Morris}, Brett M. and {Norman}, Henrik and {Parikh}, Madhura and {Persson}, Magnus V. and {Robitaille}, Thomas P. and {Segovia}, Juan-Carlos and {Singer}, Leo P. and {Tollerud}, Erik J. and {de Val-Borro}, Miguel and {Valtchanov}, Ivan and {Woillez}, Julien and {Astroquery Collaboration} and {a subset of astropy Collaboration}},
        title = "{astroquery: An Astronomical Web-querying Package in Python}",
      journal = {\aj},
         year = 2019,
        month = mar,
       volume = {157},
       number = {3},
          eid = {98},
        pages = {98},
          doi = {10.3847/1538-3881/aafc33},
archivePrefix = {arXiv},
       eprint = {1901.04520},
 primaryClass = {astro-ph.IM},
       adsurl = {https://ui.adsabs.harvard.edu/abs/2019AJ....157...98G}
}

@ARTICLE{Green_2012,
       author = {{Green}, James C. and {Froning}, Cynthia S. and {Osterman}, Steve and {Ebbets}, Dennis and {Heap}, Sara H. and {Leitherer}, Claus and {Linsky}, Jeffrey L. and {Savage}, Blair D. and {Sembach}, Kenneth and {Shull}, J. Michael and {Siegmund}, Oswald H.~W. and {Snow}, Theodore P. and {Spencer}, John and {Stern}, S. Alan and {Stocke}, John and {Welsh}, Barry and {B{\'e}land}, St{\'e}phane and {Burgh}, Eric B. and {Danforth}, Charles and {France}, Kevin and {Keeney}, Brian and {McPhate}, Jason and {Penton}, Steven V. and {Andrews}, John and {Brownsberger}, Kenneth and {Morse}, Jon and {Wilkinson}, Erik},
        title = "{The Cosmic Origins Spectrograph}",
      journal = {\apj},
         year = 2012,
        month = jan,
       volume = {744},
       number = {1},
          eid = {60},
        pages = {60},
          doi = {10.1088/0004-637X/744/1/6010.1086/141956},
archivePrefix = {arXiv},
       eprint = {1110.0462},
 primaryClass = {astro-ph.IM},
       adsurl = {https://ui.adsabs.harvard.edu/abs/2012ApJ...744...60G}
}

@ARTICLE{Green_2018,
       author = {{Green}, M.~J. and {Marsh}, T.~R. and {Steeghs}, D.~T.~H. and {Kupfer}, T. and {Ashley}, R.~P. and {Bloemen}, S. and {Breedt}, E. and {Campbell}, H.~C. and {Chakpor}, A. and {Copperwheat}, C.~M. and {Dhillon}, V.~S. and {Hallinan}, G. and {Hardy}, L.~K. and {Hermes}, J.~J. and {Kerry}, P. and {Littlefair}, S.~P. and {Milburn}, J. and {Parsons}, S.~G. and {Prasert}, N. and {van Roestel}, J. and {Sahman}, D.~I. and {Singh}, N.},
        title = "{High-speed photometry of Gaia14aae: an eclipsing AM CVn that challenges formation models}",
      journal = {\mnras},
         year = 2018,
        month = may,
       volume = {476},
       number = {2},
        pages = {1663-1679},
          doi = {10.1093/mnras/sty299},
archivePrefix = {arXiv},
       eprint = {1802.00499},
 primaryClass = {astro-ph.SR},
       adsurl = {https://ui.adsabs.harvard.edu/abs/2018MNRAS.476.1663G}
}

@article{Green_2019,
   title={A 3D Dust Map Based on Gaia, Pan-STARRS 1, and 2MASS},
   volume={887},
   ISSN={1538-4357},
   url={http://dx.doi.org/10.3847/1538-4357/ab5362},
   DOI={10.3847/1538-4357/ab5362},
   number={1},
   journal={The Astrophysical Journal},
   publisher={American Astronomical Society},
   author={Green, Gregory M. and Schlafly, Edward and Zucker, Catherine and Speagle, Joshua S. and Finkbeiner, Douglas},
   year={2019},
   month=dec, pages={93} }

@ARTICLE{Green_2025,
       author = {{Green}, Matthew J. and {van Roestel}, Jan and {Wong}, Tin Long Sunny},
        title = "{A catalogue of ultracompact mass-transferring white dwarf binaries}",
      journal = {\aap},
         year = 2025,
        month = aug,
       volume = {700},
          eid = {A107},
        pages = {A107},
          doi = {10.1051/0004-6361/202554925},
archivePrefix = {arXiv},
       eprint = {2505.10535},
 primaryClass = {astro-ph.SR},
       adsurl = {https://ui.adsabs.harvard.edu/abs/2025A&A...700A.107G}
}

@ARTICLE{Green_2026,
       author = {{Green}, Matthew J. and {Marsh}, Thomas R. and {van Roestel}, Jan and {Wong}, Tin Long Sunny and {Belloni}, Diogo and {Kilic}, Mukremin and {Breedt}, Elm{\'e} and {Brown}, Alex and {Copperwheat}, Chris M. and {Chakpor}, Anurak and {Dhillon}, V.~S. and {Segura}, Noel Castro and {Dyer}, Martin J. and {Garbutt}, James and {Jarvis}, Dan and {Kengkriangkrai}, Vasu and {Kennedy}, Mark R. and {Kerry}, Paul and {Kupfer}, Thomas and {Littlefair}, S.~P. and {McCormac}, James and {Munday}, James and {Parsons}, Steven G. and {Pike}, Eleanor and {Pelisoli}, Ingrid and {Rodr{\'\i}guez-Gil}, Pablo and {Sahman}, David I. and {Yates}, Amalie},
        title = "{No period change in two long-period AM CVn binaries}",
      journal = {\mnras},
         year = 2026,
        month = may,
       volume = {548},
       number = {3},
          eid = {stag673},
        pages = {stag673},
          doi = {10.1093/mnras/stag673},
archivePrefix = {arXiv},
       eprint = {2604.06460},
 primaryClass = {astro-ph.SR},
       adsurl = {https://ui.adsabs.harvard.edu/abs/2026MNRAS.548ag673G}
}

@ARTICLE{Goliasch_2015,
       author = {{Goliasch}, Jonas and {Nelson}, Lorne},
        title = "{Population Synthesis of Cataclysmic Variables. I. Inclusion of Detailed Nuclear Evolution}",
      journal = {\apj},
         year = 2015,
        month = aug,
       volume = {809},
       number = {1},
          eid = {80},
        pages = {80},
          doi = {10.1088/0004-637X/809/1/80},
archivePrefix = {arXiv},
       eprint = {1607.06217},
 primaryClass = {astro-ph.SR},
       adsurl = {https://ui.adsabs.harvard.edu/abs/2015ApJ...809...80G}
}

@ARTICLE{Groot_2024,
       author = {{Groot}, P.~J. and {Bloemen}, S. and {Vreeswijk}, P.~M. and {van Roestel}, J.~C.~J. and {Jonker}, P.~G. and {Nelemans}, G. and {Klein-Wolt}, M. and {Lepoole}, R. and {Pieterse}, D.~L.~A. and {Rodenhuis}, M. and {Boland}, W. and {Haverkorn}, M. and {Aerts}, C. and {Bakker}, R. and {Balster}, H. and {Bekema}, M. and {Dijkstra}, E. and {Dolron}, P. and {Elswijk}, E. and {van Elteren}, A. and {Engels}, A. and {Fokker}, M. and {de Haan}, M. and {Hahn}, F. and {ter Horst}, R. and {Lesman}, D. and {Kragt}, J. and {Morren}, J. and {Nillissen}, H. and {Pessemier}, W. and {Raskin}, G. and {de Rijke}, A. and {Scheers}, L.~H.~A. and {Schuil}, M. and {Timmer}, S.~T. and {Antunes Amaral}, L. and {Arancibia-Rojas}, E. and {Arcavi}, I. and {Blagorodnova}, N. and {Biswas}, S. and {Breton}, R.~P. and {Dawson}, H. and {Dayal}, P. and {De Wet}, S. and {Duffy}, C. and {Faris}, S. and {Fausnaugh}, M. and {Gal-Yam}, A. and {Geier}, S. and {Horesh}, A. and {Johnston}, C. and {Katusiime}, G. and {Kelley}, C. and {Kosakowski}, A. and {Kupfer}, T. and {Leloudas}, G. and {Levan}, A. and {Modiano}, D. and {Mogawana}, O. and {Munday}, J. and {Paice}, J. and {Patat}, F. and {Pelisoli}, I. and {Ramsay}, G. and {Ranaivomanana}, P.~T. and {Ruiz-Carmona}, R. and {Schaffenroth}, V. and {Scaringi}, S. and {Stoppa}, F. and {Street}, R. and {Tranin}, H. and {Uzundag}, M. and {Valenti}, S. and {Veresvarska}, M. and {Vuc̆kovi{\'c}}, M. and {Wichern}, H.~C.~I. and {Wijers}, R.~A.~M.~J. and {Wijnands}, R.~A.~D. and {Zimmerman}, E.},
        title = "{The BlackGEM Telescope Array. I. Overview}",
      journal = {\pasp},
         year = 2024,
        month = nov,
       volume = {136},
       number = {11},
          eid = {115003},
        pages = {115003},
          doi = {10.1088/1538-3873/ad8b6a},
archivePrefix = {arXiv},
       eprint = {2405.18923},
 primaryClass = {astro-ph.IM},
       adsurl = {https://ui.adsabs.harvard.edu/abs/2024PASP..136k5003G}
}

@ARTICLE{Hernandez_Diaz_2026,
       author = {{Hern{\'a}ndez-D{\'\i}az}, S. and {Stelzer}, B. and {Schwope}, A. and {Mu{\~n}oz-Giraldo}, D. and {Schreiber}, M.~R. and {Brink}, J. and {Pradeep}, K.~G. and {G{\"a}nsicke}, B.~T. and {Eracleous}, M. and {Szekerczes}, K. and {Anderson}, S.~F. and {Brownstein}, J.~R.},
        title = "{Unveiling the period-bounce population of cataclysmic variables: Spectroscopic and time-domain follow-up of eROSITA-selected candidates}",
      journal = {arXiv e-prints},
         year = 2026,
        month = jul,
          eid = {arXiv:2607.27855},
        pages = {arXiv:2607.27855},
          doi = {10.48550/arXiv.2607.27855},
archivePrefix = {arXiv},
       eprint = {2607.27855},
 primaryClass = {astro-ph.SR},
       adsurl = {https://ui.adsabs.harvard.edu/abs/2026arXiv260727855H}
}

@ARTICLE{Iben_1987,
       author = {{Iben}, Jr., Icko and {Tutukov}, Alexander V.},
        title = "{Evolutionary Scenarios for Intermediate-Mass Stars in Close Binaries}",
      journal = {\apj},
         year = 1987,
        month = feb,
       volume = {313},
        pages = {727},
          doi = {10.1086/165011},
       adsurl = {https://ui.adsabs.harvard.edu/abs/1987ApJ...313..727I}
}

@ARTICLE{Ivezic_2019,
       author = {{Ivezi{\'c}}, {\v{Z}}eljko and {Kahn}, Steven M. and {Tyson}, J. Anthony and {Abel}, Bob and {Acosta}, Emily and {Allsman}, Robyn and {Alonso}, David and {AlSayyad}, Yusra and {Anderson}, Scott F. and {Andrew}, John and {Angel}, James Roger P. and {Angeli}, George Z. and {Ansari}, Reza and {Antilogus}, Pierre and {Araujo}, Constanza and {Armstrong}, Robert and {Arndt}, Kirk T. and {Astier}, Pierre and {Aubourg}, {\'E}ric and {Auza}, Nicole and {Axelrod}, Tim S. and {Bard}, Deborah J. and {Barr}, Jeff D. and {Barrau}, Aurelian and {Bartlett}, James G. and {Bauer}, Amanda E. and {Bauman}, Brian J. and {Baumont}, Sylvain and {Bechtol}, Ellen and {Bechtol}, Keith and {Becker}, Andrew C. and {Becla}, Jacek and {Beldica}, Cristina and {Bellavia}, Steve and {Bianco}, Federica B. and {Biswas}, Rahul and {Blanc}, Guillaume and {Blazek}, Jonathan and {Blandford}, Roger D. and {Bloom}, Josh S. and {Bogart}, Joanne and {Bond}, Tim W. and {Booth}, Michael T. and {Borgland}, Anders W. and {Borne}, Kirk and {Bosch}, James F. and {Boutigny}, Dominique and {Brackett}, Craig A. and {Bradshaw}, Andrew and {Brandt}, William Nielsen and {Brown}, Michael E. and {Bullock}, James S. and {Burchat}, Patricia and {Burke}, David L. and {Cagnoli}, Gianpietro and {Calabrese}, Daniel and {Callahan}, Shawn and {Callen}, Alice L. and {Carlin}, Jeffrey L. and {Carlson}, Erin L. and {Chandrasekharan}, Srinivasan and {Charles-Emerson}, Glenaver and {Chesley}, Steve and {Cheu}, Elliott C. and {Chiang}, Hsin-Fang and {Chiang}, James and {Chirino}, Carol and {Chow}, Derek and {Ciardi}, David R. and {Claver}, Charles F. and {Cohen-Tanugi}, Johann and {Cockrum}, Joseph J. and {Coles}, Rebecca and {Connolly}, Andrew J. and {Cook}, Kem H. and {Cooray}, Asantha and {Covey}, Kevin R. and {Cribbs}, Chris and {Cui}, Wei and {Cutri}, Roc and {Daly}, Philip N. and {Daniel}, Scott F. and {Daruich}, Felipe and {Daubard}, Guillaume and {Daues}, Greg and {Dawson}, William and {Delgado}, Francisco and {Dellapenna}, Alfred and {de Peyster}, Robert and {de Val-Borro}, Miguel and {Digel}, Seth W. and {Doherty}, Peter and {Dubois}, Richard and {Dubois-Felsmann}, Gregory P. and {Durech}, Josef and {Economou}, Frossie and {Eifler}, Tim and {Eracleous}, Michael and {Emmons}, Benjamin L. and {Fausti Neto}, Angelo and {Ferguson}, Henry and {Figueroa}, Enrique and {Fisher-Levine}, Merlin and {Focke}, Warren and {Foss}, Michael D. and {Frank}, James and {Freemon}, Michael D. and {Gangler}, Emmanuel and {Gawiser}, Eric and {Geary}, John C. and {Gee}, Perry and {Geha}, Marla and {Gessner}, Charles J.~B. and {Gibson}, Robert R. and {Gilmore}, D. Kirk and {Glanzman}, Thomas and {Glick}, William and {Goldina}, Tatiana and {Goldstein}, Daniel A. and {Goodenow}, Iain and {Graham}, Melissa L. and {Gressler}, William J. and {Gris}, Philippe and {Guy}, Leanne P. and {Guyonnet}, Augustin and {Haller}, Gunther and {Harris}, Ron and {Hascall}, Patrick A. and {Haupt}, Justine and {Hernandez}, Fabio and {Herrmann}, Sven and {Hileman}, Edward and {Hoblitt}, Joshua and {Hodgson}, John A. and {Hogan}, Craig and {Howard}, James D. and {Huang}, Dajun and {Huffer}, Michael E. and {Ingraham}, Patrick and {Innes}, Walter R. and {Jacoby}, Suzanne H. and {Jain}, Bhuvnesh and {Jammes}, Fabrice and {Jee}, M. James and {Jenness}, Tim and {Jernigan}, Garrett and {Jevremovi{\'c}}, Darko and {Johns}, Kenneth and {Johnson}, Anthony S. and {Johnson}, Margaret W.~G. and {Jones}, R. Lynne and {Juramy-Gilles}, Claire and {Juri{\'c}}, Mario and {Kalirai}, Jason S. and {Kallivayalil}, Nitya J. and {Kalmbach}, Bryce and {Kantor}, Jeffrey P. and {Karst}, Pierre and {Kasliwal}, Mansi M. and {Kelly}, Heather and {Kessler}, Richard and {Kinnison}, Veronica and {Kirkby}, David and {Knox}, Lloyd and {Kotov}, Ivan V. and {Krabbendam}, Victor L. and {Krughoff}, K. Simon and {Kub{\'a}nek}, Petr and {Kuczewski}, John and {Kulkarni}, Shri and {Ku}, John and {Kurita}, Nadine R. and {Lage}, Craig S. and {Lambert}, Ron and {Lange}, Travis and {Langton}, J. Brian and {Le Guillou}, Laurent and {Levine}, Deborah and {Liang}, Ming and {Lim}, Kian-Tat and {Lintott}, Chris J. and {Long}, Kevin E. and {Lopez}, Margaux and {Lotz}, Paul J. and {Lupton}, Robert H. and {Lust}, Nate B. and {MacArthur}, Lauren A. and {Mahabal}, Ashish and {Mandelbaum}, Rachel and {Markiewicz}, Thomas W. and {Marsh}, Darren S. and {Marshall}, Philip J. and {Marshall}, Stuart and {May}, Morgan and {McKercher}, Robert and {McQueen}, Michelle and {Meyers}, Joshua and {Migliore}, Myriam and {Miller}, Michelle and {Mills}, David J.},
        title = "{LSST: From Science Drivers to Reference Design and Anticipated Data Products}",
      journal = {\apj},
         year = 2019,
        month = mar,
       volume = {873},
       number = {2},
          eid = {111},
        pages = {111},
          doi = {10.3847/1538-4357/ab042c},
archivePrefix = {arXiv},
       eprint = {0805.2366},
 primaryClass = {astro-ph},
       adsurl = {https://ui.adsabs.harvard.edu/abs/2019ApJ...873..111I}
}

@ARTICLE{Koester_2010,
       author = {{Koester}, D.},
        title = "{White dwarf spectra and atmosphere models}",
      journal = {\memsai},
         year = 2010,
        month = jan,
       volume = {81},
        pages = {921-931},
       adsurl = {https://ui.adsabs.harvard.edu/abs/2010MmSAI..81..921K}
}

@ARTICLE{Kara_2026,
       author = {{K{\'a}ra}, Jan and {Rivera Sandoval}, Liliana and {Mendoza}, Wendy and {Maccarone}, Thomas and {Pichardo Marcano}, Manuel and {Salazar Manzano}, Luis E. and {Oelkers}, Ryan J. and {van Roestel}, Jan},
        title = "{A study of transients from ground-based surveys reveals new ultra-compact accreting white dwarf binaries}",
      journal = {\pasa},
         year = 2026,
        month = mar,
       volume = {43},
          eid = {e052},
        pages = {e052},
          doi = {10.1017/pasa.2026.10184},
archivePrefix = {arXiv},
       eprint = {2511.16774},
 primaryClass = {astro-ph.SR},
       adsurl = {https://ui.adsabs.harvard.edu/abs/2026PASA...43...52K}
}

@ARTICLE{Kollmeier_2026,
       author = {{Kollmeier}, Juna A. and {Rix}, Hans-Walter and {Aerts}, Conny and {Aird}, James and {Vera Alfaro}, Pablo and {Almeida}, Andr{\'e}s and {Anderson}, Scott F. and {Arseneau}, Stefan M. and {Assef}, Roberto J. and {Aviram}, Shir and {Aydar}, Catarina and {Badenes}, Carles and {Bandyopadhyay}, Avrajit and {Barger}, Kat and {Barkhouser}, Robert H. and {Bauer}, Franz E. and {Behmard}, Aida and {Bender}, Chad and {Besser}, Felipe and {Bhattarai}, Binod and {Bilgi}, Pavaman and {Bird}, Jonathan and {Bizyaev}, Dmitry and {Blanc}, Guillermo A. and {Blanton}, Michael R. and {Bochanski}, John and {Bovy}, Jo and {Brandon}, Christopher and {Brandt}, William Nielsen and {Brownstein}, Joel R. and {Buchner}, Johannes and {Burchett}, Joseph N. and {Carlberg}, Joleen and {Casey}, Andrew R. and {Castaneda-Carlos}, Lesly and {Chakraborty}, Priyanka and {Chanam{\'e}}, Julio and {Chandra}, Vedant and {Cherinka}, Brian and {Chilingarian}, Igor and {Comparat}, Johan and {Cosens}, Maren and {Covey}, Kevin and {Crane}, Jeffrey D. and {Crumpler}, Nicole R. and {Cruz-Gonzalez}, Irene and {Cunha}, Katia and {Cunningham}, Tim and {Dai}, Xinyu and {Darling}, Jeremy and {Davidson}, Jr., James W. and {Davis}, Megan C. and {De Lee}, Nathan and {Deacon}, Niall and {M{\'e}ndez Delgado}, Jos{\'e} Eduardo and {Demasi}, Sebastian and {Demianenko}, Mariia and {Derwent}, Mark and {D'Onghia}, Elena and {Di Mille}, Francesco and {Dias}, Bruno and {Donor}, John and {Dow}, Peter N. and {Drory}, Niv and {Dwelly}, Tom and {Egorov}, Oleg and {Egorova}, Evgeniya and {El-Badry}, Kareem and {Engelman}, Mike and {Eracleous}, Mike and {Fan}, Xiaohui and {Farr}, Emily and {Fries}, Logan and {Frinchaboy}, Peter and {Froning}, Cynthia S. and {G{\"a}nsicke}, Boris T. and {Garc{\'\i}a}, Pablo and {Gelfand}, Joseph and {Gentile Fusillo}, Nicola Pietro and {Glover}, Simon and {Grabowski}, Katie and {Grebel}, Eva K. and {Green}, Paul J. and {Grier}, Catherine and {Gupta}, Pramod and {Gray}, Aidan C. and {H{\"a}berle}, Maximilian and {Hall}, Patrick B. and {Hammond}, Randolph P. and {Hawkins}, Keith and {Harding}, Albert C. and {Heged{\H{u}}s}, Viola and {Herbst}, Tom and {Hermes}, J.~J. and {Rodr{\'\i}guez Hidalgo}, Paola and {Hilder}, Thomas and {Hogg}, David W. and {Holtzman}, Jon A. and {Horta}, Danny and {Huang}, Yang and {Hwang}, Hsiang-Chih and {Ibarra-Medel}, Hector Javier and {Imig}, Julie and {Inight}, Keith and {Jana}, Arghajit and {Ji}, Alexander P. and {Jim{\'e}nez-Arranz}, {\'O}scar and {Jofre}, Paula and {Johns}, Matt and {Johnson}, Jennifer and {Johnson}, James W. and {Johnston}, Evelyn J. and {Jones}, Amy M. and {Katkov}, Ivan and {Knapp}, Gillian R. and {Koekemoer}, Anton M. and {Kounkel}, Marina and {Kreckel}, Kathryn and {Krishnarao}, Dhanesh and {Krumpe}, Mirko and {Kumari}, Nimisha and {Kupfer}, Thomas and {Lacerna}, Ivan and {Laporte}, Chervin and {Lepine}, Sebastien and {Li}, Jing and {Liu}, Xin and {Loebman}, Sarah and {Long}, Knox and {Roman-Lopes}, Alexandre and {Lu}, Yuxi and {Majewski}, Steven Raymond and {Maoz}, Dan and {McKinnon}, Kevin A. and {Medan}, Ilija and {Merloni}, Andrea and {Minniti}, Dante and {Morrison}, Sean and {Myers}, Natalie and {M{\'e}sz{\'a}ros}, Szabolcs and {Nandra}, Kirpal and {Nayak}, Prasanta K. and {Ness}, Melissa K. and {Nidever}, David L. and {O'Brien}, Thomas and {Oeur}, Micah and {Oravetz}, Audrey and {Oravetz}, Daniel and {Otto}, Jonah and {Pallathadka}, Gautham Adamane and {Palunas}, Povilas and {Pan}, Kaike and {Pappalardo}, Daniel and {Pandey}, Rakesh and {Negrete Pe{\~n}aloza}, Castalia Alenka and {Pinsonneault}, Marc H. and {Pogge}, Richard W. and {Taghizadeh Popp}, Manuchehr and {Price-Whelan}, Adrian M. and {Pulatova}, Nadiia and {Qiu}, Dan and {Ramirez}, Solange and {Rankine}, Amy and {Ricci}, Claudio and {Runnoe}, Jessie C. and {Sanchez}, Sebastian and {Salvato}, Mara and {Sarbadhicary}, Sumit K. and {Sattler}, Natascha and {Saydjari}, Andrew K. and {Sayres}, Conor and {Schinnerer}, Eva and {Schlaufman}, Kevin C. and {Schneider}, Donald P. and {Schreiber}, Matthias R. and {Schwope}, Axel and {Serna}, Javier and {Shen}, Yue and {Sif{\'o}n}, Crist{\'o}bal and {Singh}, Amrita and {Sinha}, Amaya and {Smee}, Stephen and {Song}, Ying-Yi and {Souto}, Diogo and {Stassun}, Keivan G. and {Steinmetz}, Matthias and {Stone-Martinez}, Alexander and {Stringfellow}, Guy and {Stutz}, Amelia and {S{\'a}nchez-Gallego}, Jos{\'e} and {Tan}, Jonathan C. and {Tayar}, Jamie and {Thai}, Riley and {Thakar}, Ani and {Ting}, Yuan-Sen and {Tkachenko}, Andrew and {Tovmassian}, Gagik and {Trakhtenbrot}, Benny and {Fern{\'a}ndez-Trincado}, Jos{\'e} G. and {Troup}, Nicholas},
        title = "{Sloan Digital Sky Survey. V. Pioneering Panoptic Spectroscopy}",
      journal = {\aj},
         year = 2026,
        month = jan,
       volume = {171},
       number = {1},
          eid = {52},
        pages = {52},
          doi = {10.3847/1538-3881/ae0576},
archivePrefix = {arXiv},
       eprint = {2507.06989},
 primaryClass = {astro-ph.IM},
       adsurl = {https://ui.adsabs.harvard.edu/abs/2026AJ....171...52K}
}

@ARTICLE{Korol_2022,
       author = {{Korol}, Valeriya and {Hallakoun}, Na'ama and {Toonen}, Silvia and {Karnesis}, Nikolaos},
        title = "{Observationally driven Galactic double white dwarf population for LISA}",
      journal = {\mnras},
         year = 2022,
        month = apr,
       volume = {511},
       number = {4},
        pages = {5936-5947},
          doi = {10.1093/mnras/stac415},
archivePrefix = {arXiv},
       eprint = {2109.10972},
 primaryClass = {astro-ph.HE},
       adsurl = {https://ui.adsabs.harvard.edu/abs/2022MNRAS.511.5936K}
}

@misc{NIST_2024,
       author = {Kramida, A. and Ralchenko, Yu. and Reader, J. and NIST ASD Team},
        title = {NIST Atomic Spectra Database (version 5.12)},
         year = {2024},
          doi = {10.18434/T4W30F},
         note = {National Institute of Standards and Technology, Gaithersburg, MD},
          url = {https://physics.nist.gov/asd}
}

@ARTICLE{Kupfer_2015,
       author = {{Kupfer}, T. and {Groot}, P.~J. and {Bloemen}, S. and {Levitan}, D. and {Steeghs}, D. and {Marsh}, T.~R. and {Rutten}, R.~G.~M. and {Nelemans}, G. and {Prince}, T.~A. and {F{\"u}rst}, F. and {Geier}, S.},
        title = "{Phase-resolved spectroscopy and Kepler photometry of the ultracompact AM CVn binary SDSS J190817.07+394036.4}",
      journal = {\mnras},
         year = 2015,
        month = oct,
       volume = {453},
       number = {1},
        pages = {483-496},
          doi = {10.1093/mnras/stv1609},
archivePrefix = {arXiv},
       eprint = {1507.03926},
 primaryClass = {astro-ph.SR},
       adsurl = {https://ui.adsabs.harvard.edu/abs/2015MNRAS.453..483K}
}

@ARTICLE{Kupfer_2018,
       author = {{Kupfer}, T. and {Korol}, V. and {Shah}, S. and {Nelemans}, G. and {Marsh}, T.~R. and {Ramsay}, G. and {Groot}, P.~J. and {Steeghs}, D.~T.~H. and {Rossi}, E.~M.},
        title = "{LISA verification binaries with updated distances from Gaia Data Release 2}",
      journal = {\mnras},
         year = 2018,
        month = oct,
       volume = {480},
       number = {1},
        pages = {302-309},
          doi = {10.1093/mnras/sty1545},
archivePrefix = {arXiv},
       eprint = {1805.00482},
 primaryClass = {astro-ph.SR},
       adsurl = {https://ui.adsabs.harvard.edu/abs/2018MNRAS.480..302K}
}

@ARTICLE{Kupfer_2024,
       author = {{Kupfer}, Thomas and {Korol}, Valeriya and {Littenberg}, Tyson B. and {Shah}, Sweta and {Savalle}, Etienne and {Groot}, Paul J. and {Marsh}, Thomas R. and {Le Jeune}, Maude and {Nelemans}, Gijs and {Pala}, Anna F. and {Petiteau}, Antoine and {Ramsay}, Gavin and {Steeghs}, Danny and {Babak}, Stanislav},
        title = "{LISA Galactic Binaries with Astrometry from Gaia DR3}",
      journal = {\apj},
         year = 2024,
        month = mar,
       volume = {963},
       number = {2},
          eid = {100},
        pages = {100},
          doi = {10.3847/1538-4357/ad2068},
archivePrefix = {arXiv},
       eprint = {2302.12719},
 primaryClass = {astro-ph.SR},
       adsurl = {https://ui.adsabs.harvard.edu/abs/2024ApJ...963..100K}
}

@article{Lallement_2022,
   title={Updated Gaia-2MASS 3D maps of Galactic interstellar dust},
   volume={661},
   ISSN={1432-0746},
   url={http://dx.doi.org/10.1051/0004-6361/202142846},
   DOI={10.1051/0004-6361/202142846},
   journal={Astronomy \& Astrophysics},
   publisher={EDP Sciences},
   author={Lallement, R. and Vergely, J. L. and Babusiaux, C. and Cox, N. L. J.},
   year={2022},
   month=may, pages={A147} }

@ARTICLE{Liu_2021,
       author = {{Liu}, Wei-Min and {Jiang}, Long and {Chen}, Wen-Cong},
        title = "{On the Progenitors of AM CVn Stars as LISA Sources: The Evolved Donor Star Channel}",
      journal = {\apj},
         year = 2021,
        month = mar,
       volume = {910},
       number = {1},
          eid = {22},
        pages = {22},
          doi = {10.3847/1538-4357/abdfc7},
archivePrefix = {arXiv},
       eprint = {2101.09850},
 primaryClass = {astro-ph.SR},
       adsurl = {https://ui.adsabs.harvard.edu/abs/2021ApJ...910...22L}
}

@ARTICLE{Luo_2016,
       author = {{Luo}, Jun and {Chen}, Li-Sheng and {Duan}, Hui-Zong and {Gong}, Yun-Gui and {Hu}, Shoucun and {Ji}, Jianghui and {Liu}, Qi and {Mei}, Jianwei and {Milyukov}, Vadim and {Sazhin}, Mikhail and {Shao}, Cheng-Gang and {Toth}, Viktor T. and {Tu}, Hai-Bo and {Wang}, Yamin and {Wang}, Yan and {Yeh}, Hsien-Chi and {Zhan}, Ming-Sheng and {Zhang}, Yonghe and {Zharov}, Vladimir and {Zhou}, Ze-Bing},
        title = "{TianQin: a space-borne gravitational wave detector}",
      journal = {Classical and Quantum Gravity},
         year = 2016,
        month = feb,
       volume = {33},
       number = {3},
          eid = {035010},
        pages = {035010},
          doi = {10.1088/0264-9381/33/3/035010},
archivePrefix = {arXiv},
       eprint = {1512.02076},
 primaryClass = {astro-ph.IM},
       adsurl = {https://ui.adsabs.harvard.edu/abs/2016CQGra..33c5010L}
}

@ARTICLE{Luo_2021,
       author = {{Luo}, Ziren and {Wang}, Yan and {Wu}, Yueliang and {Hu}, Wenrui and {Jin}, Gang},
        title = "{The Taiji program: A concise overview}",
      journal = {Progress of Theoretical and Experimental Physics},
         year = 2021,
        month = may,
       volume = {2021},
       number = {5},
          eid = {05A108},
        pages = {05A108},
          doi = {10.1093/ptep/ptaa083},
       adsurl = {https://ui.adsabs.harvard.edu/abs/2021PTEP.2021eA108L}
}

@INPROCEEDINGS{Marsh_2007,
       author = {{Marsh}, T.~R. and {Dhillon}, V.~S. and {Littlefair}, S.~P. and {Groot}, P.~J. and {Hakala}, P. and {Nelemans}, G. and {Ramsay}, G. and {Roelofs}, G.~H.~A. and {Steeghs}, D.},
        title = "{SDSS J0926+3624, the First Eclipsing AM CVn Star, as Seen with ULTRACAM}",
    booktitle = {15th European Workshop on White Dwarfs},
         year = 2007,
       editor = {{Napiwotzki}, R. and {Burleigh}, M.~R.},
       series = {Astronomical Society of the Pacific Conference Series},
       volume = {372},
        month = sep,
        pages = {431},
          doi = {10.48550/arXiv.astro-ph/0610414},
archivePrefix = {arXiv},
       eprint = {astro-ph/0610414},
 primaryClass = {astro-ph},
       adsurl = {https://ui.adsabs.harvard.edu/abs/2007ASPC..372..431M}
}

@ARTICLE{Maoz_2014,
       author = {{Maoz}, Dan and {Mannucci}, Filippo and {Nelemans}, Gijs},
        title = "{Observational Clues to the Progenitors of Type Ia Supernovae}",
      journal = {\araa},
         year = 2014,
        month = aug,
       volume = {52},
        pages = {107-170},
          doi = {10.1146/annurev-astro-082812-141031},
archivePrefix = {arXiv},
       eprint = {1312.0628},
 primaryClass = {astro-ph.CO},
       adsurl = {https://ui.adsabs.harvard.edu/abs/2014ARA&A..52..107M}
}

@ARTICLE{Mauche_1988,
       author = {{Mauche}, Christopher W. and {Raymond}, John C. and {Cordova}, France A.},
        title = "{Interstellar Absorption Lines in High-Resolution IUE Spectra of Cataclysmic Variables}",
      journal = {\apj},
         year = 1988,
        month = dec,
       volume = {335},
        pages = {829},
          doi = {10.1086/166971},
       adsurl = {https://ui.adsabs.harvard.edu/abs/1988ApJ...335..829M}
}

@ARTICLE{Munday_2023,
       author = {{Munday}, James and {Marsh}, T.~R. and {Hollands}, Mark and {Pelisoli}, Ingrid and {Steeghs}, Danny and {Hakala}, Pasi and {Breedt}, Elm{\'e} and {Brown}, Alex and {Dhillon}, V.~S. and {Dyer}, Martin J. and {Green}, Matthew and {Kerry}, Paul and {Littlefair}, S.~P. and {Parsons}, Steven G. and {Sahman}, Dave and {Somjit}, Sorawit and {Sukaum}, Boonchoo and {Wild}, James},
        title = "{Two decades of optical timing of the shortest-period binary star system HM Cancri}",
      journal = {\mnras},
         year = 2023,
        month = feb,
       volume = {518},
       number = {4},
        pages = {5123-5139},
          doi = {10.1093/mnras/stac3385},
archivePrefix = {arXiv},
       eprint = {2211.09834},
 primaryClass = {astro-ph.SR},
       adsurl = {https://ui.adsabs.harvard.edu/abs/2023MNRAS.518.5123M}
}

@ARTICLE{Munday_2024,
       author = {{Munday}, James and {Pelisoli}, Ingrid and {Tremblay}, P.-E. and {Marsh}, T.~R. and {Nelemans}, Gijs and {B{\'e}dard}, Antoine and {Toonen}, Silvia and {Breedt}, Elm{\'e} and {Cunningham}, Tim and {O'Brien}, Mairi W. and et al.},
        title = "{The DBL Survey I: discovery of 34 double-lined double white dwarf binaries}",
      journal = {\mnras},
         year = 2024,
        month = aug,
       volume = {532},
       number = {2},
        pages = {2534-2556},
          doi = {10.1093/mnras/stae1645},
archivePrefix = {arXiv},
       eprint = {2407.02594},
 primaryClass = {astro-ph.SR},
       adsurl = {https://ui.adsabs.harvard.edu/abs/2024MNRAS.532.2534M}
}

@ARTICLE{Nelemans_2001,
       author = {{Nelemans}, G. and {Steeghs}, D. and {Groot}, P.~J.},
        title = "{Spectroscopic evidence for the binary nature of AM CVn}",
      journal = {\mnras},
         year = 2001,
        month = sep,
       volume = {326},
       number = {2},
        pages = {621-627},
          doi = {10.1046/j.1365-8711.2001.04614.x},
archivePrefix = {arXiv},
       eprint = {astro-ph/0104220},
 primaryClass = {astro-ph},
       adsurl = {https://ui.adsabs.harvard.edu/abs/2001MNRAS.326..621N}
}

@ARTICLE{Nelemans_2004,
       author = {{Nelemans}, G. and {Yungelson}, L.~R. and {Portegies Zwart}, S.~F.},
        title = "{Short-period AM CVn systems as optical, X-ray and gravitational-wave sources}",
      journal = {\mnras},
         year = 2004,
        month = mar,
       volume = {349},
       number = {1},
        pages = {181-192},
          doi = {10.1111/j.1365-2966.2004.07479.x},
archivePrefix = {arXiv},
       eprint = {astro-ph/0312193},
 primaryClass = {astro-ph},
       adsurl = {https://ui.adsabs.harvard.edu/abs/2004MNRAS.349..181N}
}

@INPROCEEDINGS{Nelemans_2005,
       author = {{Nelemans}, G.},
        title = "{AM CVn stars}",
    booktitle = {The Astrophysics of Cataclysmic Variables and Related Objects},
         year = 2005,
       editor = {{Hameury}, J.-M. and {Lasota}, J.-P.},
       series = {Astronomical Society of the Pacific Conference Series},
       volume = {330},
        month = aug,
        pages = {27},
          doi = {10.48550/arXiv.astro-ph/0409676},
archivePrefix = {arXiv},
       eprint = {astro-ph/0409676},
 primaryClass = {astro-ph},
       adsurl = {https://ui.adsabs.harvard.edu/abs/2005ASPC..330...27N}
}

@ARTICLE{Nelemans_2010,
       author = {{Nelemans}, G. and {Yungelson}, L.~R. and {van der Sluys}, M.~V. and {Tout}, Christopher A.},
        title = "{The chemical composition of donors in AM CVn stars and ultracompact X-ray binaries: observational tests of their formation}",
      journal = {\mnras},
         year = 2010,
        month = jan,
       volume = {401},
       number = {2},
        pages = {1347-1359},
          doi = {10.1111/j.1365-2966.2009.15731.x},
archivePrefix = {arXiv},
       eprint = {0909.3376},
 primaryClass = {astro-ph.SR},
       adsurl = {https://ui.adsabs.harvard.edu/abs/2010MNRAS.401.1347N},
}

@ARTICLE{Paczynski_1967,
       author = {{Paczy{\'n}ski}, B.},
        title = "{Gravitational Waves and the Evolution of Close Binaries}",
      journal = {\actaa},
         year = 1967,
        month = jan,
       volume = {17},
        pages = {287},
       adsurl = {https://ui.adsabs.harvard.edu/abs/1967AcA....17..287P}
}

@ARTICLE{Pala_2017,
       author = {{Pala}, A.~F. and {G{\"a}nsicke}, B.~T. and {Townsley}, D. and {Boyd}, D. and {Cook}, M.~J. and {De Martino}, D. and {Godon}, P. and {Haislip}, J.~B. and {Henden}, A.~A. and {Hubeny}, I. and {Ivarsen}, K.~M. and {Kafka}, S. and {Knigge}, C. and {LaCluyze}, A.~P. and {Long}, K.~S. and {Marsh}, T.~R. and {Monard}, B. and {Moore}, J.~P. and {Myers}, G. and {Nelson}, P. and {Nogami}, D. and {Oksanen}, A. and {Pickard}, R. and {Poyner}, G. and {Reichart}, D.~E. and {Rodriguez Perez}, D. and {Schreiber}, M.~R. and {Shears}, J. and {Sion}, E.~M. and {Stubbings}, R. and {Szkody}, P. and {Zorotovic}, M.},
        title = "{Effective temperatures of cataclysmic-variable white dwarfs as a probe of their evolution}",
      journal = {\mnras},
         year = 2017,
        month = apr,
       volume = {466},
       number = {3},
        pages = {2855-2878},
          doi = {10.1093/mnras/stw3293},
archivePrefix = {arXiv},
       eprint = {1701.02738},
 primaryClass = {astro-ph.SR},
       adsurl = {https://ui.adsabs.harvard.edu/abs/2017MNRAS.466.2855P}
}

@ARTICLE{Pala_2022,
       author = {{Pala}, A.~F. and {G{\"a}nsicke}, B.~T. and {Belloni}, D. and {Parsons}, S.~G. and {Marsh}, T.~R. and {Schreiber}, M.~R. and {Breedt}, E. and {Knigge}, C. and {Sion}, E.~M. and {Szkody}, P. and {Townsley}, D. and {Bildsten}, L. and {Boyd}, D. and {Cook}, M.~J. and {De Martino}, D. and {Godon}, P. and {Kafka}, S. and {Kouprianov}, V. and {Long}, K.~S. and {Monard}, B. and {Myers}, G. and {Nelson}, P. and {Nogami}, D. and {Oksanen}, A. and {Pickard}, R. and {Poyner}, G. and {Reichart}, D.~E. and {Rodriguez Perez}, D. and {Shears}, J. and {Stubbings}, R. and {Toloza}, O.},
        title = "{Constraining the evolution of cataclysmic variables via the masses and accretion rates of their underlying white dwarfs}",
      journal = {\mnras},
         year = 2022,
        month = mar,
       volume = {510},
       number = {4},
        pages = {6110-6132},
          doi = {10.1093/mnras/stab3449},
archivePrefix = {arXiv},
       eprint = {2111.13706},
 primaryClass = {astro-ph.SR},
       adsurl = {https://ui.adsabs.harvard.edu/abs/2022MNRAS.510.6110P}
}

@ARTICLE{Paxton_2011,
       author = {{Paxton}, Bill and {Bildsten}, Lars and {Dotter}, Aaron and {Herwig}, Falk and {Lesaffre}, Pierre and {Timmes}, Frank},
        title = "{Modules for Experiments in Stellar Astrophysics (MESA)}",
      journal = {\apjs},
         year = 2011,
        month = jan,
       volume = {192},
       number = {1},
          eid = {3},
        pages = {3},
          doi = {10.1088/0067-0049/192/1/3},
archivePrefix = {arXiv},
       eprint = {1009.1622},
 primaryClass = {astro-ph.SR},
       adsurl = {https://ui.adsabs.harvard.edu/abs/2011ApJS..192....3P}
}

@ARTICLE{Paxton_2013,
       author = {{Paxton}, Bill and {Cantiello}, Matteo and {Arras}, Phil and {Bildsten}, Lars and {Brown}, Edward F. and {Dotter}, Aaron and {Mankovich}, Christopher and {Montgomery}, M.~H. and {Stello}, Dennis and {Timmes}, F.~X. and {Townsend}, Richard},
        title = "{Modules for Experiments in Stellar Astrophysics (MESA): Planets, Oscillations, Rotation, and Massive Stars}",
      journal = {\apjs},
         year = 2013,
        month = sep,
       volume = {208},
       number = {1},
          eid = {4},
        pages = {4},
          doi = {10.1088/0067-0049/208/1/4},
archivePrefix = {arXiv},
       eprint = {1301.0319},
 primaryClass = {astro-ph.SR},
       adsurl = {https://ui.adsabs.harvard.edu/abs/2013ApJS..208....4P}
}

@ARTICLE{Paxton_2015,
       author = {{Paxton}, Bill and {Marchant}, Pablo and {Schwab}, Josiah and {Bauer}, Evan B. and {Bildsten}, Lars and {Cantiello}, Matteo and {Dessart}, Luc and {Farmer}, R. and {Hu}, H. and {Langer}, N. and {Townsend}, R.~H.~D. and {Townsley}, Dean M. and {Timmes}, F.~X.},
        title = "{Modules for Experiments in Stellar Astrophysics (MESA): Binaries, Pulsations, and Explosions}",
      journal = {\apjs},
         year = 2015,
        month = sep,
       volume = {220},
       number = {1},
          eid = {15},
        pages = {15},
          doi = {10.1088/0067-0049/220/1/15},
archivePrefix = {arXiv},
       eprint = {1506.03146},
 primaryClass = {astro-ph.SR},
       adsurl = {https://ui.adsabs.harvard.edu/abs/2015ApJS..220...15P}
}

@ARTICLE{Paxton_2018,
       author = {{Paxton}, Bill and {Schwab}, Josiah and {Bauer}, Evan B. and {Bildsten}, Lars and {Blinnikov}, Sergei and {Duffell}, Paul and {Farmer}, R. and {Goldberg}, Jared A. and {Marchant}, Pablo and {Sorokina}, Elena and {Thoul}, Anne and {Townsend}, Richard H.~D. and {Timmes}, F.~X.},
        title = "{Modules for Experiments in Stellar Astrophysics (MESA): Convective Boundaries, Element Diffusion, and Massive Star Explosions}",
      journal = {\apjs},
         year = 2018,
        month = feb,
       volume = {234},
       number = {2},
          eid = {34},
        pages = {34},
          doi = {10.3847/1538-4365/aaa5a8},
archivePrefix = {arXiv},
       eprint = {1710.08424},
 primaryClass = {astro-ph.SR},
       adsurl = {https://ui.adsabs.harvard.edu/abs/2018ApJS..234...34P}
}

@ARTICLE{Paxton_2019,
       author = {{Paxton}, Bill and {Smolec}, R. and {Schwab}, Josiah and {Gautschy}, A. and {Bildsten}, Lars and {Cantiello}, Matteo and {Dotter}, Aaron and {Farmer}, R. and {Goldberg}, Jared A. and {Jermyn}, Adam S. and {Kanbur}, S.~M. and {Marchant}, Pablo and {Thoul}, Anne and {Townsend}, Richard H.~D. and {Wolf}, William M. and {Zhang}, Michael and {Timmes}, F.~X.},
        title = "{Modules for Experiments in Stellar Astrophysics (MESA): Pulsating Variable Stars, Rotation, Convective Boundaries, and Energy Conservation}",
      journal = {\apjs},
         year = 2019,
        month = jul,
       volume = {243},
       number = {1},
          eid = {10},
        pages = {10},
          doi = {10.3847/1538-4365/ab2241},
archivePrefix = {arXiv},
       eprint = {1903.01426},
 primaryClass = {astro-ph.SR},
       adsurl = {https://ui.adsabs.harvard.edu/abs/2019ApJS..243...10P}
}

@ARTICLE{Pichardo_Marcano_2021,
       author = {{Pichardo Marcano}, Manuel and {Rivera Sandoval}, Liliana E. and {Maccarone}, Thomas J. and {Scaringi}, Simone},
        title = "{TACOS: TESS AM CVn Outbursts Survey}",
      journal = {\mnras},
         year = 2021,
        month = dec,
       volume = {508},
       number = {3},
        pages = {3275-3289},
          doi = {10.1093/mnras/stab2685},
archivePrefix = {arXiv},
       eprint = {2106.15104},
 primaryClass = {astro-ph.HE},
       adsurl = {https://ui.adsabs.harvard.edu/abs/2021MNRAS.508.3275P}
}

@ARTICLE{Podsiadlowski_2002,
       author = {{Podsiadlowski}, Ph. and {Rappaport}, S. and {Pfahl}, E.~D.},
        title = "{Evolutionary Sequences for Low- and Intermediate-Mass X-Ray Binaries}",
      journal = {\apj},
         year = 2002,
        month = feb,
       volume = {565},
       number = {2},
        pages = {1107-1133},
          doi = {10.1086/324686},
archivePrefix = {arXiv},
       eprint = {astro-ph/0107261},
 primaryClass = {astro-ph},
       adsurl = {https://ui.adsabs.harvard.edu/abs/2002ApJ...565.1107P}
}

@ARTICLE{Podsiadlowski_2003,
       author = {{Podsiadlowski}, Ph. and {Han}, Z. and {Rappaport}, S.},
        title = "{Cataclysmic variables with evolved secondaries and the progenitors of AM CVn stars}",
      journal = {\mnras},
         year = 2003,
        month = apr,
       volume = {340},
       number = {4},
        pages = {1214-1228},
          doi = {10.1046/j.1365-8711.2003.06380.x},
       adsurl = {https://ui.adsabs.harvard.edu/abs/2003MNRAS.340.1214P}
}

@ARTICLE{Postnov_2014,
       author = {{Postnov}, Konstantin A. and {Yungelson}, Lev R.},
        title = "{The Evolution of Compact Binary Star Systems}",
      journal = {Living Reviews in Relativity},
         year = 2014,
        month = dec,
       volume = {17},
       number = {1},
          eid = {3},
        pages = {3},
          doi = {10.12942/lrr-2014-3},
archivePrefix = {arXiv},
       eprint = {1403.4754},
 primaryClass = {astro-ph.HE},
       adsurl = {https://ui.adsabs.harvard.edu/abs/2014LRR....17....3P}
}

@ARTICLE{Predehl_2021,
       author = {{Predehl}, P. and {Andritschke}, R. and {Arefiev}, V. and {Babyshkin}, V. and {Batanov}, O. and {Becker}, W. and {B{\"o}hringer}, H. and {Bogomolov}, A. and {Boller}, T. and {Borm}, K. and {Bornemann}, W. and {Br{\"a}uninger}, H. and {Br{\"u}ggen}, M. and {Brunner}, H. and {Brusa}, M. and {Bulbul}, E. and {Buntov}, M. and {Burwitz}, V. and {Burkert}, W. and {Clerc}, N. and {Churazov}, E. and {Coutinho}, D. and {Dauser}, T. and {Dennerl}, K. and {Doroshenko}, V. and {Eder}, J. and {Emberger}, V. and {Eraerds}, T. and {Finoguenov}, A. and {Freyberg}, M. and {Friedrich}, P. and {Friedrich}, S. and {F{\"u}rmetz}, M. and {Georgakakis}, A. and {Gilfanov}, M. and {Granato}, S. and {Grossberger}, C. and {Gueguen}, A. and {Gureev}, P. and {Haberl}, F. and {H{\"a}lker}, O. and {Hartner}, G. and {Hasinger}, G. and {Huber}, H. and {Ji}, L. and {Kienlin}, A. v. and {Kink}, W. and {Korotkov}, F. and {Kreykenbohm}, I. and {Lamer}, G. and {Lomakin}, I. and {Lapshov}, I. and {Liu}, T. and {Maitra}, C. and {Meidinger}, N. and {Menz}, B. and {Merloni}, A. and {Mernik}, T. and {Mican}, B. and {Mohr}, J. and {M{\"u}ller}, S. and {Nandra}, K. and {Nazarov}, V. and {Pacaud}, F. and {Pavlinsky}, M. and {Perinati}, E. and {Pfeffermann}, E. and {Pietschner}, D. and {Ramos-Ceja}, M.~E. and {Rau}, A. and {Reiffers}, J. and {Reiprich}, T.~H. and {Robrade}, J. and {Salvato}, M. and {Sanders}, J. and {Santangelo}, A. and {Sasaki}, M. and {Scheuerle}, H. and {Schmid}, C. and {Schmitt}, J. and {Schwope}, A. and {Shirshakov}, A. and {Steinmetz}, M. and {Stewart}, I. and {Str{\"u}der}, L. and {Sunyaev}, R. and {Tenzer}, C. and {Tiedemann}, L. and {Tr{\"u}mper}, J. and {Voron}, V. and {Weber}, P. and {Wilms}, J. and {Yaroshenko}, V.},
        title = "{The eROSITA X-ray telescope on SRG}",
      journal = {\aap},
         year = 2021,
        month = mar,
       volume = {647},
          eid = {A1},
        pages = {A1},
          doi = {10.1051/0004-6361/202039313},
archivePrefix = {arXiv},
       eprint = {2010.03477},
 primaryClass = {astro-ph.HE},
       adsurl = {https://ui.adsabs.harvard.edu/abs/2021A&A...647A...1P}
}

@ARTICLE{Smak_1967,
       author = {{Smak}, J.},
        title = "{Light variability of HZ 29.}",
      journal = {\actaa},
         year = 1967,
        month = jan,
       volume = {17},
        pages = {255},
       adsurl = {https://ui.adsabs.harvard.edu/abs/1967AcA....17..255S}
}

\appendix
\section{Comparison of YZ\,LMi with the DA star WD\,0920$+$363}\label{app:spectral_diagnostics}

\begin{figure}
        \centering
        \includegraphics[width=\columnwidth]{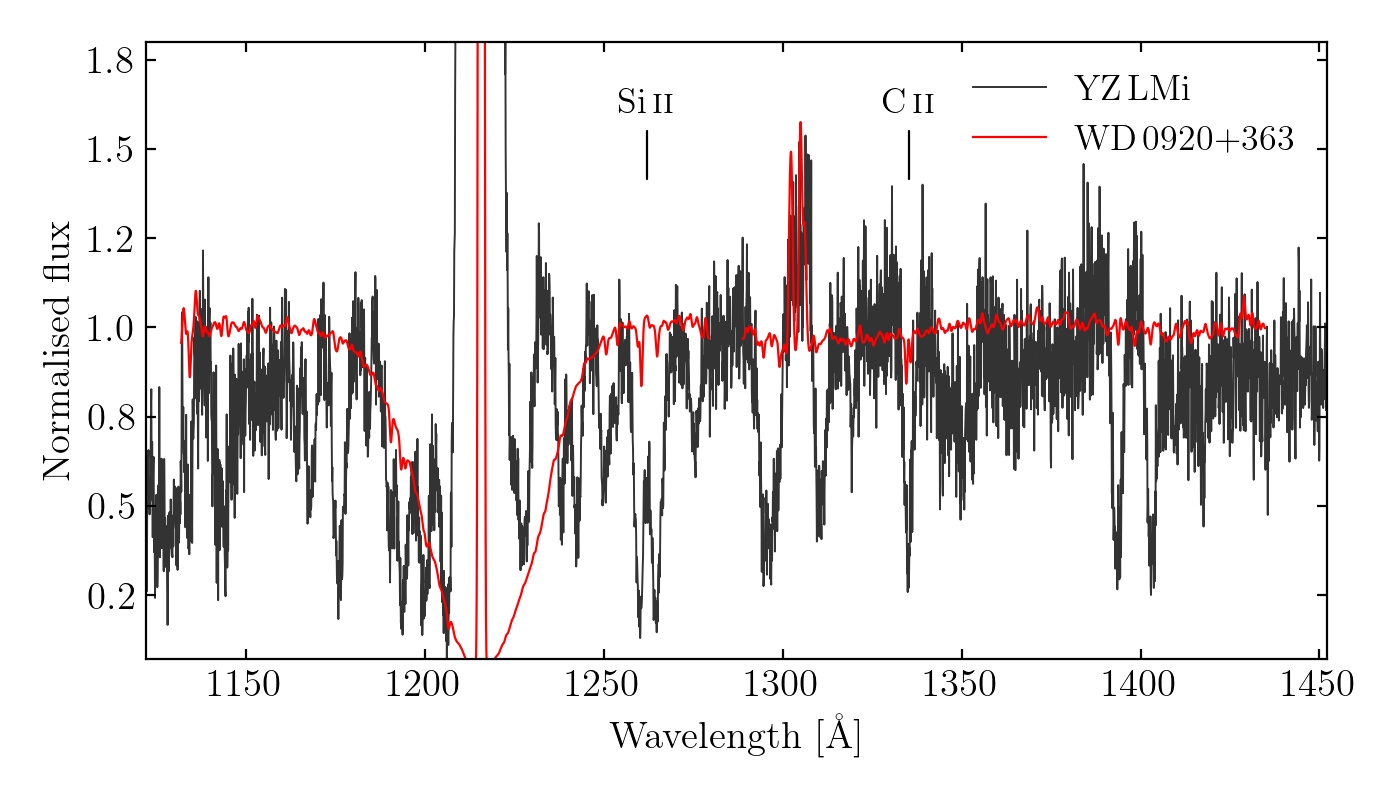}
        \caption{Spectral comparison of YZ\,LMi and WD\,0920$+$363. Normalised spectra of YZ\,LMi (black) and WD\,0920$+$363 (red). WD\,0920$+$363 has a pure-hydrogen atmosphere and lies $36\arcmin$ from YZ\,LMi on the sky. Its spectrum was obtained with COS/G130M and degraded to the spectral resolution of COS/G140L to match the YZ\,LMi observations. Candidate ISM absorption features of \ion{Si}{ii}~1259/1264\,\AA\ and \ion{C}{ii}~1335\,\AA\ are labelled.}
        \label{fig:spectra_compare}
    \end{figure}

    To identify interstellar absorption, we compared the \textit{HST}/COS spectra with published UV ISM line lists \citep[e.g.][]{Mauche_1988,Danforth_2016}. To further assess the contribution of interstellar absorption, we searched the \textit{HST} archive for nearby hydrogen-atmosphere (DA) WDs. Figure~\ref{fig:spectra_compare} compares the spectrum of YZ\,LMi with that of WD\,0920$+$363, located at RA $=09{:}23{:}57.21$, Dec $=+36{:}08{:}37.23$, at a distance of $183.0 \pm 4.0$\,pc \citep{Bailer-Jones_2021}. WD\,0920$+$363 is the closest DA WD with a nearly pure-hydrogen atmosphere to YZ\,LMi on the sky for which archival UV spectroscopy is available, with an angular separation of $36\arcmin$ \citep{Sahu_2023}. It was observed with COS using the G130M grating for 1230\,s during Cycle~19 (Proposal ID 12474; PI: B.~G\"ansicke). For comparison, its spectrum was degraded to the spectral resolution of the G140L observations.

    The spectrum of WD\,0920$+$363 shows absorption from \ion{C}{ii} near 1335\,\AA\ and the \ion{Si}{ii} doublet near 1260 and 1265\,\AA. The weaker, redder component near 1265\,\AA\ is not commonly included among the prominent UV ISM features \citep{Mauche_1988, Danforth_2016}. However, its lower-level excitation energy is only $E_i \simeq 0.036$\,eV, corresponding to a characteristic excitation temperature of approximately 400\,K, and the transition may therefore arise in the warm ISM. Because YZ\,LMi is also substantially more distant than WD\,0920$+$363, we classify the \ion{Si}{ii} feature near 1265\,\AA\ as potentially contaminated by ISM absorption. Consistent with its greater distance, YZ\,LMi also shows more interstellar absorption features (Table~\ref{tab:line_id}). No suitable comparison target with UV spectroscopy was found within $100\arcmin$ of ZTF\,J0407$-$00. Nevertheless, given their similar distances, we expect comparable contributions from interstellar absorption and therefore consider the same potential contamination of the carbon and silicon features in ZTF\,J0407$-$00.

\section{Identified ultraviolet absorption features and their origins}\label{app:uv_features}

    Table~\ref{tab:line_id} lists the absorption features identified in the UV spectra of YZ\,LMi and ZTF\,J0407$-$00, together with their likely origins. We distinguish features associated with the white dwarf photosphere from those that may arise from, or be contaminated by, absorption in the interstellar medium.

    \begin{table}
        \caption{Identified ultraviolet absorption features in the spectra of YZ\,LMi and ZTF\,J0407$-$00, together with their likely origins.}
        \label{tab:line_id}
        \centering
        \normalsize
        \begin{tabular}{lll|lll}
            \toprule
            Ion & $\lambda$ [$\AA$] & Origin & Ion & $\lambda$ [$\AA$] & Origin\\
            \midrule
            \ion{Si}{III} & 1110        & ?    & \ion{Si}{II}  & 1350        & WD   \\
            \ion{Fe}{III} & 1128        & ?    & \ion{Si}{IV}  & 1393        & ISM  \\
            \ion{N}{I}    & 1134        & ISM  & \ion{Si}{IV}  & 1403        & ISM  \\
            \ion{Fe}{II}  & 1142        & ISM? & \ion{Fe}{II}  & 1460$^{*}$  & WD   \\
            \ion{Fe}{II}  & 1150$^{*}$  & WD   & \ion{Si}{II}  & 1485        & WD   \\
            \ion{N}{I}    & 1164        & WD   & \ion{N}{I}    & 1493        & WD   \\
            \ion{N}{I}    & 1168        & WD   & \ion{Si}{II}  & 1509        & WD   \\
            \ion{C}{III}  & 1175        & ?    & \ion{Si}{II}  & 1512        & WD   \\
            \ion{Si}{II}  & 1190        & ISM  & \ion{Si}{II}  & 1526        & ISM  \\
            \ion{Si}{II}  & 1193        & ISM  & \ion{Si}{II}  & 1533        & ISM  \\
            \ion{N}{V}    & 1239        & ISM? & \ion{C}{IV}   & 1550        & ISM  \\
            \ion{N}{V}    & 1242        & ISM? & \ion{Fe}{II}  & 1580$^{*}$  & WD   \\
            \ion{S}{II}   & 1251        & ISM  & \ion{Fe}{II}  & 1608        & ISM  \\
            \ion{S}{II}   & 1254        & ISM  & \ion{Fe}{II}  & 1611        & ISM  \\
            \ion{Si}{II}  & 1260        & ISM  & \ion{He}{II}  & 1640        & ?    \\
            \ion{Si}{II}  & 1265        & ISM? & \ion{C}{I}    & 1657        & ISM? \\
            \ion{Fe}{II}  & 1282$^{*}$  & WD   & \ion{Fe}{II}  & 1690$^{*}$  & WD   \\
            \ion{O}{I}    & 1302        & ISM  & \ion{Al}{II}  & 1723        & ISM  \\
            \ion{N}{I}    & 1319        & WD   & \ion{N}{I}    & 1744        & WD   \\
            \ion{C}{II}   & 1335        & ISM  & \ion{Al}{II}  & 1765        & ISM  \\
            \bottomrule
        \end{tabular}
        \tablefoot{($^{*}$) indicates a blend of multiple lines. Features marked "WD" are of WD photospheric origin, while those marked "ISM(?)" are of possible interstellar origin or may be contaminated by interstellar absorption. The identification of ISM features is based on \citet{Mauche_1988} and \citet{Danforth_2016}. Features marked with a question mark are unidentified non-photospheric lines and are discussed in Sect.~\ref{sub:line_identification}.
        }
    \end{table}

\section{MCMC analysis}\label{app:mcmc}

\begin{figure*}
        \centering
        \includegraphics[width=0.45\textwidth]{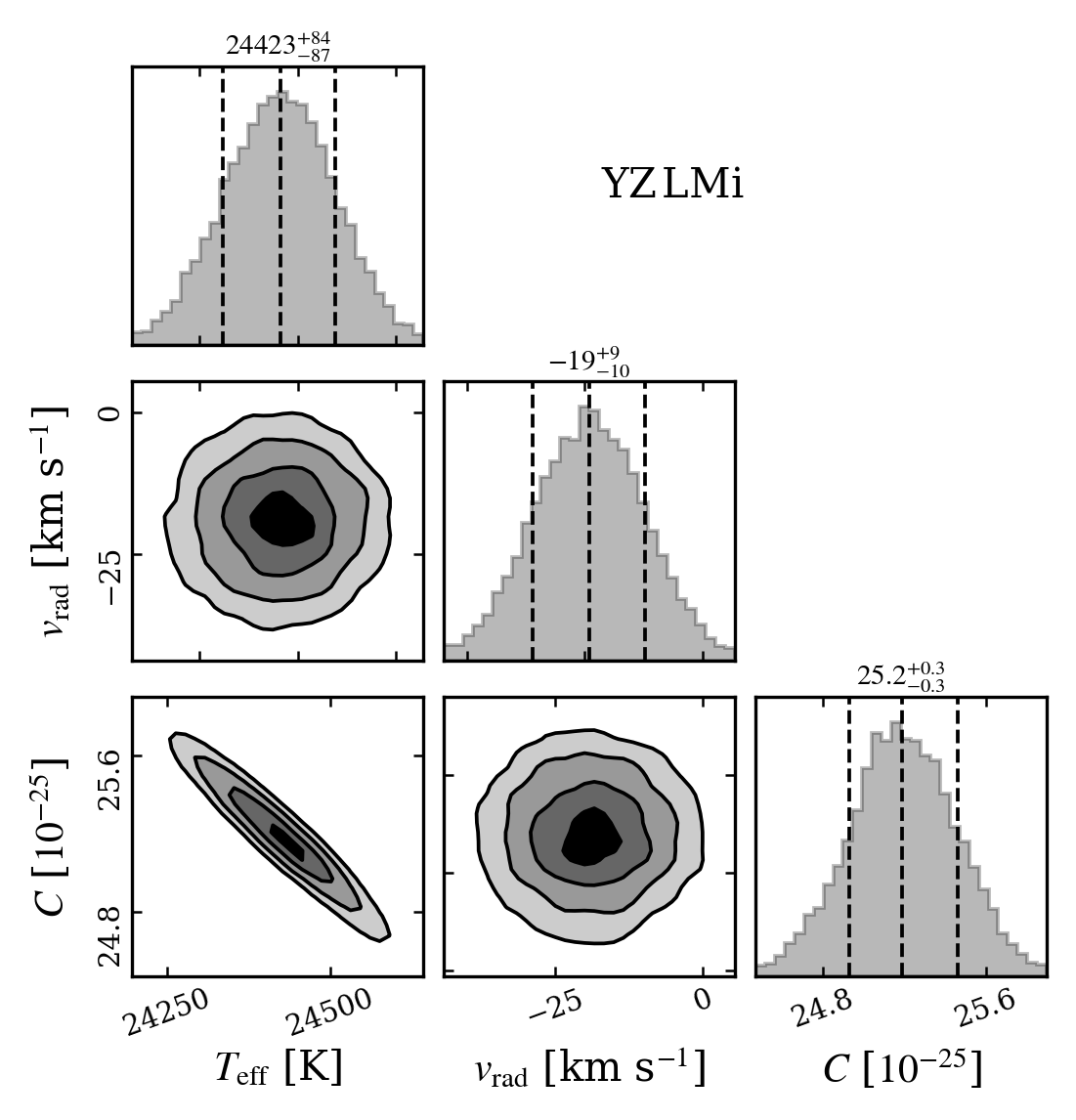}
        \includegraphics[width=0.45\textwidth]{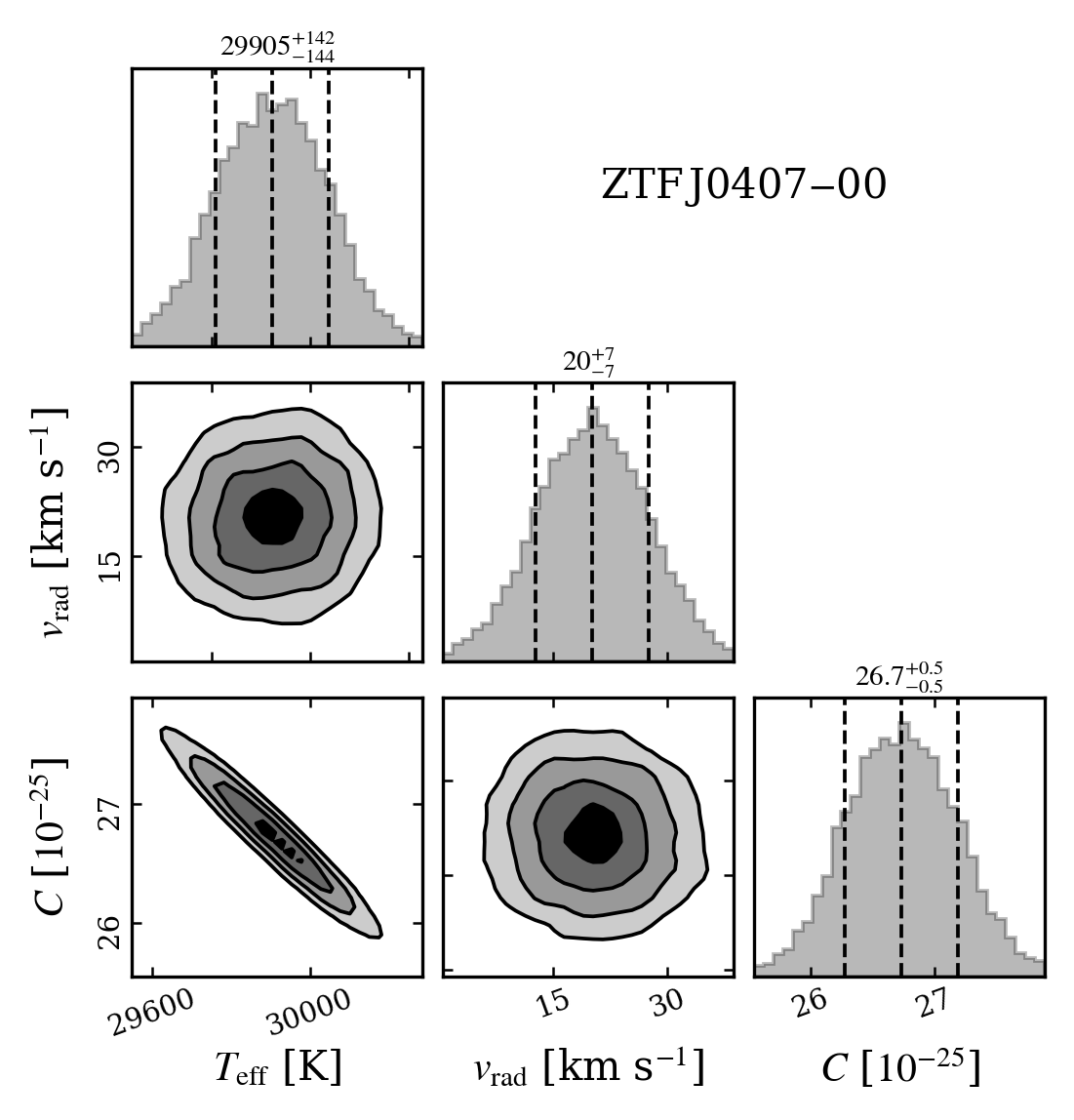}
        \caption{Posterior distributions from the final MCMC iteration of the global spectral fit for YZ\,LMi (left) and ZTF\,J0407$-$00 (right).}
        \label{fig:global_corner_maps}
    \end{figure*}

    The spectral fits described in Sect.~\ref{sub:spectral_fit} were performed using the Python package \texttt{emcee}, which implements an affine-invariant Markov chain Monte Carlo (MCMC) sampler \citep{Foreman-Macke_2013}. We adopted uniform priors of $10\,000 < T_{\mathrm{eff}} < 40\,000$\,K, $-10^3 < v_{\mathrm{rad}} < 10^3$\,km\,s$^{-1}$, $C > 0$, and $-7 < X_i < -2$ for each identified element. Each MCMC run used at least 64 walkers and 4000 steps. The posterior distributions from the final iteration of the global fit for both systems are shown in Fig.~\ref{fig:global_corner_maps}.

\end{document}